\documentclass[%
 reprint,
superscriptaddress,
amsmath,amssymb,
]{revtex4-2}
\usepackage{xcolor}
\usepackage{graphicx}% Include figure files
\usepackage{dcolumn}% Align table columns on decimal point
\usepackage{bm}% bold math
\usepackage{lipsum}
\usepackage{floatrow}

\usepackage{xcolor}
\usepackage{braket}
\usepackage{graphicx}% Include figure files
\graphicspath{{FinalFigs/}} % Path to figure files
\usepackage{dcolumn}% Align table columns on decimal point
\usepackage{bm}% bold math
\usepackage{siunitx}

\expandafter\let\csname equation*\endcsname\relax
\expandafter\let\csname endequation*\endcsname\relax
\usepackage{amsmath}
\usepackage{bbm}
\usepackage{mathtools}
\usepackage{gensymb}
\usepackage{amssymb}

\usepackage[breaklinks=true,colorlinks=true,linkcolor=blue,urlcolor=blue,citecolor=blue]{hyperref}
\usepackage{dcolumn}
\usepackage{bm}
\usepackage{float}
\usepackage[T1]{fontenc}
\usepackage[utf8]{inputenc}

\DeclareMathSymbol{\shortminus}{\mathbin}{AMSa}{"39}

\def \siExperimentalControl{I}
\def \siElectronQubit{II}
\def \siHamiltonian{III}
\def \siDetectionNuclear{IV} 
\def \siNucGates{V}
\def \siGenerality{VI}
\def \siMagneticField{VII}
\def \siStandardCircuits{VIII}
\def \siMQC{IX}
\def \siEntangleGateFid{X}
\def \siQEC{XI}

\begin{document}

% Possible title ideas:
%\title{Single gate, multipartite entanglement of diamond quantum registers}
%\title{Single gate, multipartite entanglement of room temperature diamond quantum register}
%\title{Parallelized entanglement of room temperature diamond quantum register}

% Other title suggestions:
% Efficient, multipartite entanglement generation in room-temperature quantum registers
\title{Single-gate, multipartite entanglement on a room-temperature quantum register} 
% Controlling multi-qubit entanglement in room-temperature quantum registers

\author{Joseph D. Minnella}
\affiliation{ 
Quantum Engineering Laboratory, Department of Electrical and Systems Engineering, University of Pennsylvania, 200 S. 33rd St. Philadelphia, Pennsylvania, 19104, USA
}
\affiliation{
Department of Physics and Astronomy, University of Pennsylvania, 209 S. 33rd St. Philadelphia, Pennsylvania 19104, USA
}

\author{Mathieu Ouellet}
\altaffiliation[Present address: ]{School of Applied and Engineering Physics, Cornell University, Ithaca, NY 14853, USA}
\affiliation{ 
Quantum Engineering Laboratory, Department of Electrical and Systems Engineering, University of Pennsylvania, 200 S. 33rd St. Philadelphia, Pennsylvania, 19104, USA
}

\author{Amelia R. Klein}
\affiliation{ 
Quantum Engineering Laboratory, Department of Electrical and Systems Engineering, University of Pennsylvania, 200 S. 33rd St. Philadelphia, Pennsylvania, 19104, USA
}

\author{Lee C. Bassett}
\email[Correspondence to: ]{lbassett@seas.upenn.edu}
\affiliation{ 
Quantum Engineering Laboratory, Department of Electrical and Systems Engineering, University of Pennsylvania, 200 S. 33rd St. Philadelphia, Pennsylvania, 19104, USA
}

\begin{abstract} 
Multipartite entanglement is an essential aspect of quantum systems, needed to execute quantum algorithms, implement error correction, and achieve quantum-enhanced sensing. 
In solid-state quantum registers such nitrogen-vacancy (NV) centers in diamond, entangled states are typically created using sequential, pairwise gates between the central electron and individual nuclear qubits.
This sequential approach is slow and suffers from crosstalk errors.
Here, we demonstrate a parallelized multi-qubit entangling gate to generate a four-qubit GHZ state using a room-temperature NV center in only $14.8\mu$s --- 10 times faster than using sequences of two-qubit gates and close to the fundamental limit set by the hyperfine coupling frequencies.
Parallel three-qubit gates are also realized with all nuclear-qubit subsets.
The entangled states are verified by measuring multiple quantum coherences. 
% Two-qubit entangling gates have an average fidelity of 0.96(1), and the four-qubit parallel gate has a fidelity of 0.92(4), whereas the sequential four-qubit gate fidelity is only 0.69(3).
The four-qubit parallel gate has a fidelity of 0.92(4), whereas the sequential four-qubit gate fidelity is only 0.69(3).
The approach is generalizable to other solid-state platforms, and it lays the foundation for scalable generation and control of entanglement in practical devices.
\end{abstract}

\maketitle

%==================================================================
% Introduction
%==================================================================

Quantum registers in solid-state materials are the basis for quantum networking \cite{pompili2021realization, hermans2022qubit, knaut2024entanglement, chang2025hybrid, bersin2024telecom, stas2022robust, rugar2021quantum}, quantum information processing \cite{van2012decoherence, rong2015experimental, bradley2019ten, abobeih2022fault, nagy2019high, bourassa2020entanglement, harris2025high, beukers2025control, waldherr2014quantum,reiner2024high, thorvaldson2025grover, edlbauer202511} and quantum sensing \cite{du2024single, abobeih2019atomic, van2024mapping, breitweiser2024quadrupolar, ajoy2015atomic, zaiser2016enhancing, degen2017quantum, xie2021beating}.
The nitrogen-vacancy (NV) center in diamond is the most well-established solid-state quantum register, and it is especially notable for featuring room-temperature spin coherence.  
The central electron spin, initialized and measured optically, serves as an interface for detecting and controlling surrounding spins, including $^{13}$C nuclei \cite{taminiau2012detection, taminiau2014universal, cujia2022parallel}.
Owing to their small gyromagnetic ratios, nuclear spins interact weakly with their environment and have long coherence times, making them attractive memory qubits. 
Generating high-fidelity entanglement between the electron and multiple nuclei is essential for error correction \cite{taminiau2014universal,abobeih2022fault} and enhanced sensing beyond the standard quantum limit \cite{zaiser2016enhancing, degen2017quantum, xie2021beating}.

To extend the electron spin coherence time to be comparable to that of nuclear qubits, it must be decoupled from its environment; this is achieved using dynamical decoupling (DD) control sequences. 
Crucially, DD sequences can be designed to initialize, control, and read out individual nuclear qubits, all while decoupling the electron from other noise sources \cite{van2012decoherence, taminiau2014universal}.
At cryogenic temperatures, where coherent optical transitions facilitate high-fidelity readout and spin lifetimes extend beyond one minute \cite{bradley2019ten, bartling2022entanglement}, DD sequences have facilitated the realization of quantum networking nodes \cite{pompili2021realization, hermans2022qubit, knaut2024entanglement, chang2025hybrid} and preliminary fault tolerance using nuclear quantum registers \cite{abobeih2022fault}.
At room temperature, where readout is less efficient \cite{hopper2018spin} and coherence times are shorter, DD extends coherence times and enables entanglement-assisted enhanced sensing protocols suitable for practical use cases \cite{xie2021beating, kenny2025quantum}.
The performance of such networking, computing, and sensing schemes remains limited by the speed and fidelity of generating multipartite entangled states.

Using DD sequences to control multipartite entanglement presents two critical challenges. 
First, each DD sequence is calibrated to generate bipartite entanglement with a single nucleus, thus requiring long, sequential gates to entangle multiple nuclear qubits.
Second, due to the always-on spin-spin interactions within the quantum register, executing sequences of nuclear gates leads to unwanted rotations of all other qubits (\textit{i.e.}, crosstalk).
Parallelized entangling gates address both of these issues.
Recent approaches in trapped ion and neutral atom platforms include specialized multi-qubit gates \cite{levine2019parallel, lu2019global, evered2023high} and the simultaneous implementation of multiple two-qubit gates \cite{zhu2023pairwise, grzesiak2020efficient, figgatt2019parallel}. 
% For solid-state registers operating in the strong-coupling regime (\textit{i.e.}, with hyperfine coupling frequencies larger than $1/T_2^\ast$, where $T_2^\ast$ is the electron dephasing time), multi-qubit entangling gates can be achieved using frequency-selective conditional electron pulses \cite{waldherr2014quantum, reiner2024high, thorvaldson2025grover, edlbauer202511}.
For solid-state registers operating in the strong-coupling regime (\textit{i.e.}, with hyperfine coupling frequencies larger than the electron spin's inhomogeneous linewidth), multi-qubit entangling gates can be achieved using frequency-selective conditional electron pulses \cite{waldherr2014quantum, reiner2024high, thorvaldson2025grover, edlbauer202511}.
%However, strongly-coupled registers are rare in nature and their scalability is limited by the need for strong, widely spaced couplings to a finite set of nuclei. 
However, strongly-coupled registers are rare in nature,
so they either need to be engineered \cite{reiner2024high,thorvaldson2025grover,edlbauer202511} or identified through screening \cite{waldherr2014quantum}. 
Their scalability is limited by the need for strong, widely spaced couplings for individual addressability. 
%In the case of typical, weakly-coupled registers, the sequential application of long DD sequences leads to impractically long gate times for multi-qubit control.
Weakly-coupled registers typically offer larger register sizes that enable more robust error correction and networking capabilities \cite{bradley2019ten, abobeih2022fault, chang2025hybrid}, but the sequential application of DD sequences leads to impractical gate times and errors for multi-qubit control.

In this work, we design, implement, and benchmark a multi-qubit gate using a single DD sequence that facilitates the efficient generation of multipartite entanglement, even at room temperature.
We use the entanglement metric framework developed by Takou \textit{et al.,} \cite{takou2023precise, takou2024generation} to design DD sequences that address multiple weakly coupled $^{13}$C nuclear qubits surrounding an NV center (Fig. \ref{fig:over_view}a). 
Essentially, this approach leverages the crosstalk inherent to DD sequences to simultaneously create conditional rotations of multiple nuclear qubits, where each is locally equivalent to a CNOT gate.
We experimentally generate four-qubit Greenberger–Horne–Zeilinger (GHZ) entangled states that include the electron and three weakly coupled $^{13}$C nuclear qubits, and we verify them by measuring multiple quantum coherences (MQC).
The largest parallel gates are an order of magnitude faster, and significantly higher fidelity, than their sequential counterparts.

%==================================================================
\section{Multipartite Entanglement with Dynamical Decoupling}\label{sec:Ent_w_DD}
%==================================================================

\begin{figure}[th!]
\centering
\includegraphics{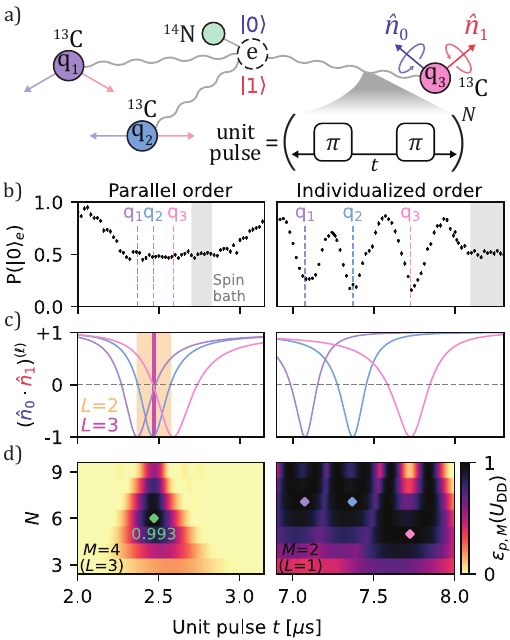}
\caption{
\textbf{Entanglement through dynamical decoupling in NV quantum registers} 
\textbf{(a)} Schematic of the NV center including nearby $^{13}$C nuclear qubits $q_{\ell}$. 
Electron-nuclear entangling gates were implemented with XY8 DD, symbolically shown in the inset. 
Entangling gates rotate nuclear qubits about distinct axes $\hat{n}_{0}$ (dark blue) and $\hat{n}_{1}$ (red), conditioned on the two states of the electron.
\textbf{(b)} DD spectroscopy measurements at $k=1,N=6$ and $k=2,N=12$; resonances associated with nuclear qubits $q_1$, $q_2$, and $q_3$ are marked with dashed lines. 
Error bars on points denote photon shot noise.
% At first ``parallel'' order, the resonances are unresolved due to overlap between the qubits and with the spin bath, while at second ``individualized'' order, the resonances are resolved.
% These spectra are acquired with $N=6$ and 12, respectively.
Additional spectroscopy data can be found in SI Sec.~\siDetectionNuclear.
\textbf{(c)} Alignment of the electron-spin-dependent nuclear rotation axes, where $-1$ ($+1$) indicates perfectly (un)conditional rotations. 
Orange and purple shaded regions indicate the intersection of unit pulse times where $(\hat{n}_0 \cdot \hat{n}_1)^{(\ell)}<0$ for two and three nuclear qubits, respectively.
\textbf{(d)} $M$-qubit entangling power metric $\varepsilon_{p,M}$ as a function of DD unit pulse parameters. 
At first-order, the maximum in $\varepsilon_{p,4}$ (green diamond) indicates an optimal DD sequence to create maximal four-way entanglement.
At second-order, diamond markers indicate the optimum parameters for two-qubit entangling gates, as the maximum value of each nuclear qubit's bipartite entangling power, $\text{max}_{\ell}(\varepsilon_{p,2})$.
}
\label{fig:over_view}
\end{figure}

DD control sequences consist of $\pi$-pulses applied to the electron, interleaved with precisely-timed free-evolution periods. 
Typical DD protocols are constructed using unit sequences where an even number of $\pi$-pulses are spaced by an equal delay, $t/2$, and the unit is repeated $N$ times (Fig. \ref{fig:over_view}a).
Due to the weak hyperfine interactions, individual nuclear qubits can be coherently controlled through the calibration of $t$ and $N$ \cite{van2012decoherence, taminiau2014universal}. 
Conceptually, $t$ sets the period of electron state flips; when chosen to be in resonance with a target nuclear qubit, the DD sequence creates $X$-axis nuclear rotations.
The resonance order, $k$, corresponds to the number of nuclear precession rotations in each period.
As $k$ increases, resonances separate due to differences in their hyperfine coupling \cite{taminiau2012detection}. 
Depending on the type of resonance, the nuclear rotations can be conditioned on the electron's spin state, allowing for two-qubit entangling gates, or they can be unconditional, allowing for local gates in the register.
% When the sequence is far from resonance, the rotation is about the $Z$ axis.

DD sequences can also be used to sense nearby nuclear qubits \cite{taminiau2012detection, abobeih2019atomic, van2024mapping, breitweiser2024quadrupolar}. 
In DD spectroscopy, the electron is initialized to an equal superposition state, and sequences with varying $t$ are applied to identify resonances associated with conditional nuclear rotations.
The equation used to determine these resonance times can be found in Methods \S\ref{sec:methods_DD}.
This is typically performed at $k\geq2$ so that individual resonances are resolved and can be fit to extract hyperfine couplings (see Fig. \ref{fig:over_view}b, ``individualized order,'' corresponding to $k=2$).
At first order (Fig. \ref{fig:over_view}b, ``parallel order,'' corresponding to $k=1$), the nuclear qubits' resonances generally overlap with each other and with the spin bath. 
For these reasons, first-order resonances are usually overlooked, yet this region is ideal for realizing parallelized entangling gates.

The nature of DD spectroscopy resonances, and correspondingly the associated DD entangling gates, is captured by the relative orientation of the electron-state-dependent nuclear rotation axes, $\hat{n}_{m_s}^{(\ell)}$.
Traditionally, entangling gates are achieved by tuning $t$ to a $k$th-order resonance where the axes for $\ell$th nuclear qubit are antiparallel, or $(\hat{n}_0 \cdot \hat{n}_1)^{(\ell)}=-1$ (Fig. \ref{fig:over_view}c).
With $N$ chosen to set the rotation angle to $\pi/2$, the two-qubit entangling gate takes the form $C_{e}X_{\ell}(\pm \pi/2)=\ket{0}\bra{0}_e\otimes X_{\ell}(\pi/2)+\ket{1}\bra{1}_e\otimes X_{\ell}(-\pi/2)$, hereon denoted $CX_{\pi/2}$.
Likewise, unconditional $X$ gates occur when $(\hat{n}_0 \cdot \hat{n}_1)^{(\ell)}=+1$.
Other nuclear qubits that are not in resonance for a given $t$ experience unconditional rotations about $Z$.
This crosstalk is an inherent characteristic of quantum register control using DD sequences and must be considered in quantum circuit design.
Additional details regarding DD can be found in Methods, \S\ref{sec:methods_DD}.

Generalizing the traditional approach, Takou \textit{et al.,} \cite{takou2023precise, takou2024generation} showed that certain near-resonance DD sequences can still act as maximal entangling gates, as long as $(\hat{n}_0 \cdot \hat{n}_1)^{(\ell)}<0$.
Crucially, if this condition is satisfied for multiple nuclear qubits for the same $t$, then a single DD sequence can act as a maximal, multipartite entangler.
The multipartite entangling ability of a DD sequence is quantified by the $M$-qubit entangling power $\varepsilon_{p,M}$.
Here, $M$ includes the electron qubit together with $L$ nuclear qubits, so $M=L+1$.
As shown in Fig. \ref{fig:over_view}c, at parallel order, the entangling condition is simultaneously satisfied for three $^{13}$C qubits ($q_1$, $q_2$, and $q_3$) proximal to the NV center ($e$) in our experiment, and the 4-qubit entangling power reaches a maximum $\varepsilon_{p,4}=0.993$ (Fig. \ref{fig:over_view}d).
Additional details regarding the entanglement metrics can be found in Methods \S\ref{sec:ent_metrics_methods}. 

%==================================================================
\section{Verifying Entanglement with Multiple Quantum Coherences}
%==================================================================

In general, the characterization of arbitrary multi-qubit states requires complete state tomography, with measurement requirements that scale exponentially with the number of qubits.
Moreover, full tomography involves measurements of correlated observables, which demands high fidelity readout of all qubits.
In the room-temperature NV-quantum-register system, where limited-fidelity readout is only available for the electron, tomography quickly becomes impractical.

Entanglement verification using multiple quantum coherences (MQC) offers an alternative approach (Fig. \ref{fig:mqc_L1}a).
MQC verification leverages the symmetry of GHZ entangled states to amplify each qubit's phase accumulation onto a single qubit, which then carries information about the number of qubits in the entangled state \cite{baum1985multiple, wei2020verifying}. 
When $M$ entangled qubits each acquire a phase $\phi$, the resulting MQC signal is given by:
\begin{equation}\label{eq:MQC_prob_ideal}
    P(\ket{0}^{\otimes M}) = \frac{1}{2}(1+\cos (M\phi)).
\end{equation}
Additional details can be found in Methods \S\ref{sec:mqc_methods}.
 
%==================================================================
\subsection{Bipartite entanglement}\label{sec:bipartite}
%==================================================================

\begin{figure}[th!] 
\centering
\includegraphics{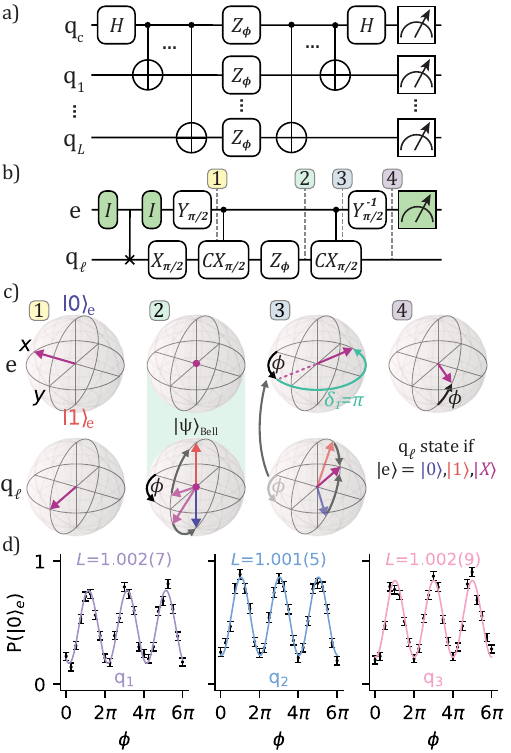}
\caption{
\textbf{Verifying entangled states using multiple quantum coherences}
\textbf{(a)} Quantum circuit to efficiently characterize entangled states using the MQC method \cite{wei2020verifying}.
\textbf{(b)} DD implementation of the MQC method used in this work (shown here for bipartite entanglement), in which only the nuclear qubits are subject to $Z_{\phi}$ gates. 
\textbf{(c)} Bloch-sphere visualization of the stages labeled in (b). 
First, both spins are initialized to $\ket{0}$ and rotated to equatorial superposition states (step 1). 
The conditional rotation that follows creates a Bell state, and a variable phase $\phi$ is added to the nuclear qubit (step 2). 
When the qubits are disentangled, $\phi$ is added to the electron phase (step 3) and the final electron rotation maps $\phi$ to an angle from $\hat{z}$, which is then measured through spin-dependent fluorescence (step 4).
\textbf{(d)} Results of MQC experiments for bipartite entanglement ($M=2$ total qubits, $L=1$ nuclear qubit) with each nuclear qubit, $q_\ell$. 
% Solid lines are fits to equation (\ref{eq:MQC_prob_NV}) with a variable amplitude.
Error bars on points denote photon shot noise, and uncertainties on $L$ are 68\% confidence intervals from  from fits to equation (\ref{eq:MQC_prob_NV}) with a variable amplitude, midpoint and phase.
}
\label{fig:mqc_L1}
\end{figure}

As an example of this approach, we adapt the MQC circuit to use the DD gates for creation and verification of two-qubit Bell states (Fig. \ref{fig:mqc_L1}b).
In contrast to the original implementation \cite{wei2020verifying}, we measure only the electron qubit; this is sufficient to quantify the size of the entangled state, albeit not its fidelity.
The calibration of all DD sequences, including local and entangling gates as well as nuclear initialization \textit{via} electron-nuclear swaps \cite{taminiau2014universal} was experimentally verified using single qubit nuclear state tomography (see the Supplementary Information (SI), Sec.~\siNucGates).

\begin{figure*}[th!]
\centering
\includegraphics{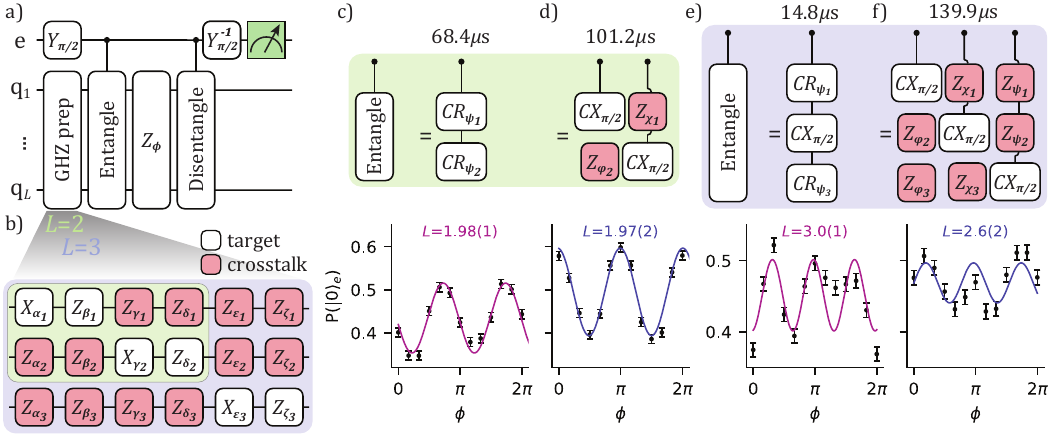}
\caption{
\textbf{Generation and verification of multipartite GHZ states}
\textbf{(a)} Block-diagram quantum circuit for entanglement verification using multiple nuclear qubits.
Qubits are initialized sequentially as shown in Fig. \ref{fig:mqc_L1}b.
\textbf{(b)} Local gates used to prepare for GHZ state generation for $L=2$ or 3 nuclear qubits (green and blue shaded regions, respectively).
Red shaded gates represent unwanted crosstalk effects, which manifest as $Z$-axis rotations on each nuclear qubit in the register that is not targeted with a white gate.
The angles of the local gates (associated with the unit pulse repeats) were determined using numerical optimization, with the gate axes (associated with unit pulse times) remaining fixed as shown.
\textbf{(c,d)} Results of three-qubit ($L=2$) MQC verification experiments using nuclear qubits $q_1$ and $q_2$ (data points in lower panels) corresponding to the parallel and sequential entangling gates shown in the upper panel. 
The corresponding disentangling gates in (a) are identical.
Red (parallel) and blue (sequential) curves are sinusoidal fits to measure the number of entangled nuclear qubits $L$.
% This parallel entangling gate takes longer than the $L=3$ parallel gate since a larger $N$ was need to minimize residual entanglement with the non-targeted, uninitialized nuclear qubit ($q_3$ in this experiment).
See the SI, Sec.~\siMQC~for results corresponding to all other combinations of qubits.
\textbf{(e,f)} Results of four-qubit ($L=3$) MQC verification experiments. 
Error bars on points denote photon shot noise, and uncertainties on $L$ are 68\% confidence intervals from from fits to equation (\ref{eq:MQC_prob_NV}).
% \textbf{(c,d)} Results of four-qubit MQC verification experiments (data points in lower panels) corresponding to the parallel and sequential entangling gates shown in the upper panel. 
% The corresponding disentangling gates in (a) are identical.
% Blue curves are sinusoidal fits to measure the number of entangled nuclear qubits $L$.
% \textbf{(e,f)} Results of three-qubit MQC verification experiments using nuclear qubits $q_1$ and $q_2$. 
% See the SI, Sec.~\siMQC~for results corresponding to all other combinations of qubits.
% This parallel entangling gate takes longer than the $L=3$ parallel gate since a larger $N$ was need to minimize residual entanglement with the uninitialized nuclear qubit ($q_3$ in this experiment).
}
\label{fig:mqc_L2_L3}
\end{figure*}

Fig. \ref{fig:mqc_L1}c highlights the key steps of the DD gate MQC process.
Given the form of the $CX_{\pi/2}$ entangling gate, local gates are required to prepare a $Z$-basis Bell state, $\ket{\Psi^+}=\frac{1}{\sqrt{2}}(\ket{01}+i\ket{10})$. 
The local gates affect the relative phase of the entangled state and shift the phase of the MQC signal in equation (\ref{eq:MQC_prob_ideal}) without affecting the frequency.
Additionally, we only apply the variable phase gate $Z_\phi$ to the nuclear qubits.
As a result, when the electron is entangled with $L$ nuclear qubits, we expect to measure a signal of the form:
\begin{equation}\label{eq:MQC_prob_NV}
    P(\ket{0}_e) = \frac{1}{2}(1+\cos (L\phi - \delta_L) ),
\end{equation}
where $\delta_L$ is the phase shift that depends on local gates; for the bipartite case, $\delta_{1}=\pi$. 
As shown in Fig. \ref{fig:mqc_L1}d, the results of this experiment are consistent with the generation of bipartite entanglement between qubit $e$ and each of $q_1$, $q_2$, and $q_3$.
Notably, the single-frequency MQC signal verifies the existence of bipartite ($L=1$) entanglement and precludes the unintentional presence of additional entangled qubits. 
This is expected from resolved DD resonances at the individualized order (Fig. \ref{fig:over_view}).

%==================================================================
\subsection{Multipartite entanglement}
%==================================================================

We consider and compare two methods to generate multipartite ($L\geq2$) entangled states: through \emph{sequential} one- and two-qubit gates at individualized order, and through a single \emph{parallel} gate at first order.
While the approach to verifying multipartite entanglement is similar to the bipartite setting, the quantum circuits become more complicated (Fig. \ref{fig:mqc_L2_L3}a).
Neither gate takes the form $CX_{\pi/2}^{\otimes L}$ on the register.
In the parallel case, this is because the rotation geometry is unique for each qubit, with some nuclear qubits conditionally rotated off resonance according to $CR_{\theta}= \ket{0}\bra{0}_e \otimes R_{\hat{n}_0}(\theta) + \ket{1}\bra{1}_e \otimes R_{\hat{n}_1}(\theta)$.
In the sequential case, this is because each individualized DD sequence creates crosstalk errors on the other qubits. 

In either case, it is necessary to apply local gates prior to entanglement to prepare the desired GHZ state (see Fig \ref{fig:mqc_L2_L3}b).
These gates set the basis of each qubit, compensating for the crosstalk effects of the sequential protocol or matching the parallel entangling gate's geometry.
Although the necessary local gates can in principle be predicted based on analytical theory, in practice it is more effective to optimize them numerically, especially since the local gates themselves suffer from crosstalk.
% See Sec.~\siMQC~of the SI for details of this optimization method together with simulations of its efficacy; additional experimental considerations of the MQC method can be found in Methods \S\ref{sec:experimental_MQC_methods}.
See SI Sec.~\siMQC~for optimization details and Methods \S\ref{sec:experimental_MQC_methods} for additional experimental considerations.

\begin{figure*}[th!]
\centering
\includegraphics{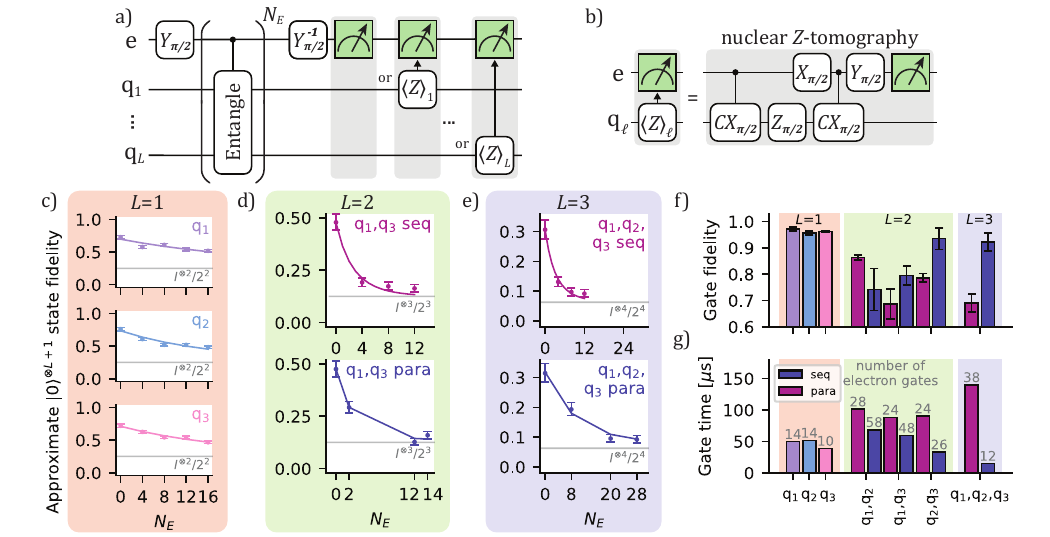}
\caption{
\textbf{Entangling-gate fidelities}
\textbf{(a)} Quantum circuit to efficiently measure entangling-gate fidelity. 
After repeating the entangling gate $N_E$ times, the $Z$ projection of either the electron or an individual nuclear qubit is measured, and the combined results are used to approximate the $\ket{0}^{\otimes M}$ state fidelity.
\textbf{(b)} Sub-circuit to measure nuclear-qubit $Z$ projections. 
\textbf{(c)} Experimental results (data points) for bipartite entangling gates associated with each nuclear qubit.
%, where each multiple of four entangling gates is expected to return the state $\ket{00}$.
Solid curves are fits using a noisy quantum channel model to extract the entangling-gate fidelity.
\textbf{(d,e)} Experimental results (data points) for three-way and four-way entanglement generated either sequentially (top panel) or using a parallel gate (bottom panel). 
% Multiples of four sequential gates ideally returns $\ket{0000}$; 
% here, $N_E$ was limited to 12 due to the $T_2$ coherence time of the electron qubit.
Repeating the parallel gate does not generally return exactly $\ket{0}^{\otimes M}$ due to its more complicated geometry; values of $N_E$ were chosen to maximize the overlap with $\ket{0}^{\otimes M}$.
Fits to the data (solid curves) account for this effect along with gate errors and are used determine the fidelity of each gate.
Error bars on points in (c-e) are propagated from each measurement's photon shot noise.
\textbf{(f)} Entangling-gate fidelities determined from fits to the data in (c-e). 
Error bars denote 68\% confidence intervals from fits to the noisy quantum channel model.
\textbf{(g)} Entangling-gate durations. 
The number of electron pulses (equivalent to 2$N$) are printed above each bar.
}
\label{fig:gate_fids} 
\end{figure*}

Starting with $L=2$ (two nuclear qubits), we can design a parallel three-qubit entangling gate with any subset of $(q_1,q_2,q_3)$; Fig. \ref{fig:mqc_L2_L3}c shows the example of $e$, $q_1$, and $q_2$.
Since any residual entanglement with the spectator nuclear qubit will lower the entangling gate fidelity, the $L=2$ gate parameters must be selected such that crosstalk on the spectator qubit is unconditional (see SI Sec.~\siNucGates~ for details).
%Any residual entanglement generated with non-targeted nuclear qubits will directly lower the entangling gate's fidelity \cite{takou2024generation}.
The MQC method is also sensitive to such effects; residual entanglement manifests as multi-frequency beating in the signal (see SI Sec.~\siMQC~ for theoretical analysis and exemplary data).
The single-frequency oscillation observed in Fig. \ref{fig:mqc_L2_L3}c, with $L=2$, confirms that the gate creates a tripartite entangled state as desired.
The sequential approach (with optimized crosstalk corrections) also generates a tripartite state with the expected MQC signal (Fig. \ref{fig:mqc_L2_L3}d); however, the sequential gate duration is almost 50\% longer than the parallel gate.

Figures~\ref{fig:mqc_L2_L3}e and \ref{fig:mqc_L2_L3}f show the results of MQC verification experiments for 4-qubit GHZ states ($L=3$).
%The parallel gate, consisting of a single, first-order DD sequence, is $\approx$10 times faster than the corresponding sequential gate.
The advantages of parallel gates increase with system size; the parallel $L=3$ gate is $\approx$10 times faster than the corresponding sequential gate.
Moreover, we observe the expected MQC frequency at $L=3$ from the parallel gate, whereas the sequential approach results in a shifted frequency of $L=2.6(2)$.
This discrepancy reflects the lower fidelity of the sequential gate (shown in the next section); essentially, the electron accumulates unwanted phase contributions from crosstalk as each nuclear qubit is sequentially entangled and disentangled in the MQC protocol.

%==================================================================
\section{Entangling gate fidelities}\label{sec:ent_gate_fids}
%==================================================================

In order to quantify the entangling-gate fidelity, it is necessary to distinguish the errors associated with the entangling gates from those due to state preparation and measurement (SPAM).
We adapt a method presented by Evered \textit{et al.}~\cite{evered2023high} for this purpose. 
The key idea is to repeatedly apply the entangling gate and measure the state fidelity after each iteration. 
Gate errors accumulate with each repetition and decrease the state fidelity, whereas SPAM errors remain constant.
In our case, repeating the entangling gate by certain multiples ($N_E$) ideally results in a disentangled state of the form $\ket{0}^{\otimes M}$, the fidelity of which can be quantified using only a small number of measurements (Fig. \ref{fig:gate_fids}a).
If one further makes the approximation that each final experimental state is separable, \textit{e.g.} $\langle Z_e \otimes Z_{q_{\ell}} \rangle \approx \langle Z_e \rangle \langle Z_{q_{\ell}} \rangle$ in the bipartite case, then the state fidelity can be approximated from independent measurements of each qubit's $Z$ projection. 
The electron is measured directly, and the nuclear qubits are measured using $Z$-axis state tomography \cite{taminiau2014universal} (Fig. \ref{fig:gate_fids}b).

In the bipartite case, applying the $CX_{\pi/2}$ gate in multiples of four returns the system to its initial state of $\ket{00}$, up to gate errors. 
Fig. \ref{fig:gate_fids}c shows the results of this experiment using each of $q_1$, $q_2$, and $q_3$, where the state fidelity steadily decays towards the classically mixed asymptote.
The multipartite sequential entangling gates have the analogous property of returning the ideal state $\ket{0}^{\otimes M}$ when $N_E$ is a multiple of four (see Fig. \ref{fig:gate_fids}d and \ref{fig:gate_fids}e, top panel).
In the parallel approach, the gates' complicated geometry makes the analysis more difficult. 
We simulate the application of repeated parallel gates to identify which $N_E$ values most closely approximate $\ket{0}^{\otimes M}$ (see SI Sec.~\siEntangleGateFid). 
The simulated density matrices for each $N_E$ were then used as starting points in our fitting model, described below.

The state fidelity data of Fig. \ref{fig:gate_fids}c-e were fit using a quantum depolarizing channel model to calculate the constant SPAM error, $\varepsilon_{\text{SPAM}}$, and entangling-gate error, $\varepsilon_{\text{gate}}$; see the SI, Sec.~\siEntangleGateFid~for details and additional $L=2$ data sets.
The entangling-gate fidelity, $G_M=(1-\varepsilon_{\text{gate}})^M$, is shown in Fig. \ref{fig:gate_fids}f.
We observe an average (standard deviation) bipartite gate fidelity of $G_2^{\text{avg}}= 0.96(1)$.
The three-qubit gate fidelities vary significantly: the parallel gate fidelities are anti-correlated with how many electron operations are needed (Fig. \ref{fig:gate_fids}g), whereas the sequential gates are further limited by their intrinsic crosstalk.
Nonetheless, the average fidelity for the parallel gates, $G_3^{\text{para}}= 0.83(9)$, exceeds that of the sequential approach, $G_3^{\text{seq}}= 0.78(8)$, and the parallel gates are $\sim2$ times faster.
For $M=4$, the advantages of the parallel approach become dramatically apparent.
The four-qubit parallel gate is the fastest gate overall, having a similar number of pulses as the two-qubit gates but a shorter $t$; this contributes to its high fidelity of $G_4^{\text{para}} = 0.92(4)$, despite the large number of qubits involved.
The sequential gate, on the other hand, exhibits a low fidelity of $G_4^{\text{seq}} = 0.69(3)$, due to the combined impacts of crosstalk, its longer duration, and a large number of electron pulses. 

%==================================================================
\section{Generality and extensions} 
%==================================================================

\begin{figure}[th]
\centering
\includegraphics{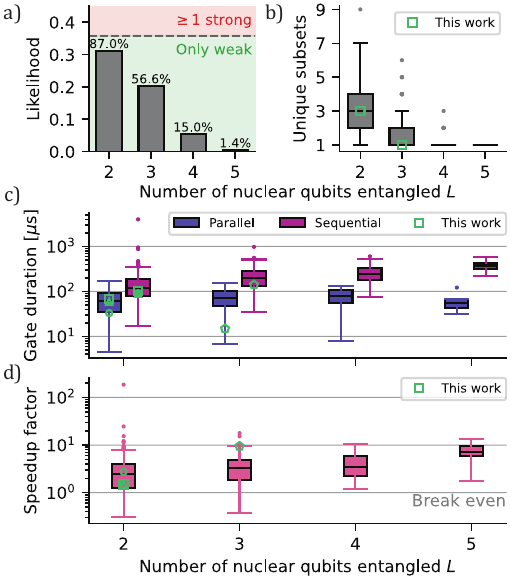}
\caption{
\textbf{Generality of parallel entanglement in NV–nuclear registers}
\textbf{(a)} Likelihood of a register supporting parallel entanglement. 
In this analysis, we exclude registers that contain at least one strongly coupled nuclear qubit (red region, corresponding to 64\% of configurations).
Percentages listed above each bar correspond to the likelihood of weakly-coupled registers (green region) supporting parallel entanglement with at least one set of $L$ nuclear qubits.
On average, the weakly-coupled registers contain 5.7 (2.5) individually-addressable nuclear qubits in total.
\textbf{(b)} Number of unique subsets of $L$ nuclear qubits in registers hosting parallel entanglement.
The middle black line represents the median, the box spans the interquartile range (IQR), the whiskers extend to $1.5 \times$ the IQR, and points indicate outliers.
\textbf{(c)} Parallel and $k=2$ sequential entangling gate durations.
\textbf{(d)} Parallel entanglement speedup factor. 
}
\label{fig:generality_main} 
\end{figure}

To contextualize these results, we simulated 500 random, weakly-coupled NV–nuclear qubit registers to evaluate their capacity for parallel entanglement.
We excluded configurations with one or more strongly-coupled qubits, although such mixed systems could potentially be addressed as well.
For each remaining configuration, we determined which combinations of nuclear qubits could be entangled in parallel, and we calculated the corresponding gate parameters for both parallel and sequential operations. 
The results, summarized in Fig.~\ref{fig:generality_main}, illustrate that $L=2$ and $L=3$ parallel entanglement gates are available in the majority of configurations (Fig. \ref{fig:generality_main}a).
In some cases, entangled states of four or even five nuclear qubits can be created.
Furthermore, the registers generally support parallel entanglement of multiple unique qubit combinations (Fig. \ref{fig:generality_main}b).
The parallel gate duration remains approximately constant as $L$ increases, whereas the sequential gates steadily increase (Fig. \ref{fig:generality_main}c).
Thus, the speedup of the parallel approach becomes more pronounced with increasing $L$ (Fig. \ref{fig:generality_main}d).
Further simulation details are provided in Methods Sec. \ref{sec:generality_methods}, with additional simulations in SI Sec.~\siGenerality.

The experimental results reported here fall within the simulated distributions.
In future work, registers can reasonably be identified that enable $L=4$ or $L=5$ parallel entangling gates.
The $L=5$ case is of particular importance for fault tolerant operation \cite{abobeih2022fault}; see SI Sec.~\siQEC~for further discussion and simulations of quantum error correction.

%==================================================================
\section{Conclusions} 
%==================================================================

In demonstrating a method for generating large entangled states quickly, efficiently, and with high fidelity, we have addressed a principal challenge of quantum control in central spin systems like diamond NV centers. 
The always-on nature of qubit-coupling interactions in such systems leads to unavoidable crosstalk, and operations composed only of pairwise gates become impractically long.
Rather, our approach directly harnesses the star topology of the central-spin system to achieve parallelized entangling gates that are an order of magnitude faster than their sequential counterparts, and more immune to crosstalk.
Notably, and in contrast to approaches developed for strongly-coupled registers \cite{waldherr2014quantum, reiner2024high, thorvaldson2025grover, edlbauer202511}, the parallel gate speed is not limited by the need to resolve differences in parallel hyperfine couplings, $A_\parallel$.
In fact, the four-qubit gate duration of 15 $\mu$s is near the fundamental interaction limit given by the inverse of the perpendicular hyperfine couplings ($A_\perp\approx 60$~kHz for all three qubits).  

The parallel entanglement framework and experimental methods presented are readily applicable to other defects in diamond \cite{sukachev2017silicon, knaut2024entanglement, bersin2024telecom, rugar2021quantum, harris2025high, beukers2025control} and other host materials like silicon carbide \cite{nagy2019high, bourassa2020entanglement}, or silicon \cite{reiner2024high, thorvaldson2025grover, edlbauer202511}.
The density of nuclear spins can also be tuned using isotopic engineering \cite{itoh2014isotope} to optimize the distribution of similarly coupled nuclear qubits to be entangled in parallel.
% For systems with symmetric central spin projections (\textit{e.g.}, $S=1/2$), the entangling gate more closely approximates $CX_{\pi/2}^{\otimes L}$, allowing for even easier implementation \cite{takou2023precise}.
% Utilizing multi-frequency control to manipulate the NV center's $m_s=\pm 1$ electronic states would also enable this simplification.
% Likewise, the experimental methods we present to detect and benchmark multipartite entangling gates are naturally suited to these systems, and they scale to larger register sizes without incurring significant increases in experimental overhead. 
% The density of nuclear spins can also be tuned using isotopic engineering \cite{itoh2014isotope} to optimize the distribution of similarly coupled nuclear qubits to be entangled with a single DD sequence.

With a fidelity of 0.92(4), the four-qubit parallel entangling gate is comparable to the state of the art in highly engineered quantum computing systems such as trapped ions \cite{lu2019global} and neutral atoms \cite{evered2023high}. 
The entangled-state fidelities are primarily limited by SPAM errors associated with the NV center's room-temperature properties. 
These can be solved by working at cryogenic temperatures \cite{robledo2011high}. 
Alternatively, the room-temperature SPAM errors can be reduced using repetitive readout at high magnetic fields \cite{neumann2010single}, spin-to-charge conversion \cite{hopper2020real}, or dynamical nuclear polarization methods \cite{schwartz2018robust}.

These results directly impact entanglement-enhanced quantum sensing protocols, which generally rely on the generation of high-fidelity GHZ states to reach the Heisenberg sensing limit \cite{degen2017quantum, kenny2025quantum}.
The potential of generating such states at room temperature will facilitate their use in practical applications such as nano-NMR and scanning magnetometry.
Additionally, the dramatic speedup from paralleled entangling gates opens the door to new quantum error correcting possibilities \cite{takou2023precise, abobeih2022fault}, to the point that fault tolerance could be achieved at room temperature for the first time.
In the low-temperature regime, where entanglement can be distributed between quantum registers using photons \cite{hermans2022qubit} and preliminary fault tolerance has been demonstrated \cite{abobeih2022fault}, replacing sequences of two-qubit gates with parallelized multi-qubit gates will increase the circuit efficiency and enable the implementation of larger quantum codes.

%==================================================================
\section{Methods}
%==================================================================

\subsection{Experimental system}

The sample used in this work is a type-IIa electronic-grade synthetic diamond (Element Six) with natural abundance of $^{13}$C impurities.
The NV center is at the focus of a solid immersion lens encircled by an antenna for microwave (mw) frequency control. 
All experiments are performed at room temperature in ambient conditions.
A permanent magnet was aligned to the NV symmetry axis using pulsed electron-spin resonance experiments and positioned to create a magnetic field strength of 338G.
The magnetic field strength was chosen to minimize nuclear-qubit gate durations and angular errors.
Further details of the field alignment and simulations to determine the field strength can be found in the SI Sec.~\siMagneticField.

Green (532nm) laser pulses of 2$\mu$s were used to (re)initialize the electron spin and charge state through optical pumping, and shorter 300ns pulses were used to measure the spin-state photoluminescence contrast.
Synchronization of the optics and mw signals was achieved using two different configurations.
The first used two arbitrary waveform generators (AWGs), one (Tektronix AWG520) dedicated to optical control and the other (Tektronix AWG7102) for mw control.
The second configuration used a Swabian Instruments PulseStreamer 8/2 for both optical and mw control. 
Additional details can be found in the SI Sec.~\siExperimentalControl.
Electron gate errors were quantified using Pulse Bootstrap Tomography \cite{dobrovitski2010bootstrap}; see the SI Sec.~\siElectronQubit~for details.

\subsection{Dynamical decoupling}\label{sec:methods_DD}

The Hamiltonian governing the central spin electron interacting with $L$ nuclear qubits is given by:
\begin{equation}\label{eq:Hamiltonian_methods}
    H = \mathbbm{1} \otimes \frac{\omega_{\text{Lar}}}{2} \sum_{\ell=1}^{L} \sigma_{z}^{(\ell)} + \frac{Z_e}{2} \otimes \sum_{\ell=1}^{L} (A_{||}^{(\ell)} \sigma_{z}^{(\ell)} + A_{\perp}^{(\ell)} \sigma_{x}^{(\ell)}),
\end{equation}
where $\omega_{\text{Lar}}$ is the nuclear Larmor frequency; $Z_e=s_0\ket{0}\bra{0}+s_1\ket{1}\bra{1}$ is the electron spin operator, where $s_j$ are the two electron spin projections chosen as the computational basis ($s_0=0$ and $s_1=-1$ for this work); and $A_{||,\perp}^{(\ell)}$ are the parallel and perpendicular hyperfine couplings between the electron and the $\ell$th nuclear qubit.
This can be rewritten as \cite{takou2023precise}:
\begin{equation}
    H = \sum_{j \in \{0,1\}} \ket{j}\bra{j}_e \otimes \sum_{\ell}^{L} H_{j}^{(\ell)},
\end{equation}
where each $H_{j}^{(\ell)}$ is given by:
\begin{equation}\label{eq:H_j_methods}
    H_{j}^{(\ell)} = \frac{\omega_L + s_{j}A_{||}^{(\ell)}}{2} \sigma_{z}^{(\ell)} + \frac{s_{j}A_{\perp}^{(\ell)}}{2} \sigma_{x}^{(\ell)}.
\end{equation}
The notation $\sigma_{i}^{(\ell)}$ in equation (\ref{eq:H_j_methods}) means the $i$th Pauli matrix on the $\ell$th component of the $L$-nuclear-qubit Hilbert space and the identity on all other components.
This form of the Hamiltonian highlights how the electron state conditions different and unique dynamics for each nuclear qubit. 
This is further made apparent by the free evolution operator $U_f(t)$ for the system:
\begin{equation}\label{eq:U_free_methods}
    U_f(t) = \sum_{j \in \{0,1\}} \ket{j}\bra{j}_e \bigotimes_{\ell}^{L} \exp \bigg(-it H_{j}^{(\ell)} \bigg),
\end{equation}
from which each $\exp(-it H_{j}^{(\ell)})$ term can be viewed as a rotation operator acting on the $\ell$th nuclear qubit.
Note a subtle shift in notation from equation (\ref{eq:H_j_methods}) to (\ref{eq:U_free_methods}), where each index $\ell$ no longer implies identity operators on the other qubits, and each two-dimensional $H_j^{(\ell)}$ can viewed as acting on a distinct subspace. 
Additional details and derivations can be found in the SI Sec.~\siHamiltonian.

The free evolution periods of DD sequences leverage equation (\ref{eq:U_free_methods}) to control the rotational effects of each nuclear qubit, as well as extend the electron coherence time.
The net unitary operator $U_{\text{DD}}$ from performing a time-symmetric DD sequence of unit pulse time $t$ and with $N$ repeats is given by:
\begin{equation}\label{eq:U_DD}
    U_{\text{DD}} = \sum_{j \in \{0,1\}} \ket{j}\bra{j}_e \bigotimes_{\ell}^{L} R_{\hat{n}_j^{(\ell)}(t)}(N\phi^{(\ell)}(t)),
\end{equation}
where $R$ is a spin 1/2 rotation operator about the axis $\hat{n}_j^{(\ell)}$ and by an angle of $N\phi^{(\ell)}$ for the $\ell$th nuclear qubit.
See the SI Sec.~\siNucGates~for details in calculating each rotation operator based on the hyperfine couplings of the register.
This formulation highlights the conditional nature of each nuclear qubit's rotation depending on the electron state $\ket{j}_e$. 

Resonant $X$-axis control of a target nuclear qubit is achieved with the proper choice of unit pulse time $t_m$ that creates $(\hat{n}_0 \cdot \hat{n}_1)^{(\ell)}(t_m)=\pm1$.
Such a choice of $t_m$ occurs periodically and is given by:
\begin{equation}
    t_m^{(\ell)} = \frac{4 \pi m}{\omega_0^{(\ell)} + \omega_1^{(\ell)}},
\end{equation}
for $m \in \mathbbm{Z^+}$ and $\omega_j^{(\ell)} =\sqrt{(s_j A_{\perp}^{(\ell)})^2 + (\omega_L +s_j A_{||}^{(\ell)})^2}$ \cite{takou2023precise}. 
For odd $m=2k+1$, $(\hat{n}_0 \cdot \hat{n}_1)^{(\ell)}=-1$ and for even $m=2k$, $(\hat{n}_0 \cdot \hat{n}_1)^{(\ell)}=+1$.
Here, the integer $k$ specifies the DD order as discussed in the introduction.
When the electron-state-dependent nuclear rotation axes are maximally anti-aligned, or $(\hat{n}_0 \cdot \hat{n}_1)^{(\ell)}=-1$, the nuclear rotations are maximally dependent on the state of the electron.
With $N$ set to create the correct rotation angle, the net gate is $C_{e}X_{\ell}(\pm \pi/2)=\ket{0}\bra{0}_e\otimes X_{\ell}(\pi/2)+\ket{1}\bra{1}_e\otimes X_{\ell}(-\pi/2)$ between the electron and the target nuclear qubit $q_{\ell}$.
For all other spins, the choice of $t$ is off-resonance and the resulting rotation is unconditional and about the $Z$ axis.
Similarly, when the unit pulse time is on resonance and $(\hat{n}_0 \cdot \hat{n}_1)^{(\ell)}=+1$, the resulting $\ell$th nuclear qubit's rotation is an unconditional $X$ axis rotation, with all other nuclear rotations being off-resonance and about the $Z$ axis.

For example, when attempting to rotate the first nuclear qubit unconditionally about the $X$ axis by $\pi/2$, the net unitary acting on the register would take the form $U=I_e \otimes X_{\pi/2} \otimes Z_{\theta_{(2)}}...\otimes Z_{\theta_{(L)}}$ where each crosstalk rotation angle $\theta_{(\ell)}$ depends on the choice of $t$ and $N$ that were used to achieve the desired $X_{\pi/2}$ rotation of $q_1$, and the specific hyperfine couplings of the $\ell$th nuclear qubit.
The goal of the parallelized entangling gate is to leverage this crosstalk in such a way that each nuclear qubit can be maximally entangled for a single choice of $t$ and $N$.
Further information on the $t$ and $N$ parameter choices for each nuclear qubit's gate, together with their experimental verification, can be found in the SI Sec.~\siStandardCircuits.

\subsection{Entanglement metrics}\label{sec:ent_metrics_methods}

To quantify the bipartite entangling ability of a DD sequence with a particular nuclear qubit $q_{\ell}$, one can calculate the first Makhlin invariant, which takes the form:
\begin{equation}
    G_1^{(\ell)} = \bigg(\cos^2 \frac{N\phi^{(\ell)}}{2} + (\hat{n}_0 \cdot \hat{n}_1)^{(\ell)} \sin^2 \frac{N\phi^{(\ell)}}{2} \bigg)^2,
\end{equation}
for time-symmetric DD sequences such as XY8 \cite{takou2023precise}. 
This entanglement metric (bounded from 0 to 1) is minimal when bipartite entanglement is maximal. 
Using this form of $G_1^{(\ell)}$, it was shown that, with the proper choice of $N$, $G_1^{(\ell)}=0$ if $(\hat{n}_0 \cdot \hat{n}_1)^{(\ell)}<0$. 
Finding the unit pulse times that satisfies this condition for each target nuclear qubit is the first step in calibrating the parallel entangling gate.
Furthermore, to quantify the multipartite entangling ability of a DD sequence with $L$ target nuclear qubits, one can use the $M$-qubit entangling power:
\begin{equation}
    \varepsilon_{p,M}(U_{DD}) = \bigg( \frac{d}{d+1}\bigg)^M \prod_{\ell}^{L}(1-G_1^{(\ell)}),
\end{equation}
where $M=L+1$ is the total number of qubits in targeted for entangling, including the electron, and $d=2$ is the dimension of the qubit subspace \cite{takou2024generation}. 
Often, as in Fig. \ref{fig:over_view}, the normalized version of this metric is most useful, without the constant coefficient in front of the product.
The normalized metric ranges from 0 (the DD sequence creates no entanglement) to 1 (the DD sequence is a maximal multipartite entangler).
Owing to the central spin nature of solid-state defect systems, $\varepsilon_{p,M}(U_{DD})$ depends only on each bipartite entanglement invariant $G_1^{(\ell)}$. 
Calculating $\varepsilon_{p,M}(U_{DD})$ with each of the target nuclear qubits in the range of unit pulse times that satisfy $(\hat{n}_0 \cdot \hat{n}_1)^{(\ell)}<0$ reveals the optimum $(t,N)$ combination to generate maximal multipartite entanglement.

We utilize the non-unitary entangling power to account for the impact of residual entanglement generated with non-targeted nuclear spins  \cite{takou2024generation}.
This entanglement metric is derived using the partial trace quantum channel $\mathcal{E}$ over non-targeted nuclear qubits.
The set of all nuclear qubits is partitioned into a subset that is targeted (size $L$) and the rest that are not targeted (size $L_{\text{total}}-L$).
A simple approximate form for this entanglement metric is given by 
\begin{equation}
    \varepsilon_{p,M}(\mathcal{E}) =
    \frac{\varepsilon_{p,M}(U_{DD})}{2}
    \left(1 + \prod_{\substack{\ell \in \text{not} \\ \text{targeted}}}^{L_{\text{total}}-L}
    G^{(\ell)}_1 \right).
\end{equation}
The non-unitary entangling power is bounded above by the unitary entangling power; $\varepsilon_{p,M}(\mathcal{E}) \leq \varepsilon_{p,M}(U_{DD})$, with equality holding when no residual entanglement is generated ($G^{(\ell)}_1=1$ for all non-targeted nuclear qubits) \cite{takou2024generation}.
Therefore, any residual entanglement generated leads to a Makhlin invariant less than 1, lowering this entanglement metric. 
This metric is essential when designing parallel entangling gates with subsets of known nuclear qubits, \textit{e.g.}, in the case of $L=2$ parallel entangling gates in this work.
Additional details regarding how these metrics were used to calibrate each parallel entangling gate can be found in SI, Sec.~\siNucGates.

\subsection{Multiple quantum coherences}\label{sec:mqc_methods}

In the original MQC circuit proposed by Wei \textit{et al.} \cite{wei2020verifying} (Fig. \ref{fig:mqc_L1}a), the $M$ qubit register is first initialized to $\ket{0}^{\otimes M}$. 
The control (top) qubit $q_c$ is then placed into an equal superposition state so that the following CNOT gates create a GHZ state. 
Once entangled, each qubit's relative phase is shifted by an equal amount $\phi$ yielding $\ket{\text{GHZ}_{\phi}^{M}}=\frac{1}{\sqrt{2}}(\ket{0}^{\otimes M} + e^{-iM\phi}\ket{1}^{\otimes M})$. 
The system is then disentangled back to the original state by reversing the first half of the circuit. 
The result (before the last Hadamard that projects the control qubit phase onto the measurement axis) is that the control qubit's phase is amplified based on how many qubits it was entangled with: 
\begin{equation}\label{eq:MQC_final_state}
    \ket{\psi_f}=\frac{1}{\sqrt{2}}(\ket{0} + e^{-iM\phi}\ket{1})\otimes\ket{0}^{\otimes M-1}.
\end{equation}
Thus the final probability of the entire system returning to the initial state is given by:
\begin{equation}\label{eq:MQC_prob_ideal_methods}
    P(\ket{0}^{\otimes M}) = \frac{1}{2}(1+\cos (M\phi)),
\end{equation}
which crucially carries a frequency equal to the number of qubits in the entangled state. 
%If every qubit's $Z$ projection is subsequently measured, a bound on the state fidelity can be established through Fourier analysis of the signal \cite{wei2020verifying}.
%Such analysis was not performed in this work because our implementation of the protocol did not disentangle the system back to its initial $\ket{0}^{\otimes M}$ state and thus full state tomography would have been required.
%Rather, we use this protocol to sense the size of the GHZ state through a measurement of the number of nuclear qubits in it.

\subsection{Experimental MQC considerations}\label{sec:experimental_MQC_methods}

Nuclear phase gates $Z_{\phi}$ were implemented using off-resonant DD sequences.
Since such gates are realized for any off-resonant $t$, the optimum parameters can chosen strategically. 
The off-resonance region prior to the first order resonances not only offers fast pulse times ($t<1\,\mu s$) but also can parallelize the phase gate, turning the sequence into an unconditional $M$-qubit gate.
Experimentally, the finite pulse duration of electron gates sets a lower limit on $t$; this restriction in turn sets a lower bound on the angular resolution $\Delta\phi = \phi_{N=1}$ of the phase gate.
With $\Delta\phi$ specified, $t$ was optimized to minimize the angular error for each nuclear qubit in the register. 
Then, in order to increase the phase, the unit pulse was repeated $N$ times, leading to $\phi=N\Delta\phi$.
The simulated four-qubit process fidelities for a parallelized gate of the form $I_e \otimes Z_{\phi}^{^{\otimes 3}}$ were $\approx$99$\%$ for $\phi=\pi/2$.
Further details and a table of pulse parameters are provided in the SI, Sec.~\siNucGates.

% In the case of the $L=2$ parallel entangling gate, the conditional cross talk of the entangling gate, as well as the parallelized nature of the phase gate, lead to a more complicated MQC signal. 
% Considering conditional crosstalk of the form $CR_{\theta}=\ket{0}\bra{0}_e \otimes R_{n_0}(\theta) + \ket{1}\bra{1}_e \otimes R_{n_1}(\theta)$ between the electron and the remaining (untargeted) nuclear qubit leads to an altered MQC signal of:
% \begin{equation}\label{eq:MQC_mixed_signal}
%     P(\ket{0}_e) = \frac{1}{2}\bigg(1 + \frac{1}{2}\text{Re}(e^{i(L\phi - \delta_L)} \text{tr} [A(\phi,\theta)]) \bigg),
% \end{equation}
% where $A(\phi,\theta)=R_{n_0} Z_{\phi} R_{n_0} R_{n_1}^{\dagger} Z_{\phi}^{\dagger} R_{n_1}^{\dagger}$.
% Expanding this expression using explicit forms of each rotation operator reveals two tone beating at frequencies $L\pm1$.
% Thus, the $L=2$ parallel entanglement gate's MQC data in Fig. \ref{fig:mqc_L2_L3}(e) were fit using a sum of two sinusoids, where the expected frequencies are $L_1=1$ and $L_2=3$.
% A derivation of equation (\ref{eq:MQC_mixed_signal}) and the two-tone beating can be found in the SI Sec.~\siMQC.

\subsection{Entangling gate fidelities}

$M$-qubit state fidelities are calculated according to the trace overlap of the quantum state $\rho$ with the target state $\rho_{\text{target}}$; $F_{M}=\text{tr}(\rho \cdot\rho_{\text{target}})$.
Based on the form of the bipartite and sequential entangling gates, when $N_E$ is a multiple of four, the target state is $\ket{0}^{\otimes M}$ --- same as the initial state.
The repeats of the parallel gate were chosen to maximize overlap with $\ket{0}^{\otimes M}$ as a target state.
This simplifies the state fidelities to be given by only a single component of $\rho$;
$F_{M}=\bra{0}^{\otimes M} \rho \ket{0}^{\otimes M}$.
The fidelity of this separable target state can further be approximated by independent $Z$ axis measurements of each qubit:
\begin{equation}
\label{eq:state_fid_M}
    F_M \approx \frac{1}{2^M} \prod_{\ell=1}^{M}(1 + \langle Z_{\ell} \rangle). 
\end{equation}
This approximation ignores correlations between qubits, which is reasonable since the initial and final state are separable with vanishing pairwise covariances and cumulants.
Experimentally, the electron $Z$-axis projection is measured directly using spin-dependent fluorescence, whereas nuclear qubits are measured using $Z$-axis tomography, as shown in Fig. \ref{fig:mqc_L2_L3}(b).
See the SI, Sec.~\siEntangleGateFid, for a derivation of equation (\ref{eq:state_fid_M}) and additional details regarding the fidelity measurements.

\subsection{Generality and extensions}\label{sec:generality_methods}

Random NV–nuclear qubit registers were generated by uniformly sampling nuclear spin positions within a spherical volume surrounding an NV center. 
%The radius of this sphere was set by the atomic density of diamond and abundance (1.1\% in this work) of $^{13}$C nuclear spins.
%For example, the radius of the cut-off sphere needed to encapsulate 100 nuclear qubits is approximately 2.3nm.
%Therefore, the majority of random nuclear qubit generated contribute to the spin bath (see below).
The radius of this sphere was set to 2.3nm, which encapsulates approximately 100 nuclear spins at natural abundance (1.1\%).
Nuclei at the surface of this sphere all contribute to the spin bath.
From the location of each nuclear qubit, the hyperfine matrix was calculated using the dipole–dipole interaction.
The boundary between strongly and weakly coupled was set by the inhomogeneous linewidth of the electron spin transitions, $\sqrt{2}/\pi T_2^*\approx200$\,kHz in this work.
If a register contained any strongly coupled nuclear qubits, the case was not considered further (red region of Fig. \ref{fig:generality_main}a).
Approximately 64\% of randomly generated registers contained at least one nuclear qubit with a hyperfine component larger than this cutoff.
The remaining 36\% of registers (green region of Fig. \ref{fig:generality_main}a) were evaluated for parallel entanglement.
We further applied lower bounds to separate addressable nuclei from the spin bath: $|A_{\parallel}|>15$\,kHz (based on the location of the spin-bath resonance) and $A_{\perp}>10$\,kHz (so that a sufficiently small $N$ can address the qubit).
With these cutoffs, the average number of weakly coupled, addressable nuclear qubits per register is 5.7, with a standard deviation of 2.5 at natural $^{13}$C concentration.

For each of these registers, we searched for parallel entangling gates following the algorithm of Takou \textit{et al.} \cite{takou2024generation}.
Additional details are provided in SI Sec.~\siNucGates~and Sec.~\siGenerality.
Because a large number of registers were generated to improve statistical significance, a conservative parallel entangling gate search was employed.
Specifically, a maximum of $N\leq50$ and a minimum non-unitary entangling power of $\varepsilon_{p,M}(\mathcal{E}) \geq 0.8$ were imposed.
Hence, these results represent a lower bound on the available gates.
%Larger searches increase computational complexity, becoming prohibitive for the many cases considered here.
%The bound on $N$ may exclude additional parallel entanglement cases, while the constraint on $\varepsilon_{p,M}(\mathcal{E})$ ensures that identified gates have fidelities comparable to those explored in this work.
%Lowering the bound on $\varepsilon_{p,M}(\mathcal{E})$ can enable larger entangled states to be generated in parallel, albeit at reduced fidelity \cite{takou2024generation}.
See SI Sec.~\siGenerality~for additional simulations of gate durations, comparisons to $k=2$ and $k=3$ sequential two-qubit gates, and the infidelity arising from residual entanglement.

\begin{acknowledgments}
This work was primarily supported by the National Science Foundation under the ExpandQISE program (award No. OSI-2427091).
M.O. acknowledges support from the Natural Sciences and Engineering Research Council of Canada (NSERC).
A.R.K. acknowledges support from the National Science Foundation (award No. ECCS-2129183). 
The authors are grateful to Edwin Barnes, Sophia Economou and Evangelia Takou for vibrant discussions regarding the entanglement metric framework and other theoretical considerations.
We also thank Jordan Gusdorff and Zhechun Ding for helpful discussions and critical reading of this manuscript.
\end{acknowledgments}

\bibliography{Main}

\end{document}

\title{Supplementary Information: \\ Single-gate, multipartite entanglement on a room-temperature quantum register}

\author{Joseph D. Minnella}
\affiliation{ 
Quantum Engineering Laboratory, Department of Electrical and Systems Engineering, University of Pennsylvania, 200 S. 33rd St. Philadelphia, Pennsylvania, 19104, USA
}
\affiliation{
Department of Physics and Astronomy, University of Pennsylvania, 209 S. 33rd St. Philadelphia, Pennsylvania 19104, USA
}

\author{Mathieu Ouellet}
\altaffiliation[Present address: ]{School of Applied and Engineering Physics, Cornell University, Ithaca, NY 14853, USA}
\affiliation{   
Quantum Engineering Laboratory, Department of Electrical and Systems Engineering, University of Pennsylvania, 200 S. 33rd St. Philadelphia, Pennsylvania, 19104, USA
}

\author{Amelia R. Klein}
\affiliation{ 
Quantum Engineering Laboratory, Department of Electrical and Systems Engineering, University of Pennsylvania, 200 S. 33rd St. Philadelphia, Pennsylvania, 19104, USA
}

\author{Lee C. Bassett}
\email[Correspondence to: ]{lbassett@seas.upenn.edu}
\affiliation{ 
Quantum Engineering Laboratory, Department of Electrical and Systems Engineering, University of Pennsylvania, 200 S. 33rd St. Philadelphia, Pennsylvania, 19104, USA
}

\date{\today}

\maketitle

%\onecolumngrid

\tableofcontents

\newpage

%=====================================================
\section{Experimental Control}
%=====================================================

\begin{figure}[h!]
\centering
\includegraphics{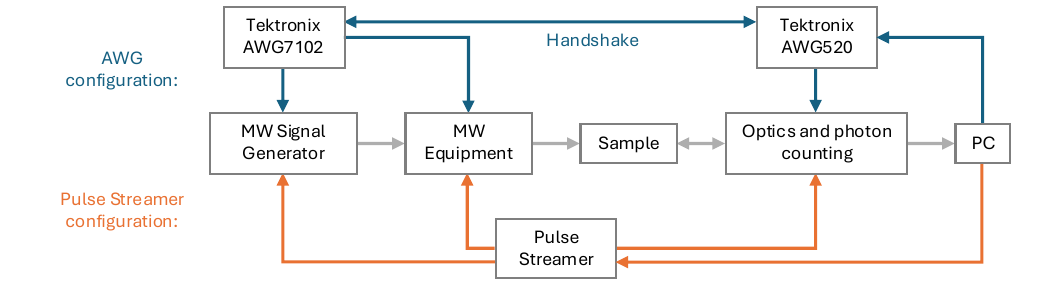}\label{fig:setup}
\caption{
\textbf{Experimental control configurations} 
Equipment in the middle is common to both configurations. 
In both configurations, the two analog channels of either the AWG7102 or the Pulse Streamer were connected to the IQ modulation of the MW signal generator.
The "MW equipment" block contains the mixer, high isolation switch, amplifier and attenuator. 
The "Optics and photon counting" block contains the APD, switch and DAQ.
The PC block contains a very tired PhD student.
} 
\label{fig:e_control}
\end{figure}

Optical and MW control synchronization was handled using two distinct configurations, each of which provided data for this work. 
In the first configuration (AWG configuration of Fig. \ref{fig:setup}), two arbitrary waveform generators (AWG) were used to handle the MW and optical control separately, coordinating control stages via handshaking.
The lead AWG (AWG520, Tektronix) controlled an acousto-optic modulator (AOM) (1250C-848, ISOMET) to modulate a green laser, routed photon counts from an avalanche photodiode (APD) (COUNT-T100 SPCM, Laser Components) to the data acquisition (DAQ) unit (PCIe-6323, National Instruments) using a switch (ZYSWA-2-50DR, Mini-Circuits), and acted as an external clock to the DAQ.
After electron initialization, the higher time resolution (typically 1ns sampling) MW control AWG (AWG7102, Tektronix) was triggered.
The two analog outputs of this AWG controlled the IQ modulation of a signal generator (SG384, Stanford Research Systems, 750-4050MHz LO bandwidth) for phase control of the MWs.
The signal generator output fed into a high bandwidth mixer (ZX05-63LH+, Mini-Circuits) for fast amplitude modulation (controlled by an AWG digital channel).
For high isolation during periods without MWs, the signal then fed into a high isolation switch (ZASWA-2-50DRA+, Mini-Circuits) to prevent leakage.
From there, a fixed amplifier (ZHL-15W-422-S+, Mini-Circuits) and variable attenuator (Rudat 6000-60, Mini-Circuits) were used to control the power of the MW signal before reaching the sample. 
Signals reached the NV using a custom SMA-connected PCB with wire bond traces to the antenna around the solid immersion lens. 
At the end of MW control, the AWG7102 triggered the AWG520 to perform optical readout of, or reinitialize, the electron.

In the second control configuration (Pulse Streamer configuration of Fig. \ref{fig:setup}), rather than using two AWGs to separately control optical and MW stages, a single AWG (Pulse Streamer 8/2, Swabian) was used to control all non-AWG equipment discussed above. 
The Pulse Streamer's analog channels have lower bandwidth (125MHz) than the AWG7102, so to prevent ringing when using square pulses, Gaussian rising and falling edges were optimized and applied before and after setting a constant voltage output. 
Once the constant voltage level was reached, the mixer was used to amplitude modulate the output of the signal generator, providing clean rising and falling edges to the pulse.
Digital channels of the Pulse Streamer can be modulated with 1ns resolution, comparable to the sampling of the AWG7102 used. 

The AWG configuration provided the data shown in Fig. 1 and 2 of the main text, while the Pulse Streamer configuration provided data shown in Fig. 3 and 4.
All of the data shown in this supplement was taken with the AWG configuration, except for Sec. \ref{sec:e_control}, \ref{sec:mqc_sup} and \ref{sec:ent_gate_fid_sup}.

%=====================================================
\section{Electronic qubit}
%=====================================================

\subsection{Control}\label{sec:e_control}

The logical states of the spin 1 electronic qubit were chosen to be $\ket{0} = \ket{m_s=0}$ and $\ket{1} = \ket{m_s=-1}$. 
To calibrate the transition frequency between these spin states, a pulsed electron spin resonance (ESR) experiment (Fig. \ref{fig:e_control}a) was used.
The frequency range swept was first estimated by the Zeeman splitting of the permanent magnet relative to the NV zero field splitting of $\approx2.8$GHz. 
The carrier frequency of the signal generator was detuned using IQ modulation where a low-power 1$\mu$s MW pulse was supplied at each frequency. 
The sample data set in Fig. \ref{fig:e_control}a shows the expected three resonances for each electron spin transition due to the strong 2.2MHz hyperfine interaction with the unpolarized, nuclear spin 1, Nitrogen 14 neighbor. 

From the resonance locations fitted in pulsed ESR, the center one (Nitrogen spin state $\ket{m_s = 0}$) is used as the carrier frequency in all subsequent experiments. 
To calibrate pulse duration, standard resonant Rabi oscillations of the electron were measured using higher power MWs (Fig. \ref{fig:e_control}b).
Typical Rabi frequencies used were around 150MHz for the AWG configuration and 100MHz for the Pulse Streamer configuration. 
From the Rabi fitting results, the pulse durations needed for a $\pi/2$ and $\pi$ rotation were calculated and used for further experiments. 
Due to the high power MWs used, direct time-dependent simulations showed nearly no effect from the unpolarized Nitrogen 14 spin states, which can act as further frequency detuning during Rabi oscillations. 
In lower power regimes, this detuning can prevent the electron from making full oscillations about its Bloch sphere and thus affect the gate fidelities achieved based on the pulse durations calibrated.

\begin{figure}[th!]
\centering
\includegraphics{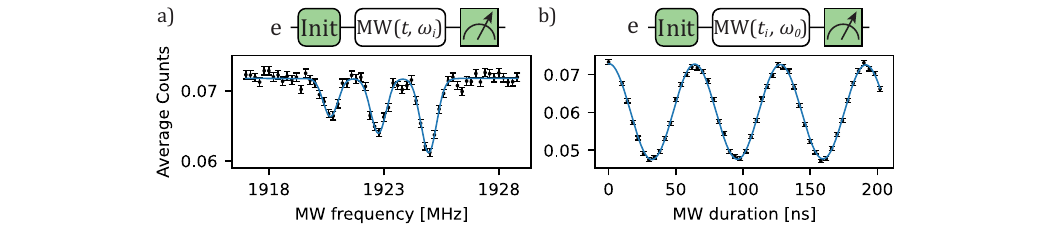}
\caption{
\textbf{Standard electron control calibration} 
\textbf{(a)} Pulsed electron spin resonance experiment diagram for frequency calibration and example results.
Splittings between the resonances match the expected Nitrogen 14 interaction strength of 2.2MHz.
\textbf{(b)} Rabi oscillation experiment diagram for gate duration calibration and example results. 
Very short pulse durations (less than 10ns) were omitted due to the limited rise and fall times of the hardware.
Both of these experiments were executed periodically to update electron gate calibration and mitigate drift over time.
} 
\label{fig:e_control}
\end{figure}

%=====================================================
\subsection{Pulse bootstrap tomography}
%=====================================================

\begin{figure}[th!]
\centering
\includegraphics{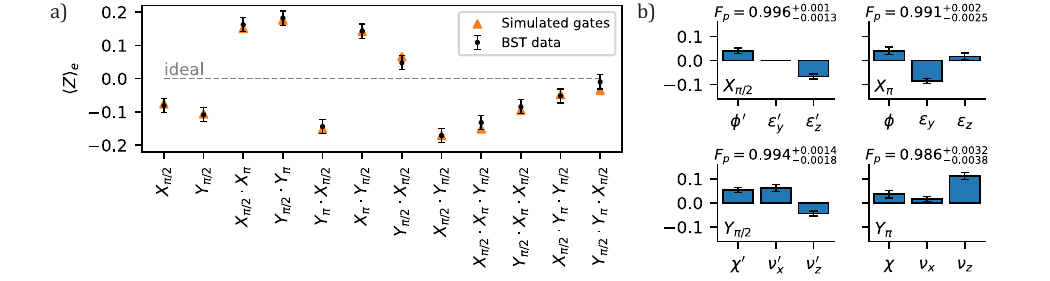}
\caption{
\textbf{Electron control pulse errors}
\textbf{(a)} Results of each pulse sequence needed for error parameter calculations. 
Each sequence provides a linear equation for the error parameters based on the gates used, which can then be solved to extract them.
\textbf{(b)} Gate errors measured from sequences in (a). 
Variable convention follows \cite{dobrovitski2010bootstrap} where $\phi,\chi$ are the angular errors and $\varepsilon_i,\nu_i$ are the off-axis errors. 
Each gate's process fidelity $F_p$ is shown in the title.
Based on errors extracted, each sequence was then simulated (orange triangular points in (a)) to confirm analysis, as shown by the agreement with the original data in (a).
} 
\label{fig:BST}
\end{figure}

Pulse bootstrap tomography (BST) was used to quantify coherent errors in the electron control pulses \cite{dobrovitski2010bootstrap}.
This protocol quantifies the angular and off-axis errors present in all electron pulses used in this work: $X_{\pi/2},X_{\pi},Y_{\pi/2}$ and $Y_{\pi}$. 
Gate errors are extracted by using multiple sequences of pulses, each of which creates a linear combination of the error parameters. 
The sequences ($x$ axis of Fig. \ref{fig:BST}a) together generate a system of equations that yield each gate's error parameters.
The results (black points with error bars) of each pulse sequence required for this method are shown in Fig. \ref{fig:BST}a and the calculated gate error parameters are shown in Fig. \ref{fig:BST}b, following the parametrization in \cite{dobrovitski2010bootstrap}. 
For example, the imperfect $X_{\pi}$ gate is parameterized as:
\begin{equation}
    X_{\pi} = \exp \big( -i \hat{\varepsilon} \cdot \vec{\sigma} (\pi + 2\phi)/2 \big),
\end{equation}
\begin{equation}
    \hat{\varepsilon} = \big(\sqrt{1-\varepsilon_y^2-\varepsilon_z^2},\varepsilon_y,\varepsilon_z \big).
\end{equation}
Based on the results of Fig. \ref{fig:BST}b, imperfect unitary operators were constructed and the sequences of Fig. \ref{fig:BST}a were simulated (orange triangular points) to verify correct analysis of Fig. \ref{fig:BST}a results. 
Additionally, each gate's process fidelity was calculated using these imperfect gates (Fig. \ref{fig:BST}b).
Most importantly, these imperfect gates were used in all subsequent gate-based simulations shown.
Despite all pulses showing approximately 99\% fidelity, one could further mitigate errors through various technique like fine-tuning the pulse voltage for each of the four pulse types to individually adjust the Rabi frequency, or pulse shaping \cite{bartling2403universal}.
However, the small errors present were enough to achieve all desired experiments with XY8 dynamical decoupling sequences further mitigating errors (see Sec. \ref{sec:DD}).

%=====================================================
\subsection{Photoluminescence calibration}
%=====================================================

In order to calibrate the photoluminescence (PL) levels of each electron spin state, two calibration AWG lines were used with every repeat of a PL-calibrated experiment. 
The first line used no MW control, thus measuring the PL of the initialized $\ket{0}$ state, and the second line applied a calibrated $\pi$ pulse to measure the PL level of $\ket{1}$. 
Adiabatic passage of the electron from $\ket{0} \rightarrow \ket{1}$ in the presence of the Nitrogen 14 detuning could have been used as well \cite{bassett2014ultrafast}.
However, at high MW powers, this influence was minimal as discussed in Sec. \ref{sec:e_control} and also in the supplement of \cite{dobrovitski2010bootstrap}. 

\begin{figure}[th!]
\centering
\includegraphics{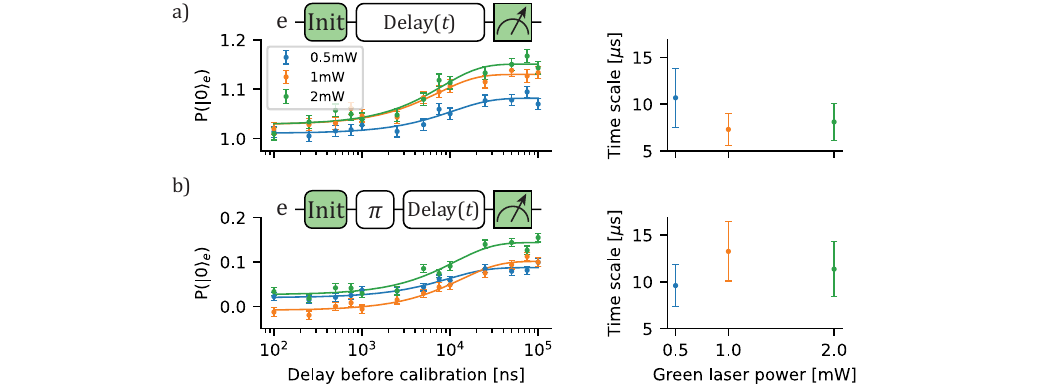}
\caption{
\textbf{Delayed PL calibration}
\textbf{(a,b)} Pulse sequence and results of measuring the time scale of the PL increase when calibrating the $\ket{0}$, $\ket{1}$ states of the electron, respectively. 
Laser power slightly affects the steady state reached, but not the fitted time scale of the increase. 
} 
\label{fig:PL}
\end{figure}

We observed an unexpected increase in PL during $\approx$10$\mu$s and greater experimental durations. 
To better understand this behavior, we added a variable delay before measurement of each calibration line (no pulse and $\pi$-pulse) (Fig. \ref{fig:PL}a,b).
The reference PL levels increase approximately equally for both spin states, with the steady state being roughly 15\% larger than immediate calibration (no delay) depending on the laser power used. 
This unexpected behavior presented issues with longer experiments involving nuclear qubits that were greater than $\approx$10$\mu$s --- the PL from no-delay calibration would consistently indicate larger than 1 probabilities. 
Therefore, all PL calibrations used in this work were delayed by 20$\mu$s to be approximately within the steady state region without considerably increasing overall experimental durations. 
These issues also complicated the use of adiabatic passage, since that method involves longer experimental sequences and the exponentially flat region used to calibrate $\ket{1}$ began to increase, affecting calibration values. 
A similar delay could be added prior to an adiabatic passage in future work, but the high power $\pi$ pulse was accurate enough for our purposes.

%=====================================================
\section{System Hamiltonian}\label{sec:Ham}
%=====================================================

Consider a central electronic spin  surrounded by $L$ nuclear spins in the presence of a static magnetic field $B$ in the $z$ direction. 
Inter-nuclear spin-spin interactions are neglected here due to their much weaker strength than electron-nuclear interactions. 
Furthermore, neglecting non-secular terms of the spin-spin interaction tensor, and in the Larmor frame of the electron, the Hamiltonian governing this spin system is given by:
\begin{equation}\label{eq:Hamiltonian}
    H = \mathbbm{1} \otimes \frac{\omega_{\text{Lar}}}{2} \sum_{\ell=1}^{L} \sigma_{z}^{(\ell)} + \frac{Z_e}{2} \otimes \sum_{\ell=1}^{L} (A_{||}^{(\ell)} \sigma_{z}^{(\ell)} + A_{\perp}^{(\ell)} \sigma_{x}^{(\ell)}),
\end{equation}
where $\omega_{\text{Lar}}=\gamma_{\text{nuc}}B$ is the nuclear spin Larmor frequency, $A_{||,\perp}^{(\ell)}$ are the parallel and perpendicular components of the hyperfine spin-spin tensor of the $\ell$th nuclear spin, and $Z_e$ is the spin operator acting on the electronic two-level subsystem. It is defined as:
\begin{equation}
     Z_e = s_0 \ket{0}\bra{0} + s_1 \ket{1}\bra{1},
\end{equation}
where the logical qubit states $\ket{0}$ and $\ket{1}$ are chosen from the spin-1 triplet as $\ket{0}=\ket{m_s=0}$ and $\ket{1}=\ket{m_s=-1}$, as is common in NV center systems. This choice set  $s_{0}=0$ and $s_{1}=-1$. 
Each Pauli operator $\sigma_i^{(\ell)}$ acts only on the $\ell$th nuclear spin and is defined as the tensor product of identity operators with a single $\sigma_i$ on the $\ell$th factor:
\begin{equation}
\sigma_i^{(\ell)} = \mathbbm{1} \otimes \mathbbm{1} \otimes \underbrace{... \sigma_i...}_{\substack{\ell\text{th} \\ \text{factor}}} \otimes \mathbbm{1}.
\end{equation}

In principle, the Nitrogen 14 spin can be considered in this Hamiltonian. 
However in this work, we restrict our attention to the $L$ surrounding Carbon 13 spins. 
The Nitrogen 14 spin strongly couples to the electron with $A_{||}\approx2.2$ MHz, while the nearest Carbon 13 spins weakly couple, with both parallel and perpendicular components in the range 10 to 100 kHz. 
No strongly coupled Carbon 13 spins were located around the NV studied.
Since our focus is on the dynamics of the Carbon 13 spins using dynamical decoupling, we can safely assume that the electron is effectively decoupled from the Nitrogen 14 spin in the parameter regions of interest. 
Therefore, $L$ will refer to the number of Carbon 13 spins, referred to as the "nuclear qubits" from now on, being considered in the register.

The Hamiltonian of equation (\ref{eq:Hamiltonian}) can be simplified to the form \cite{takou2023precise}:
\begin{equation}
H = \sum_{j \in \{0,1\}} \ket{j}\bra{j} \otimes \sum_{\ell}^{L} H_{j}^{(\ell)},
\end{equation}
where each $H_{j}^{(\ell)}$ is given by:
\begin{equation}\label{eq:H_j}
H_{j}^{(\ell)} = \frac{\omega_{\text{Lar}} + s_{j}A_{||}^{(\ell)}}{2} \sigma_{z}^{(\ell)} + \frac{s_{j}A_{\perp}^{(\ell)}}{2} \sigma_{x}^{(\ell)}.
\end{equation}
This form of the Hamiltonian facilitates the derivation of the free evolution operator $U_f(t)$:
\begin{equation}\label{eq:U_no_simp}
    U_f(t) = \exp \bigg(-it \sum_{j \in \{0,1\}} \ket{j}\bra{j} \otimes \sum_{\ell}^{L} H_{j}^{(\ell)} \bigg).
\end{equation}
The two terms in the sum over electronic spin states $\ket{j}$ commute with one another, and thus the exponential factorizes:
\begin{equation}\label{eq:U_expanded}
    U_f(t) = \exp \bigg(-it \ket{0}\bra{0} \otimes \sum_{\ell}^{L} H_{0}^{(\ell)} \bigg)\exp \bigg(-it \ket{1}\bra{1} \otimes \sum_{\ell}^{L} H_{1}^{(\ell)} \bigg).
\end{equation}
For any finite dimensional Hermitian operator $A$, the operator $\exp(-it \ket{j}\bra{j} \otimes A)$ has a alternative form given by:
\begin{equation}\label{eq:general_proj_simp}
    \exp(-it \ket{j}\bra{j} \otimes A) = \Bar{\ket{j}}\Bar{\bra{j}} \otimes \mathbbm{1} + \ket{j}\bra{j} \otimes \exp(-itA),
\end{equation}
where the barred projection operator is defined as $\Bar{\ket{j}}\Bar{\bra{j}} = \mathbbm{1} - \ket{j}\bra{j}$. 
In the two-dimensional case, these are just $\Bar{\ket{0}}\Bar{\bra{0}} = \ket{1}\bra{1}$ and $\Bar{\ket{1}}\Bar{\bra{1}} = \ket{0}\bra{0}$. 
The form of equation (\ref{eq:general_proj_simp}) is more convenient to work with because the tensor product between the electron and the nuclear qubits is outside of the exponential. 
Applying equation (\ref{eq:general_proj_simp}) to equation (\ref{eq:U_expanded}) yields:
\begin{equation}\label{eq:U_pull_down_j}
\begin{split}
    U_f(t) &= \bigg( \ket{1}\bra{1} \otimes \mathbbm{1} + \ket{0}\bra{0} \otimes \exp \bigg(-it \sum_{\ell}^{L} H_{0}^{(\ell)} \bigg) \bigg) 
            \bigg( \ket{0}\bra{0} \otimes \mathbbm{1} + \ket{1}\bra{1} \otimes \exp \bigg(-it \sum_{\ell}^{L} H_{1}^{(\ell)} \bigg) \bigg) \\
         &= \ket{0}\bra{0} \otimes \exp \bigg(-it \sum_{\ell}^{L} H_{0}^{(\ell)} \bigg) + \ket{1}\bra{1} \otimes \exp \bigg(-it \sum_{\ell}^{L} H_{1}^{(\ell)} \bigg) \\
         &= \sum_{j \in \{0,1\}} \ket{j}\bra{j} \otimes \exp \bigg(-it \sum_{\ell}^{L} H_{j}^{(\ell)} \bigg),
\end{split}
\end{equation}
which shows that the electron projection operators factor out of the exponential in equation (\ref{eq:U_no_simp}). 
Furthermore, each of the nuclear qubit sub-Hamiltonians $H_j^{(\ell)}$ commute for different nuclear qubit  subspaces:
\begin{equation}
    [H_j^{(\ell)},H_j^{(k)}] = 0,
\end{equation}
which again allows the exponential in equation (\ref{eq:U_pull_down_j}) to factorize:
\begin{equation}\label{eq:U_almost_done}
    U_f(t) = \sum_{j \in \{0,1\}} \ket{j}\bra{j} \otimes \prod_{\ell}^{L} \exp \bigg(-it H_{j}^{(\ell)} \bigg).
\end{equation}
Recall that $H_{j}^{(\ell)}$ is defined as the $2^L \times 2^L$-dimensional operator with the identity in all components except the $\ell$th. 
Because of this simple form, the identities factor down from the exponential and we are left with only an exponential of a $2\times2$ operator $H_j^{(\ell)}$ (no identities implied) for each nuclear qubit. 
Thus at this point, the index notation of $\ell$ will simply denote each nuclear qubit's $2\times2$ sub-Hamiltonian, without implied identity operators. 
In this shift in notation and the simplification described, we are left with:
\begin{equation}\label{eq:U_almost_done}
    U_f(t) = \sum_{j \in \{0,1\}} \ket{j}\bra{j} \bigotimes_{\ell}^{L} \exp \bigg(-it H_{j}^{(\ell)} \bigg),
\end{equation}
showing how each $\ell$th subspace of the Hilbert space transforms individually based on the electron state $\ket{j}$. 

To better understand the nuclear qubit transformations, define two normalized vectors $\hat{p}_j^{(\ell)}$ for each nuclear qubit dotted into the Pauli vector $\vec{\sigma}=(\sigma_x,\sigma_y,\sigma_z)$ based on the form of $H_{j}^{(\ell)}$ in equation (\ref{eq:H_j}). 
This can be done for general electron spin projections and then later specified to the logical basis often used:
\begin{equation}\label{eq:n_free}
    \hat{p}_j^{(\ell)} =(s_j A_{\perp}^{(\ell)},0,\omega_{\text{Lar}} +s_j A_{||}^{(\ell)}) / \omega_j^{(\ell)},
\end{equation}
where the normalization factor $\omega_j^{(\ell)}$ is the magnitude of the vector $\vec{p}_j^{\text{ }(\ell)}$: 
\begin{equation}\label{eq:omega_j}
    \omega_j^{(\ell)} \equiv |\vec{p}_j^{\text{ }(\ell)}| =\sqrt{(s_j A_{\perp}^{(\ell)})^2 + (\omega_{\text{Lar}} +s_j A_{||}^{(\ell)})^2}.
\end{equation}
With this definition, each $H_{j}^{(\ell)}$ can be expressed as:
\begin{equation}
    H_{j}^{(\ell)} = \omega_j^{(\ell)} \hat{p}_j^{(\ell)} \cdot \vec{\sigma} / 2.
\end{equation}
Thus, each nuclear qubit operator in equation (\ref{eq:U_almost_done}) is given by:
\begin{equation}
\begin{split}
    \exp \bigg(-it H_{j}^{(\ell)} \bigg) &= \exp \bigg(-i (t \omega_j^{(\ell)}) \hat{p}_j^{(\ell)} \cdot \vec{\sigma} / 2 \bigg) \\
    &= R_{\hat{p}_j^{(\ell)}}(\theta_j^{(\ell)}),
\end{split}
\end{equation}
which is a spin 1/2 rotation operation with rotation axis given by equation (\ref{eq:n_free}) and angle $\theta_j^{(\ell)} = t\omega_j^{(\ell)}$. 
Substituting this into equation (\ref{eq:U_almost_done}) gives the most intuitive form of free evolution for this system:
\begin{equation}\label{eq:U_free_done}
    U_f(t) = \sum_{j \in \{0,1\}} \ket{j}\bra{j} \bigotimes_{\ell}^{L} R_{\hat{p}_j^{(\ell)}}(\theta_j^{(\ell)}).
\end{equation}
Equation (\ref{eq:U_free_done}) shows that the spin state of the electron $\ket{j}$ conditions how each nuclear qubit evolves around its respective Bloch sphere.
Now consider the common choice of electron spin states $\ket{0}=\ket{m_s=0}$ and $\ket{1}=\ket{m_s=-1}$. 
If the electron is in state $\ket{0}$, nuclear qubit evolution simply obeys Larmor precession about the $\hat{z}$ axis of the magnetic field. 
However, if the electron spin is in state $\ket{1}$, it acts as a small change in the local magnetic field of each nuclear qubit. 
This leads to a unique precession axis and rate for each nuclear qubit based on equations (\ref{eq:n_free}) and (\ref{eq:omega_j}).
In the case of free evolution, $\hat{p}_0^{(\ell)}$ and $\hat{p}_1^{(\ell)}$ are nearly aligned and importantly fixed by the strength of the hyperfine coupling of the $\ell$th nuclear qubit. 
Yet the non-zero angle between $\hat{p}_0^{(\ell)}$ and $\hat{p}_1^{(\ell)}$ allows for dynamical decoupling of the electron to engineer controllable rotations of the nuclear qubits about desired axes, as discussed in the following section.

%=====================================================
\subsection{Dynamical decoupling}\label{sec:DD}
%=====================================================

Dynamical decoupling (DD) refers to the process of switching the electron state periodically to achieve two main outcomes. 
First, DD can be used to engineer controllable interactions with nuclear qubits (\textit{i.e.}, implement nuclear gates). 
Second, the process of repeatedly flipping the electronic state averages out its interactions with the larger spin environment, thereby decoupling it from unwanted noise \cite{van2012decoherence}. 
This technique is essential for extending the electron coherence time beyond its short dephasing $T_2^*$ time.
At room temperature, this is often a difference of three orders of magnitude from $\mu$s to ms. 
In what follows, we focus our attention on dynamical decoupling as a tool of sensing and controlling nuclear qubits. 
Between electron flips, the system undergoes free evolution according to $U_f(t)$ in equation (\ref{eq:U_free_done}). 
By carefully controlling the time between flips and the total amount of them, one can realize universal nuclear qubit control. 

The simplest DD sequence is commonly known as CPMG (Carr-Purcell-Meiboom-Gill), with a unit pulse sequence consisting of two $\pi$-pulses with even spaced delays and about the same axis.
The timing conventions used throughout this work is that the total CPMG unit pulse takes a duration $t$. 
This unit pulse is then repeated a total of $N$ times to achieve a particular outcome with the nuclear qubits. 
Schematically, the pulse sequence is given by:
\begin{equation}
\text{CPMG} = \bigg(\frac{t}{4} - \pi - \frac{t}{2} - \pi - \frac{t}{4} \bigg)^N,
\end{equation}
where each $\pi$ denotes a $\pi$-pulse about either the $X$ or $Y$ axis.

First, consider the effect of a single unit pulse ($N=1$) with $\pi$-pulses about the $X$ axis. 
Using equation (\ref{eq:U_free_done}), the total unitary operator is given by:
\begin{equation}
    U_{\text{CPMG}} = U_{f}(t/4) \cdot X_{\pi} \cdot U_{f}(t/2) \cdot X_{\pi} \cdot U_{f}(t/4).
\end{equation}
We only need to consider how $X_{\pi} U_{f}(t/2) X_{\pi}$ transforms the electronic subspace since it acts as the identity on all nuclear qubits.
This transformation simply swaps the projectors: 
\begin{equation}
\begin{split}
    X_{\pi} \cdot U_{f}(t/2) \cdot X_{\pi} &= \ket{1}\bra{1} \bigotimes_{\ell}^{L} R_{\hat{p}_0^{(\ell)}}(\theta_0^{(\ell)}(t/2)) + \ket{0}\bra{0} \bigotimes_{\ell}^{L} R_{\hat{p}_1^{(\ell)}}(\theta_1^{(\ell)}(t/2)) \\
    &\equiv \Bar{U}_{f}(t/2).
\end{split}
\end{equation}
This simplifies which terms appear in the product of $U_{CPMG} = U_{f}(t/4)\Bar{U}_{f}(t/2)U_{f}(t/4)$. 
The resulting expression, which matches the analysis of \cite{takou2023precise}, leads to a net unitary of very similar form to free evolution:
\begin{equation}\label{eq:U_CPMG_N1}
    U_{\text{CPMG}}(t) = \sum_{j \in \{0,1\}} \ket{j}\bra{j} \bigotimes_{\ell}^{L} R_{\hat{n}_j^{(\ell)}}(\phi^{(\ell)}),
\end{equation}
however, these rotation operators are different than the ones given in the free evolution analysis. 
Using the shorthand $R_{\hat{p}_j^{(\ell)}}(\theta_j^{(\ell)}(t)) = R_{j}^{(\ell)}(t)$ for the free evolution rotations, the net DD rotations are given by:
\begin{equation}\label{eq:R_cases}
    R_{\hat{n}_j^{(\ell)}}(\phi^{(\ell)}) =
\begin{cases} 
R_{0}^{(\ell)}(t/4)R_{1}^{(\ell)}(t/2)R_{0}^{(\ell)}(t/4) \text{ for } j=0,\\ 
R_{1}^{(\ell)}(t/4)R_{0}^{(\ell)}(t/2)R_{1}^{(\ell)}(t/4) \text{ for } j=1
\end{cases}
\end{equation}
The DD nuclear rotation angles $\phi^{(\ell)}$ are electron spin state independent, unlike the case of free evolution \cite{takou2023precise}. 
This allows for both the unit pulse times and repeats (next paragraph) to be calibrated independent of the electron state.

Repeating the unit pulse $N$ times can now be analyzed directly.
Using induction, one can show that the $N$th power of equation (\ref{eq:U_CPMG_N1}) increases the nuclear rotations angles linearly:
\begin{equation}\label{eq:U_CPMG_N}
    U^{N}_{\text{CPMG}}(t) = \sum_{j \in \{0,1\}} \ket{j}\bra{j} \bigotimes_{\ell}^{L} R_{\hat{n}_j^{(\ell)}}(N\phi^{(\ell)}).
\end{equation}
Therefore, the choice of $N$ sets the net nuclear rotation angle.
On the other hand, the choice of unit pulse time sets the rotation axes and the angular resolution $\Delta\phi=\phi^{(\ell)}$.
Most often, the unit pulse time is chosen to be on resonance with a particular nuclear qubit's free precession. 
In doing so, the two electron spin dependent rotation axes either take a parallel $(\hat{n}_0 \cdot \hat{n}_1)_{\ell} = 1$ or antiparallel $(\hat{n}_0 \cdot \hat{n}_1)_{\ell} = -1$ alignment. 
This is in stark contrast to the small, but nonzero angle between free evolution rotation axes $\hat{p}_j^{(\ell)}$, which is fixed by the physical parameters of the system. 
Parallel rotation axes correspond to nuclear rotations that are independent of the electron state (unconditional gate), while anti-parallel axes generate operations that are maximally dependent on the electron state. 
In the anti-parallel case, this leads to an entangling gate between the qubits, which can be maximally entangling with the proper choice of $N$. 
Details of these different types of rotations will be discussed more in Sec. \ref{sec:NG Calibration}. 
Together, they enable dynamical decoupling to  be a form of universal control for nuclear qubits \cite{taminiau2014universal}. 

Using the form of equation (\ref{eq:U_CPMG_N1}), \cite{taminiau2012detection} derived a closed form of $(\hat{n}_0 \cdot \hat{n}_1)_{\ell}$ for the common choice of spin projections in the NV system, which was recently generalized in \cite{takou2023precise}. 
In either case, analyzing $(\hat{n}_0 \cdot \hat{n}_1)_{\ell}$ as a function of unit pulse time allows one to derive an expression for which unit pulse times give rise to these resonant, (anti)parallel axis geometries. 
However, the resulting equation is transcendental and cannot be solved analytically \cite{taminiau2012detection}. 
Under the approximation of a strong magnetic field relative to the hyperfine couplings ($\omega_\text{Lar} \gg
 A_{||,\perp}^{(\ell)}$), an approximate resonance condition equation can be derived:
\begin{equation}\label{eq:res}
    t_m^{(\ell)} = \frac{4 \pi m}{\omega_0^{(\ell)} + \omega_1^{(\ell)}} \text{ for } m \in \mathbbm{Z},
\end{equation}
where $\omega_j^{(\ell)}$ is defined in equation (\ref{eq:omega_j}). 
The parity of $m$ gives rise to parallel or antiparallel nuclear rotation axes:
\begin{equation}\label{eq:n01_cases}
   (\hat{n}_0 \cdot \hat{n}_1)_{\ell}(t_m^{(\ell)})  =
\begin{cases} 
-1 \text{ for odd }m \rightarrow m=2k-1,\\ 
+1 \text{ for even }m \rightarrow m=2k
\end{cases}
\end{equation}
because of this, the choice of resonance will always be quoted as a value of $k$ starting at 1 and the context of the resonance (conditional or unconditional) can be determined from the context. 

Although the analysis considered thus far has analyzed CPMG DD, this form of DD is often not viable experimentally. 
Due to small pulse errors in the electron $\pi$-pulses, the error accumulates after many applications. 
Alternative DD sequences have been developed to cancel these pulse errors to leading order, preserving the state of the electron \cite{wang2012comparison}. 
To this end, XY8 DD was predominantly used throughout this work. 
The unit pulse for this sequence has four times as many $\pi$-pulses, of which the axes of each rotation is chosen symmetrically:
\begin{equation}
\text{XY8} = \frac{t}{4} - X_{\pi} - \frac{t}{2} - Y_{\pi} - \frac{t}{2} - X_{\pi} - \frac{t}{2} - Y_{\pi} - \frac{t}{2} - Y_{\pi} - \frac{t}{2} - X_{\pi} - \frac{t}{2} - Y_{\pi} - \frac{t}{2} - X_{\pi} - \frac{t}{4}.
\end{equation}
The convention of timing chosen for this work defines the XY8 unit pulse to be four times longer than the CPMG unit pulse. 
Therefore, all unit pulse times always have the meaning of $t/2$ between successive $\pi$-pulses, except for the first and last delay period of $t/4$, and $N$ unit pulse repeats corresponds to $2N$ $\pi$-pulses in total.
Importantly, all results derived for CPMG apply equally to XY8.

When using only the XY8 unit pulse, it can be difficult to achieve particular nuclear qubit rotations with high accuracy due to the discrete nature of angle accumulation through unit pulse repeats. 
Therefore, in order to strike a balance between electron and nuclear rotation errors, multiple schemes of DD were applied using as much XY8 as possible. 
To interpolate between multiples of four repeats, DD schemes with smaller unit pulses were used.
This included CPMG, XY4 and XY6 in the order of $N\equiv1,2,3(\text{mod}4)$.
Symmetrization beyond XY8 did not show an improvement in preserving the electron state.
Interpolating with these smaller unit pulses allowed for more precise nuclear qubit rotation operators to be achieved, as discussed more in Sec. \ref{sec:NG Calibration}. 

%=====================================================
\section{Detection of nearest $^{13}$C qubits}\label{sec:DD_spec}
%=====================================================

\begin{figure}[h!]
\centering
\includegraphics{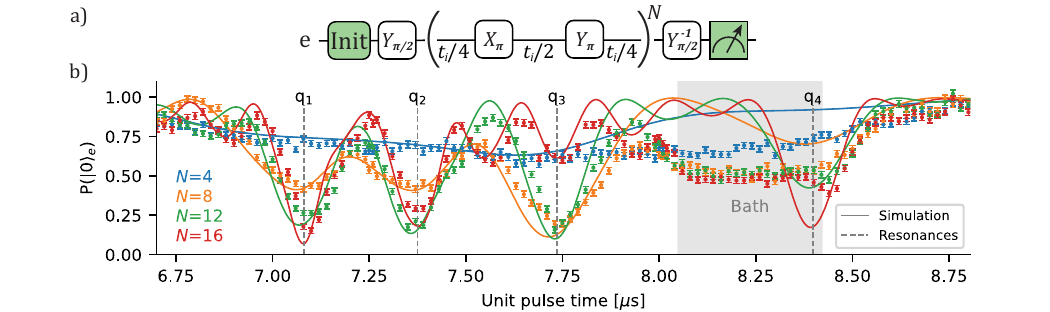}
\caption{
\textbf{Second order ($k=2$) DD spectroscopy}
\textbf{(a)} Schematic of the pulse sequence used, with the XY8 unit pulse abbreviated. 
Placing the electron in an equatorial state leads the following variable DD sequence to entangle nearby nuclear qubits in their thermal states. 
This entanglement creates off diagonal elements of the electron density matrix, which are mapped onto the final state with the final $Y_{\pi/2}^{-1}$ pulse.
\textbf{(b)} Second order resonance region of four distinct nuclear qubits.
Dashed lines denote resonance unit pulse times based on equation (\ref{eq:res}).
Note, the simulation is not fit to the data.
}
\label{fig:DDspec}
\end{figure}

DD spectroscopy \cite{taminiau2012detection} was used to sense the spin environment and determine the hyperfine parameters $A_{||,\perp}$ of nearby $^{13}$C nuclear qubits (Fig. \ref{fig:DDspec}). 
This experiment works by entangling the electron with the nuclear qubits in uninitialized thermal states.
This entanglement reduces the probability of finding the electron in its initial state, producing a resonance when entanglement is achieved.  
The unit pulse times that lead to these resonances are given by equation (\ref{eq:res}) for odd $m$.
This probability as a function of unit pulse time can be derived by considering an initial density matrix of the form:
\begin{equation}
\begin{split}
    \rho_0 &= \ket{X}\bra{X} \otimes \bigg(\frac{1}{2}I\bigg)^{\otimes L} \\
    &= \frac{1}{2^{L+1}}\bigg(\ket{0}\bra{0} \otimes I^{\otimes L} + \ket{1}\bra{1} \otimes I^{\otimes L} + \ket{0}\bra{1} \otimes I^{\otimes L} + \text{h.c.} \bigg)
\end{split}
\end{equation}
after the first $\pi/2$ pulse is applied to the electron.
Using the unitary operator of equation (\ref{eq:U_CPMG_N1}), applying the DD sequence to the system results in:
\begin{equation}
    U_{DD}\rho_0 U_{DD}^{\dagger} = \frac{1}{2^{L+1}} \bigg(\ket{0}\bra{0} \otimes I^{\otimes L} 
    + \ket{1}\bra{1} \otimes I^{\otimes L} 
    + \ket{0}\bra{1} \bigotimes_{\ell}^L R_{\hat{n}_0^{(\ell)}} R_{\hat{n}_1^{(\ell)}}^{\dagger} 
    + \text{h.c.} \bigg),
\end{equation}
which has created a variable off-diagonal element for the electron that depends on the conditional nature of the nuclear qubit rotation operators $R_{\hat{n}_0^{(\ell)}}$ and $R_{\hat{n}_1^{(\ell)}}$. 
When these are unconditional, or non-entangling, we have $R_{\hat{n}_0^{(\ell)}} R_{\hat{n}_1^{(\ell)}}^{\dagger}=\mathbbm{1}$ for each $\ell$
Therefore, the off diagonal element is unchanged and the probability of the electron remaining in the state $\ket{X}$ is just 1.
However, for conditional operators, tracing out the nuclear qubits and measuring in the $X$ basis (equivalent to applying the final $\pi/2$-pulse) leads to:
\begin{equation}
\begin{split}
    \rho_e &= \trace_{\ell} \rho \\
    &= \frac{1}{2} \bigg(\ket{0}\bra{0} + \ket{1}\bra{1} + \ket{0}\bra{1}\frac{1}{2^L} \prod_{\ell}^L\trace (R_{\hat{n}_0^{(\ell)}} R_{\hat{n}_1^{(\ell)}}^{\dagger}) + \text{h.c.} \bigg),
\end{split}
\end{equation}
and
\begin{equation}\label{eq:DD_spec}
\begin{split}
    P_e(\ket{X}) &= \trace (\ket{X}\bra{X}\rho_e) \\
    &= \frac{1}{2} \bigg(1 + \frac{1}{2^L} \text{Re} \big( \prod_{\ell}^{L} \trace( R_{\hat{n}_0^{(\ell)}} R_{\hat{n}_1^{(\ell)}}^{\dagger} ) \big) \bigg).
\end{split}
\end{equation}
Again, applying the final $\pi/2$-pulse makes equation (\ref{eq:DD_spec}) equivalent to measuring $P_e(\ket{0})$, which is what is done experimentally.
Note, the rotation angles implicitly depend both on unit pulse time and repeats, and the axes depends on unit pulse time, as explained in Sec. \ref{sec:DD}. 
In addition to the nearest nuclear qubits, this experiment also senses the larger weak spin bath, which shows up as a much broader resonance indicating entanglement with the environment (Fig. \ref{fig:DDspec}). 
As discussed in the main text, the set of first order resonances ($k=1$) are not resolvable due to overlap with the bath.
At higher orders, as low as the second order ($k=2$) as shown in Fig. \ref{fig:DDspec}, stronger coupled nuclear qubits separate from the bath and become individually resolvable. 
In addition to the three nuclear qubits considered in the main text, one can also start to see the next strongest nuclear qubit on the right side (positive parallel hyperfine coupling) of the bath, which becomes resolved at the third ($k=3$) order.
Note, the simulation shown is not a fit to the data, but a prediction based on previously measured hyperfine couplings (Table \ref{tab:HF table}).
The couplings' initial measurement did involve the fitting of a similar simulation to this experiment, but this was not repeated here. 

\begin{table}[h!]
\centering
\begin{tabular}{c|c|c}
Nuclear qubit & $A_{||}$ [kHz] & $A_{\perp}$ [kHz]\\
\hline
$q_1$ & -118.8 & 68.4 \\
$q_2$ & -86.1 & 58.3 \\
$q_3$ & -46.4 & 67.7 \\
$q_4$ & 10.1 & 25.4 \\
\end{tabular}
\caption{
\textbf{Four strongest nuclear qubits}
No resonances other than the ones observed in Fig. \ref{fig:DDspec} have been observed for this NV up to third order, indicating these are the four strongest coupled nuclear qubits.
The similar couplings, particularly the sign of the parallel components, of the first three qubits are of central importance for their ability to be entangled in parallel (more information in Sec. \ref{sec:Parallel entanglement}).
All simulations included these four nuclear qubits, even when an experiment did not intend to control some of them.
}
\label{tab:HF table}
\end{table}

%=====================================================
\section{Nuclear gates}
%=====================================================

\subsection{Calibration}\label{sec:NG Calibration}

\begin{figure}[h!]
\centering
\includegraphics{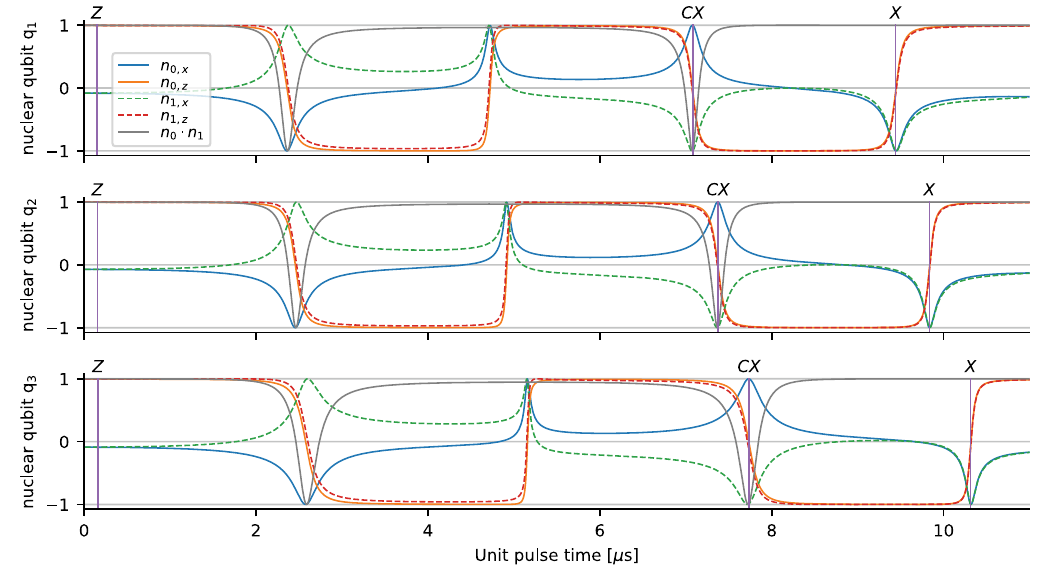}
\caption{
\textbf{Nuclear qubit rotation axis components}
Individual components of each nuclear qubit rotation axis $\hat{n}_j^{(\ell)}$ for each electron spin state $\ket{j}$ based on equation (\ref{eq:n_k}). 
$Y$ components are 0 based on Hamiltonian frame. 
Unit pulse times used for each gate (table \ref{tab:NG calib}) are marked with purple lines.
}
\label{fig:n_axes}
\end{figure}

Each nuclear gate is calibrated by first setting the unit pulse time $t$ to select the rotation axis and whether the rotation is conditional or unconditional, and second, setting the unit pulse repeats $N$ to achieve a desired rotation angle. 
Based on the choice of coordinates for the Hamiltonian in equation (\ref{eq:Hamiltonian}), the unit pulse time can be chosen to rotate nuclear qubits about $X$ and $Z$ axes. 
As discussed in Sec. \ref{sec:DD}, $X$ rotations occur resonantly, or for particular unit pulse times, given by equation (\ref{eq:res}). 
The direction of these can be conditioned on the state of the electron ($CX$ gate) or independent of it (unconditional $X$ gate).
At all other times outside of these resonances, the rotation is unconditionally about the $Z$ axis.
Each of these gates can be visualized (Fig. \ref{fig:n_axes}) by considering each component of the rotation axes using the following identities on the nuclear rotation operators in equation (\ref{eq:R_cases}):
\begin{equation}\label{eq:phi_tr}
    \trace(R_{\hat{n}}(\phi)) = 2 \cos (\phi/2) \rightarrow \phi = 2 \arccos (\frac{1}{2}\trace(R_{\hat{n}}(\phi))),
\end{equation}
\begin{equation}\label{eq:n_k}
    \trace(\sigma_k R_{\hat{n}}(\phi)) = -2 i \sin(\phi/2) n_{k} \rightarrow n_k = \frac{\trace(\sigma_k R_{\hat{n}}(\phi))}{2i \sin(\phi/2)}
\end{equation}
Except in the case of parallel entanglement, all $CX$ and $X$ unit pulse times were chosen to be at the second order resonance. 
This provides well-individualized control on each nuclear qubit, with the crosstalk on other spins being approximately a $Z$ gate. 
The flexibility in the choice of unit pulse time for $Z$ gates allowed for quick unit pulse times to be used, of which the exact values were set based on the angular calibration.
See table \ref{tab:NG calib} for the unit pulse times used for all the gates. 

Once the unit pulse times set the nuclear rotation axis, the net rotation angle can be calibrated. 
The choice of unit pulse time also affects the angular resolution $\Delta\phi=\phi^{(\ell)}$ of each rotation operator.
Since the unit pulse time is often set to a resonance time for the rotation axes, $\phi^{(\ell)}$ is also fixed, up to which order resonances are used. 
The unit pulse can then be repeated $N$ times to achieve a rotation angle $N\phi^{(\ell)}$, as highlighted in equation (\ref{eq:U_CPMG_N}).
Thus, once the $N=1$ rotation angle $\phi^{(\ell)}$ is calculated, either using equation (\ref{eq:phi_tr}) numerically or closed form expressions as in \cite{takou2023precise}, the repeats needed ${N}^{(\ell)}_{\text{cal}}$ for a desired rotation angle $\phi^{(\ell)}_{\text{target}}$ on the $\ell$th nuclear qubit can be calibrated according to:
\begin{equation}
    {N}^{(\ell)}_{\text{cal}} = \text{round}(\phi^{(\ell)}_{\text{target}} / \phi^{(\ell)} ).
\end{equation}
The discrete nature of angle calibration does introduces a finite angular error for each gate, but often this is small and can be reduced by using multiple types of DD unit pulses (see end of Sec. \ref{sec:DD}) or strategic choice of the magnetic field strength (see Sec. \ref{sec:B field} for details). 
For $Z$ gate calibration, the unit pulse time was specifically chosen so that $N=4$ (single XY8 unit pulse with minimal error) repeats created a $\pi/2$ rotation for each nuclear qubit. 
Almost all nuclear gates used in this work were rotations by $\pi/2$.
The number of repetitions needed to achieve this net angle, as well as the absolute difference from $\pi/2$, is listed in table \ref{tab:NG calib} for each nuclear qubit. 

\begin{table}[h!]
\centering
\begin{tabular}{c|c|c|c|c|c}
Gate & Nuclear qubit & Unit pulse time [\(\mu\)s] & Repeats & Gate time [\(\mu\)s] & $|$Angular error$|$ [rad] \\
\hline
\multirow{3}{*}{$CX_{\pi/2}$} 
 & $q_1$ & 7.081 & 7 & 49.6 & 0.042 \\
 & $q_2$ & 7.375 & 7 & 51.6 & 0.001 \\
 & $q_3$ & 7.736 & 5 & 38.7 & 0.009 \\
 %& 3 & 8.398 & 11 & 92.4 & 0.012 \\
\hline
\multirow{3}{*}{$X_{\pi/2}$} 
 & $q_1$ & 9.442 & 7 & 66.1 & 0.000 \\
 & $q_2$ & 9.834 & 9 & 88.5 & 0.081 \\
 & $q_3$ & 10.314 & 12 & 123.8 & 0.040 \\
 %& 3 & 11.197 & 135 & 1511.6 & 0.001 \\
\hline
\multirow{3}{*}{$Z_{\pi/2}$} 
 & $q_1$ & 0.148 & 4 & 0.6 & 0.001 \\
 & $q_2$ & 0.154 & 4 & 0.7 & 0.000 \\
 & $q_3$ & 0.162 & 4 & 0.6 & 0.003 \\
 %& 3 & 0.175 & 4 & 0.7 & 0.001 \\
\end{tabular}
\caption{
\textbf{Simulated nuclear gate calibration}
DD unit pulse parameters used for each gate for each nuclear qubit. 
Simulation results verified experimentally in Sec. \ref{sec:exp gate calib}.
}
\label{tab:NG calib}
\end{table}

%=====================================================
\subsection{Parallel $Z$ gates}\label{sec:parallel_phase}
%=====================================================

The unit pulse times for $Z$ gates were also chosen to lie in a region where they could be accurately parallelized in multi-nuclear qubit experiments. 
Prior to the first order resonances, each nuclear qubit's rotation axes are nearly identical and aligned along $Z$. 
Furthermore, a single unit pulse time can be chosen (averaged based on individual times in table \ref{tab:NG calib}) to achieve approximately equal rotation angles on each nuclear qubit, resulting in a multi-qubit gate of the form:
\begin{equation}\label{eq:Z_para}
    Z_L({\phi}) = I_e \otimes Z_{\phi}^{(1)} \otimes...\otimes Z_{\phi}^{(L)}
\end{equation}
on $L$ nuclear qubits.
These gates are useful in general, but will be particularly useful in Sec. \ref{sec:mqc_sup} for the multiple quantum coherences method.
Calculating the process fidelity $F_p$ of the DD sequences against equation (\ref{eq:Z_para}) with and without electron pulse error (table \ref{tab:Z_para_fid}) shows high ($\approx$99$\%$) fidelity for each subset of qubits.
\begin{table}[h!]
\label{tab:Z_para_fid}
\centering
\begin{tabular}{c|c|c|c|c}
Nuclear qubit subsets & Unit pulse time [$\mu$s] & Unit pulse repeats & $F_p$ without pulse error & $F_p$ with pulse error \\ 
\hline
$q_1$, $q_2$ & 0.151 & 4 & 0.994 & 0.993 \\
\hline
$q_1$, $q_3$ & 0.155 & 4 & 0.990 & 0.990 \\
\hline
$q_2$, $q_3$ & 0.158 & 4 & 0.993 & 0.993\\
\hline
$q_1$, $q_2$, $q_3$ & 0.155 & 4 & 0.988 & 0.987 \\
\end{tabular}
\caption{
\textbf{Parallelized $Z$ gate parameters and process fidelities}
Other parallelized angles were used in this work, but $\pi/2$ is shown here as a representative example.
}
\label{tab:Z_para}
\end{table}

%=====================================================
\subsection{Parallel entanglement}\label{sec:Parallel entanglement}
%=====================================================

\begin{figure}[h!]
\centering
\includegraphics{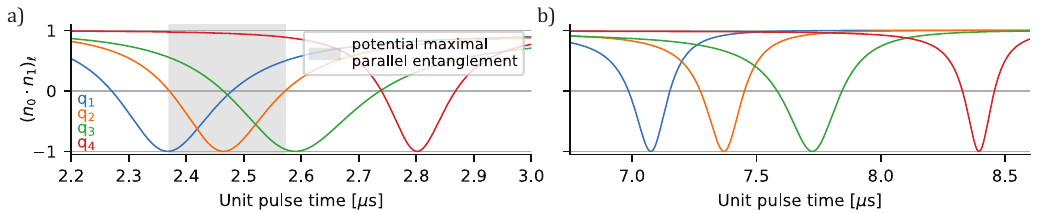}
\caption{
\textbf{Conditional rotation axis alignment}
\textbf{(a)} First order, referred to in main text as "parallel order", resonances.
The grey shaded regions is set based on the intersection of at least two of the $(\hat{n}_0 \cdot \hat{n}_1)_{\ell}$ curves that are negative and thus indicates when one DD gate can entangle multiple nuclear qubits.
\textbf{(b)} Second order, or "individual order", resonances.
Only maximal bipartite entangling gates are possible at this, and all subsequent, order for these four nuclear qubits.
}
\label{fig:n01_k1_k2}
\end{figure}

In order to calibrate a parallel entangling gate, multiple steps must be taken to ensure high fidelity operation. 
Foremost, a maximal parallel entangling gate is possible when there exists a unit pulse time such that $(\hat{n}_0 \cdot \hat{n}_1)_{\ell} < 0$ for each targeted $\ell$th nuclear qubit \cite{takou2023precise}.
Considering this quantity at first and second order (Fig. \ref{fig:n01_k1_k2}) for the four strongest nuclear qubits foremost shows there is an intersection (grey shaded region of Fig. \ref{fig:n01_k1_k2}a) between the three strongest nuclear qubits ($\qone$, $\qtwo$ and $\qthree$), and subsets of them, at first order ($k=1$) (Fig. \ref{fig:n01_k1_k2}a).
At second order ($k=2$), and therefore all subsequent orders, there does not exist any intersection for these nuclear qubits.
Fig. \ref{fig:n01_k1_k2} additionally shows that the fourth strongest nuclear qubit $\qfour$ does not satisfy this requirement with any other nuclear qubits at either order. 
$\qfour$ is only included in Fig. \ref{fig:n01_k1_k2} and this analysis to highlight its spectator role in these experiments and to quantify any non-maximal entanglement generated with it.
Thus, for these specific hyperfine couplings, at most three nuclear qubits could be entangled with a single gate with the correct choice of $N$.

Once it has been determined that a parallel entangling gate is possible, the precise calibration of the DD parameters can be determined using the algorithm of Takou \text{et. al.} \cite{takou2024generation}.
The key details of how this algorithm was used in this work are summarized here. 
For each subset of nuclear spins containing at least two nuclear spins, \textit{i.e.} $\{ \{q_1,q_2\},\{q_1,q_3\},\{q_2,q_3\},\{q_1,q_2,q_3\} \}$ in this work, scan the parameter space of $(t,N)$ pairs and calculate the $M$-way non-unitary entangling power $\varepsilon_{p,M}(\mathcal{E})$. 
Then, save which $(t,N)$ pairs generate a $\varepsilon_{p,M}(\mathcal{E})$ greater than a set threshold.
$M$ refers to the total number of qubits in the system, including the electron and $L_{\text{total}}$ will refer to the number of nuclear qubits, $M=L_{\text{total}}+1$.
$L_{\text{total}}$ is in principle fixed for a given NV-nuclear register based on the number of weakly coupled nuclear qubits resolvable from the spin bath, in this work $L_{\text{total}}=4$.
The number of targeted nuclear qubits is given by $L$, which partitions $L_{\text{total}}$ according to $L_{\text{total}} = L + L_{\text{not targeted}}$.
$L$ changes based on the number of nuclear qubits targeted for entanglement in a given experiment.
With these counting definitions, the non-unitary entanglement metric is (approximately) given by
\begin{equation}\label{eq:ep_non_unitary}
    \varepsilon_{p,M}(\mathcal{E}) \approx
    \frac{\varepsilon_{p,M}(U_{DD})}{2}
    \left(1 + \prod_{\substack{\ell \in \text{not} \\ \text{targeted}}}^{L_{\text{total}}-L}
    G^{(\ell)}_1 \right).
\end{equation}
where $\varepsilon_{p,M}(U_{DD})$ is the unitary entangling power calculated over the targeted nuclear qubits in the subset, given by
\begin{equation}\label{eq:ep_unitary}
    \varepsilon_{p,M}(U_{DD}) = \bigg( \frac{d}{d+1} \bigg)^M
    \prod_{\substack{\ell \in \text{targeted} \\ \text{subset}}}^{L} (1 - G_1^{(\ell)}),
\end{equation}
and $G_1^{(\ell)}$ is the first Makhlin invariant given by
\begin{equation}
    G_1^{(\ell)} = \bigg(\cos^2 \frac{N\phi^{(\ell)}}{2} + (\hat{n}_0 \cdot \hat{n}_1)^{(\ell)} \sin^2 \frac{N\phi^{(\ell)}}{2} \bigg)^2.
\end{equation}
This approximation is very accurate; for a comparison between the more complicated exact form of $\varepsilon_{p,M}(\mathcal{E})$ with equation \ref{eq:ep_non_unitary}, see \cite{takou2024generation}.
$G_1^{(\ell)}$ ranges from 0 (maximal entanglement between the electron and nuclear qubit $q_{\ell}$) and 1 (no entanglement). 
Ignoring the constant factor of $\left( d/(d+1) \right)^M$ in equation \ref{eq:ep_unitary} (convention adopted throughout this work), leads $\varepsilon_{p,M}(U_{DD})$ to take values between 0 (no entanglement between electron and targeted nuclear qubits) to 1 (maximal entanglement between electron and targeted nuclear qubits).
If any residual entanglement is generated with non-targeted nuclear qubits, their Makhlin invariant will be less than 1, thus lowering the non-unitary entangling power $\varepsilon_{p,M}(\mathcal{E})$.
The non-unitary entangling power is bounded from above by the unitary entangling power: $\varepsilon_{p,M}(\mathcal{E}) \leq \varepsilon_{p,M}(U_{DD})$, with equality occurring when no residual entanglement is generated \cite{takou2024generation}.
Therefore, the use of the non-unitary entangling power over the unitary version is crucial when considering subsets of targeted nuclear qubits so that residual entanglement with non-targeted nuclear qubits is properly taken into account.

\begin{figure}[h!]
\centering
\includegraphics{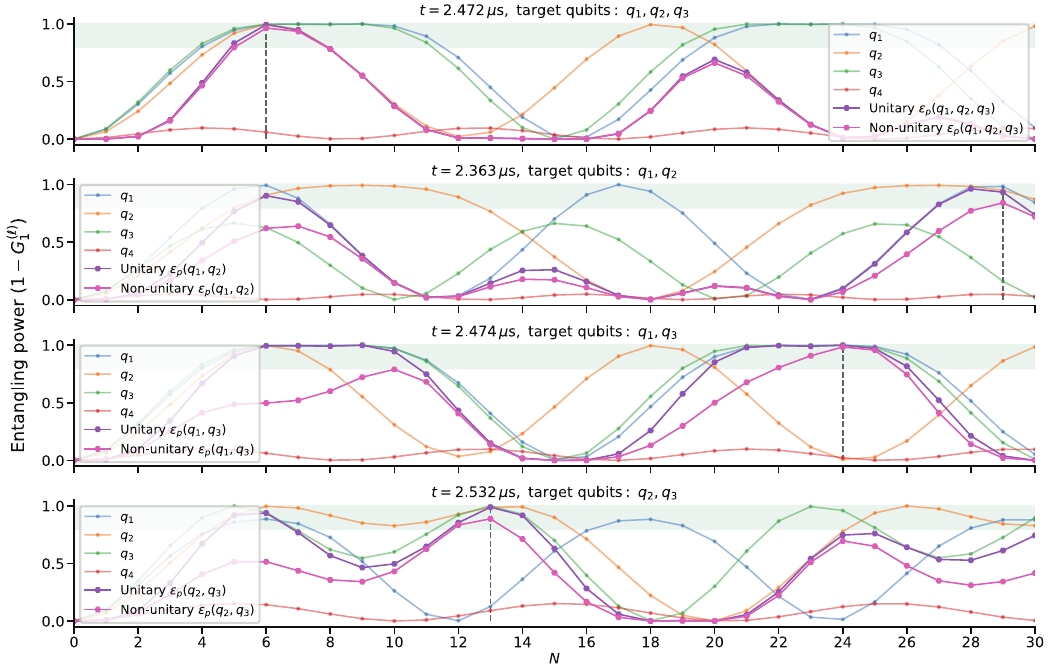}
\caption{
\textbf{Optimal parallel entanglement cases}
Green shaded region denotes the threshold of $\varepsilon_{p,M}(\mathcal{E})>0.8$ set for the gate parameter search. 
Grey dashed lines denote optimal $N$ determined.
At these values, the entangling power of the non-targeted nuclear qubits is near zero.
}
\label{fig:PE_cases}
\end{figure}

In applying this algorithm to identify optimal parallel entangling gates, the range of $t$ values can be set based on the intersection of $\hat{n}_0 \cdot \hat{n}_1$ ranges described above (shaded region of Fig. \ref{fig:n01_k1_k2}a).
If no result is found for some subset of nuclear qubits, one can consider a larger range of $N$ values until a valid case is returned, or lower the entangling power threshold value.
The former is would be preferred to generate higher fidelity gates, while the later may be preferred to generate faster gates. 
Applying this algorithm with a threshold of $\varepsilon_{p,M}(\mathcal{E})>0.8$ to the specific hyperfine couplings used in this work leads to the cases presented in Fig. \ref{fig:PE_cases} and summarized in Table \ref{tab:PE_calib}. 
Importantly, this procedure shows that it is possible to entangle all three nuclear qubits at once, as well as entangle all three subsets of nuclear qubits, without inducing considerable residual entanglement.

\begin{table}[h!]
\label{tab:PE_calib}
\centering
\begin{tabular}{c|c|c|c|c|c|c}
\makecell{Target\\nuclear qubits} &
\makecell{Non-targeted\\nuclear qubits} &
\makecell{Unit pulse \\ time [$\mu$s]} &
\makecell{Unit pulse \\ repeats} &
\makecell{Gate \\ time [$\mu$s]} &
\makecell{Target \\ $\epsilon_{p,M}(U_{DD})$} &
\makecell{Non-unitary \\ $\epsilon_{p,M}(\mathcal{E})$} \\
\hline
$\qone,\qtwo,\qthree$ & $\qfour$ & 2.472 & 6  & 14.8 & 0.9940 & 0.9635 \\
\hline
$\qone,\qtwo$ & $\qthree,\qfour$ & 2.363 & 29 & 68.4 & 0.9630 & 0.8411 \\
\hline
$\qone,\qthree$ & $\qtwo,\qfour$ & 2.474 & 24 & 59.4 & 0.9997 & 0.9842 \\
\hline
$\qtwo,\qthree$ & $\qone,\qfour$ & 2.532 & 13 & 32.9 & 0.9884 & 0.8872 \\
\end{tabular}
\caption{\textbf{Optimal parallel entanglement gate parameters from Fig. \ref{fig:PE_cases}}}.
\end{table}

% It is helpful to note here that in the case of $L=3$ nuclear qubits, the choice of unit pulse time creates complicated conditional rotation axes for some of the nuclear qubits, causing some nuclear qubits to rotate more than $\pi/2$.
% Since the unit pulse time is effectively on resonance for nuclear qubit $\qtwo$, its conditional rotation axes lie along $\pm X$ and the value of $N$ calibrated in Fig. \ref{fig:G1_v_N} corresponds to an angle of $\pi/2$.
% However, for nuclear qubits $\qone$ and $\qthree$ the conditional rotation axes are  oriented approximately at a $\pi/4$ angle from the $\pm Z$ axis, lying in the $\pm X,\pm Z$ planes.
% As a result, while the chosen $N$ sets a $\pi/2$ rotation for $\qtwo$, a larger rotation angle is required for $\qone$ and $\qthree$ to reach full anti-alignment and thus achieve maximal entanglement.
% This subtlety, discussed in \cite{takou2023precise}, is reiterated here due to its considerable effect on the specific nuclear qubits  studied.
% The complicated geometry of this multi-qubit gate complicates both its characterization and practical use, and is addressed in later sections. 
% See Sec. \ref{sec:ideal_MQC} and Fig. \ref{fig:ent_axes} for a visualization of this multi-qubit gate's geometry.

In addition to these optimized cases, we also consider the effects of parallel entangling gates that generate considerable residual entanglement with non-targeted nuclear qubits.
For these gates, parameters are chosen more arbitrarily and will serve as examples showing the experimental effects of residual entanglement. 
Targeting entanglement with the subset of nuclear qubits $\qone,\qtwo$ and $\qtwo,\qthree$, the unit pulse times were chosen at the intersection of each respective $(\hat{n}_0 \cdot \hat{n}_1)_{\ell}$ curve and $N$ was set to 6.
This value of $N$ leads to considerable residual entanglement with the non-targeted nuclear qubits, as quantified by significantly reduced non-unitary entangling power (Table \ref{tab:PE_residual_calib}). 
Note these faulty gates, and the experiments using them, are included only to highlight how residual entanglement can manifest in measurable signals (see Sec. \ref{sec:residual} for additional details). 

\begin{table}[h!]
\label{tab:PE_residual_calib} 
\centering
\begin{tabular}{c|c|c|c|c|c}
\makecell{Target\\nuclear qubits} &
\makecell{Non-targeted\\nuclear qubits} &
\makecell{Unit pulse \\ repeats} &
\makecell{Gate \\ time [$\mu$s]} &
\makecell{Target \\ $\epsilon_{p,M}(U_{DD})$} &
\makecell{Non-unitary \\ $\epsilon_{p,M}(\mathcal{E})$} \\
\hline
$\qone,\qtwo$ & $\qthree,\qfour$ & 2.418 & 6 & 0.9789 & 0.5429 \\
\hline
$\qone,\qthree$& $\qtwo,\qfour$ & 2.520 & 6 & 0.8878 & 0.4690 \\
\end{tabular}
\caption{
\textbf{Parallel entanglement gate parameters that generate residual entanglement}
Note how it is possible that the targeted entangling power to be large (near 1) but the non-unitary entangling power to be much smaller due to residual entanglement.
This effect is due to the next strongest nuclear qubit that was not targeted ($\qthree$ in the case of targeting $\qone$ and $\qtwo$), while $\qfour$ only slightly lowered the metric.
}
\end{table}

\section{Generality and extensions}\label{sec:generality} 
%=====================================================

This section contextualizes the results of this work by simulating and averaging over many nuclear-qubit configurations to determine the probability that a given configuration can be entangled in parallel, along with the associated gate times and fidelities.
To do so, we randomly generate other spatial distributions of $^{13}$C nuclear qubits around a NV center.
To simplify the model, nuclear qubits are placed at continuous locations uniformly randomly sampled within fixed sphere of radius $R_c$.
The radius of the cut-off sphere is fixed based on the atomic density of diamond, the $^{13}$C abundance and the total number of nuclear qubits to be considered.
For example, at the natural abundance of 1.1\%, the radius of the cut-off sphere needed to encapsulate $N_{\text{nuc}}=100$ nuclear qubits is $R_c \approx 2.3$nm.
Then, $N_{\text{nuc}}$ random locations with distances less than $R_{c}$ are uniformly sampled within the sphere.
For each configuration of nuclear qubits, the hyperfine coupling to the electron is calculated based on the physical location relative to the NV center at the origin.
Since our focus is on weakly coupled nuclear qubits—those farther from the NV center—the influence of discrete lattice site distances on the hyperfine coupling is negligible.

\begin{figure}[h!]
\centering
\includegraphics{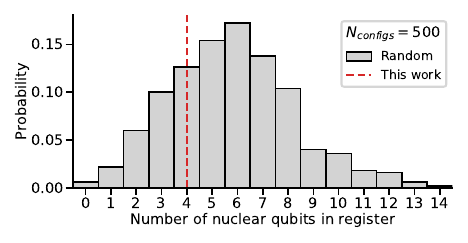}
\caption{
\textbf{Sampled distribution of number of weakly coupled nuclear qubits in register}
Each configuration sampled here already passed the pre-filter of containing no strongly coupled nuclear qubits.
The number of register nuclear qubits $L_{\text{total}}$ is therefore based on the number of weakly coupled nuclear qubits stronger than $|A_{\parallel}|>15$kHz and $A_{\perp}>10$kHz, ensuring they could be resolved from the spin bath.
}
\label{fig:N_spins_in_register}
\end{figure}

Far fewer than the total number of nuclear qubits $N_{\text{nuc}}$ included in the initial model are considered part of the NV-nuclear register.
Only nuclear qubits with hyperfine parameters between a certain, physically motivated, range are considered. 
First, each configuration is pre-filtered so that no strongly coupled nuclear qubits are present. 
Only configurations where all hyperfine components (parallel and perpendicular) are less than 200 kHz are considered further. 
This cutoff is based on the electron spin transition inhomogeneous linewidth of $\sqrt{2}/\pi T_2^* \approx 200$kHz.
This pre-filter process passes about 1 in every 3 configurations ($\approx$1400 configurations were examined to generate 500 passable configurations, hence $\approx$36\% pass).
Configurations that include one or more strongly coupled nuclear qubits are more difficult to entangle in parallel since DD is primarily intended to control weakly coupled nuclear qubits.
After passing this pre-filter, the configuration is included in the averaging analysis that follows. 
To ensure register nuclear qubits are not part of the spin bath of very weakly and similarly coupled spins, only parallel hyperfine couplings greater than 15 kHz are considered for parallel entanglement.
While the parallel component largely determines separation from the spin bath, a lower bound on the perpendicular component was also placed at 10 kHz to ensure the nuclear qubits included can be rotated sufficient fast (reasonable $N$).
Together, this ensures that only the nuclear qubits that could reasonably have their hyperfine couplings measured with DD spectroscopy are included. 
In other words, nuclear qubits that would contribute to the spin bath are not considered because they could not be experimentally measured using DD spectroscopy.
Note, techniques such as DD-radio frequency can be employed to also utilize nuclear qubits with effectively no perpendicular hyperfine coupling, extending the register size further \cite{bradley2019ten}. 

At natural abundance, these bounds lead to on average $\bar{L}_{\text{total}} \approx 6$ nuclear qubits (standard deviation of $\sigma_{L_\text{total}} \approx 2.5$) in the register and therefore considered for parallel entanglement.
This was averaged over 500 pre-filtered configurations and is the sample size for all simulations that follow, unless stated otherwise.
The exact distribution of register sizes sampled is shown in Fig. \ref{fig:N_spins_in_register}.
Foremost, this indicates that the nuclear qubit configuration studied in this work is on the smaller end of the expected value, minus the spread. 
For reference, the nuclear register studied in \cite{taminiau2012detection} did not contain any nuclear qubits with hyperfine coupling (parallel or perpendicular) greater than even 80kHz and was able to resolve 6 nuclear qubits from the spin bath. 
The weakest nuclear qubits resolved had components of $|A_{\parallel}|,A_{\perp}\approx15$kHz; comparable to the weakest nuclear qubit ($q_4$) observed for the configuration studied in this work.

Of the configurations that did not contain any strongly coupled nuclear qubits, the register qubits are then run through the parallel entanglement calibration algorithm of Takou \textit{et al.} \cite{takou2024generation} described in the previous section \ref{sec:Parallel entanglement}.
Note that the parameters of the algorithm (Table \ref{tab:generality_config}) used were the same as those used to calibrate the parallel gates of this work. 
Due to the presence of spin bath experimentally, additional care needs to be taken when sweeping potential unit pulse times.
Unit pulse times the lie within the spin bath region will lead to significant residual entanglement with bath spins that are not being included in this model.
To account for this, unit pulse times corresponding to the bath region were removed from the calibration algorithm. 
This avoided region was set based on the resonance times of the lower bound hyperfine coupling used.

\begin{table}[h!]
\label{tab:generality_config}
\centering
\begin{tabular}{c|c|c}
Parameter & Value & Primary effect \\ 
\hline
$^{13}$C concentration & 1.1\% & Register size $L_{\text{total}}$ \\
\hline
Max hyperfine & $|A_{\parallel}|,A_{\perp}<200$kHz & Fraction of usable configurations \\
\hline
Min hyperfine & $|A_{\parallel}|>15$kHz, $A_{\perp}>10$kHz & Register size $L_{\text{total}}$ \\
\hline Unit pulse time $t$ resolution & 1ns & Probability of parallel entanglement \\
\hline
Max $N$ & 50 & Probability of parallel entanglement \\
\hline
Min non-unitary entangling power & 0.8 & Probability of parallel entanglement \\
\end{tabular}
\caption{
\textbf{Simulation settings for parallel entanglement of random nuclear qubit configurations}
In addition to the primary effect listed, there can be other secondary effects.
For example, while the $^{13}$C concentration primarily effects how large the nuclear qubit register is, the register size can have an effect on the probability of parallel entanglement being possible.
Overall, these are conservative parameters --- increasing the maximum $N$ or lowering the entangling threshold has a large effect on the probability of a given parallel entangled state to be formed.
Assuming a unit pulse times in the range of $\approx$2 to $4\mu$s, an $N$ as large as 500 to 250 may be considered before the entangling operation takes 1ms (roughly the electron $T_2$ time).
Given the computational complexity of needing to average over many configurations here, such large parameter were not explored.
}
\label{tab:Z_para}
\end{table}

\begin{figure}[th!]
\centering
\includegraphics{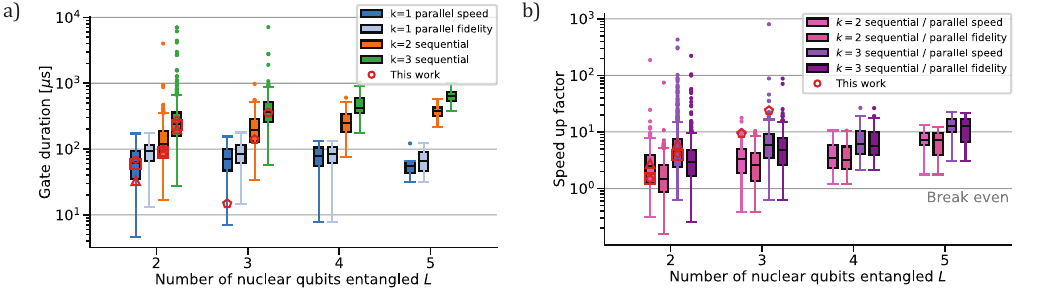}
\caption{
\textbf{Expected parallel entanglement durations}
% \textbf{(a)} Probability of an NV center without strongly coupled nuclear qubits exhibiting parallel entanglement.
% Statistical outliers are denoted with small point markers.
% \textbf{(b)} Durations and speed-up ratios for parallel entanglement at first order, sequential entanglement using two-qubit gates at second and third order. 
\textbf{(a)} Box plots of the distributions of durations for each entangling gate.
"parallel speed" refers to parallel gates calibrated to minimize duration and "parallel fidelity" refers to parallel gates calibrated to maximize non-unitary entangling power.
The middle black line represents the median, the box spans the interquartile range (IQR), the whiskers extend to $1.5 \times$ the IQR, and points indicate outliers.
Red markers of different shapes denote the specific cases observed in this work.
\textbf{(b)}
Parallel entangling gate speedup calculated as the ratio of durations in each specific register. 
}
\label{fig:generality_num_duration}
\end{figure}

For a given subset of nuclear qubits that can be entangled in parallel, there can be multiple DD gate parameters that satisfy the threshold conditions set. 
However, only one parallel gate for each subset should be saved, which can be chosen a number of ways. 
Here, we consider picking the parallel entangling gate with the shortest duration and also the gate with the largest non-unitary entangling power, and compare both outcomes. 
The trade off in choosing the fastest gate may be a loss of fidelity (although the entangling power threshold dictates how much can be lost), while higher fidelity gates may take longer to execute. 
When considering the parallel entanglement of larger numbers of nuclear qubits, there is often only one possible parallel gate and thus no trade off to be made.

Averaging over the weakly coupled registers reveals gate duration distributions for each type of parallel entangling gate, as well as sequences of two-qubit gates at any order $k$ (Fig. \ref{fig:generality_num_duration}a).
From this, the speedup of the parallel gate over each the sequential gates can be calculated (Fig. \ref{fig:generality_num_duration}b).
The only slightly irregular feature of our results is the speed up factor for the $L=3$ gate, which is on the high end or slightly outside of $1.5\times$ the interquartile range. 
This is due to the fact that our $L=3$ parallel gate was able to be performed with high fidelity on the first oscillation of each nuclear qubit (small $N$ value). 
This placed the absolute duration on the lower end of $1.5\times$ the interquartile range, while the sequential gate durations were near the median. 
Consequently, the ratio of durations is large, but other cases sampled were still much larger. 

While there are many factors that can impact the fidelity of a gate, we analyze one of the predominate effects here --- residual entanglement generated with nuclear qubits in mixed states.
An expression for the gate fidelity due to residual entanglement was derived in Takou \textit{et al.} \cite{takou2023precise} and the key details are summarized here.
The entangling gate fidelity due to residual entanglement $F_r$ is given by
\begin{equation}\label{eq:gate_fid_residual_ent}
    F_r = \frac{1}{2^{L+1}+1}\left(1 + 2^{L-1} \sum_{k}^{2^{L_{\text{total}}-L}} \left| \sum_{j=0,1}c_j^{(k)} p_j^{(k)} \right|^2 \right)
\end{equation}
where the summation over $k$ runs over all possible bit strings of length $L_{\text{total}}-L$ (the number of nuclear qubits in the environment).
For each bit string $k$, let $M$ of the bits be 0 and $P = L_{\text{total}}-L-M$ of the bits be 1.
Then, the terms $c_j^{(k)}$ and $p_j^{(k)}$ are calculated based on tracing out the environmental nuclear qubits.
Given the known DD rotation operators acting on the environment, these have explicit expressions given by:
\begin{equation}\label{eq:c_j}
\begin{split}
    c_j^{(k)} &= \prod_{m\in \{\text{bits=0}\}}^{m_M} \langle 0 | R_{\hat{n}_j^{(m)}}(\phi^{(m)}) | 0 \rangle  \\
    &=\prod_{m\in \{\text{bits=0}\}}^{m_M} \left[ \cos\left( \phi^{(m)}/2 \right) - i n_{z,j}^{(m)} \sin\left( \phi^{(m)}/2 \right) \right]
    \end{split}
\end{equation}

\begin{equation}\label{eq:p_j}
\begin{split}
    p_j^{(k)} &= \prod_{s\in \{\text{bits=1}\}}^{s_P} \langle 1 | R_{\hat{n}_j^{(s)}}(\phi^{(s)}) | 0 \rangle \\
    &= \prod_{s\in \{\text{bits=1}\}}^{s_P} \left[ - i \left( n_{x,j}^{(s)} + n_{y,j}^{(s)}\right) \sin\left( \phi^{(s)}/2 \right) \right]
\end{split}
\end{equation}
where $R_{\hat{n}_j^{(m)}}(\phi^{(m)})$ are the DD rotation operators given in equation \ref{eq:R_cases}.
Note, one can assume $n_{y,j}^{(s)}=0$ for the frame of the Hamiltonian used here.
In the ideal case of no residual entanglement generated, $c_j^{(k)}$ and $p_j^{(k)}$ are 1 and thus $F_r =1$.

\begin{figure}[th!]
\centering
\includegraphics{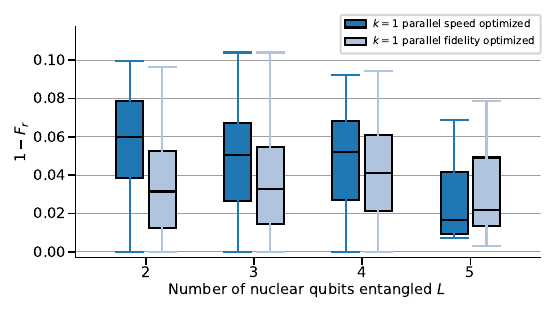}
\caption{
\textbf{Expected parallel entangling gate error rates due to residual entanglement}
Error rates ($1 - F_r$) simulated here are directly related to the non-unitary entangling power threshold set in the calibration process.
Stricter (larger) thresholds will lead to lower error rates, but fewer potential parallel entanglement cases.
The middle black line represents the median, the box spans the interquartile range (IQR), the whiskers extend to $1.5 \times$ the IQR, and points indicate outliers.
}
\label{fig:generality_fidelity}
\end{figure}

Applying equation \ref{eq:gate_fid_residual_ent} to the configurations that hosted parallel entanglement highlights low error rates ($1-F_r$), all less than $\approx0.1$ (Fig. \ref{fig:generality_fidelity}).
The excellent performance of these gates is based on the considerably large (near 1) non-unitary entangling power threshold set during the calibration process.
As expected, the gates that were chosen for their shorter duration have a slightly higher error rate than the gates chosen for the larger non-unitary entangling power. 
For additional details on the relationship between the entangling power and gate fidelity, see \cite{takou2023precise} 
Note, the cases where the fidelity exactly equals 1 indicates that all of the nuclear qubits in the register were used in the parallel entanglement process (no environment qubits).
Experimentally, there are other sources of error in the entangling gate.
The leading factor for the loss of fidelity observed in this work likely stems from pulse errors in controlling the electron. 
This accentuates the dephasing of the electron due to the spin bath, which also lowers fidelity. 
Additional analysis of these effects could be incorporated, following the results of \cite{takou2024generation}.

\begin{figure}[th!]
\centering
\includegraphics{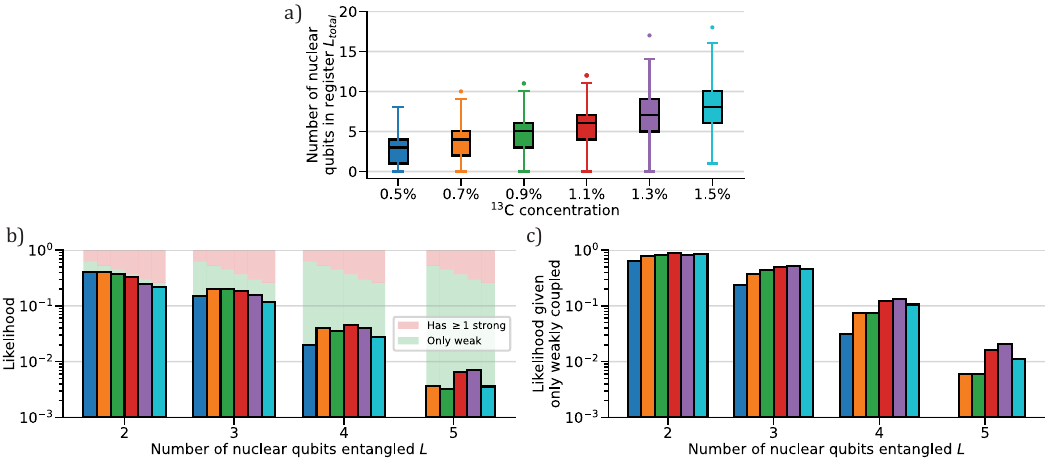}
\caption{
\textbf{Nuclear qubit concentration effects on parallel entanglement}
\textbf{(a)} Total number of weakly coupled nuclear qubits in register.
\textbf{(b)} Probability of any configuration hosting parallel entanglement of different sizes $L$.
The red shaded regions denote the portion of configurations that have at least one strongly coupled nuclear qubit and are not considered for parallel entanglement.
The green shaded regions denote the rest of the configurations that are considered for parallel entanglement.
As concentration increases, the probability of having at least one strongly coupled nuclear qubit increases.
Bar colors correspond to the concentrations in (a).
\textbf{(c)} Probability of a weakly coupled configuration hosting parallel entanglement.
}
\label{fig:generality_enrichment}
\end{figure}

One can also consider diamond samples that have different concentrations of $^{13}$C nuclei using this framework. 
On one side, with higher concentrations, there are more weakly coupled nuclear qubits in the register to participate in the parallel entanglement (Fig. \ref{fig:generality_enrichment}a), but on the other side, this may lead to increased residual entanglement and in turn lower probabilities of such a parallel gate existing.
Similarly, with lower concentrations, the register size decreases but so does the potential residual entanglement.
Owing to the computational complexity associated with including more nuclear qubits in a register, smaller sample sizes (200 configurations) and unit pulse time steps (10ns) were used when considering different concentrations.
A primary bottleneck in running the parallel entanglement algorithm with a large number configurations, each with a large number of nuclei in the system, is the evaluation of every possible subset of nuclear qubits to target entanglement with.
This power set of $N_{\text{qubits}}$ nuclear qubits has an exponentially large size of $2^{N_{\text{qubits}}}$ so for very large system sizes nearing $\approx20$ nuclear qubits, there are $\approx 1$ million subsets to check over all gate parameters. 
In this case, pre-filtering for nuclear qubits that have an intersection of $(\hat{n}_0 \cdot \hat{n}_1)_{(\ell)}<0$ before iterating though all gate parameters would be essential for large scale calculations. 
At lower concentrations, there tend to be lower likelihoods to form larger parallel entangled states (Fig. \ref{fig:generality_enrichment}b), which can be attributed to the smaller register sizes. 
Similarly, as concentration increases, there tends to be a loss of likelihood. 
However, in these cases, fewer configurations pass the pre-filtering process for containing a strongly coupled nuclear qubit. 
If one conditions the likelihood on the register being only weakly coupled (Fig. \ref{fig:generality_enrichment}c), then at increased concentrations there is a similar, if not slightly larger likelihood to form larger entangled states in parallel. 
Note, a finer-grained parameter space of DD gates may further accentuate these changes.
Future research taking into these computational complexities can be dedicated to finding an optimal concentration, where optimality can be defined according to likelihood of parallel entanglement, size of the entangled state, duration or fidelity of the gate. 

%=====================================================
\section{Magnetic field optimization}\label{sec:B field}
%=====================================================

%=====================================================
\subsection{Alignment}\label{sec:B_alignment}
%=====================================================

The external magnetic field was mounted on a set of two micrometer translation stages in the plane normal to the NV axis.
One controlled the height from the optical table and the other controlled right-left translation. 
Translation along the NV symmetry axis to and from the sample was adjusted using a threaded casing around the magnet attached to an optical kinematic mount. 
The kinematic mount has two angular micrometers to adjust the polar and azimuthal angles of the field. 

The alignment and field strength of the magnet in a given position were measured using the resonant frequencies of pulsed electron spin resonance experiments for both $\ket{m_s=0} \rightarrow \ket{m_s= \pm1}$ transitions \cite{stanwix2010coherence}.
The translation micrometers were adjusted to optimize field strength, then the angular micrometers were adjusted to optimize field alignment. 
The process converged after 2-3 rounds of optimization. 
The final misalignment angle was measured to be 0.4$^{\circ}$$_{-0.4}^{+1.6}$, and the field strength was $338.19 \pm 0.14$G.
The alignment and field strength were periodically checked to correct for drift over extended periods of time.

%=====================================================
\subsection{Field strength}
%=====================================================

\begin{figure}[th!]
\centering
\includegraphics{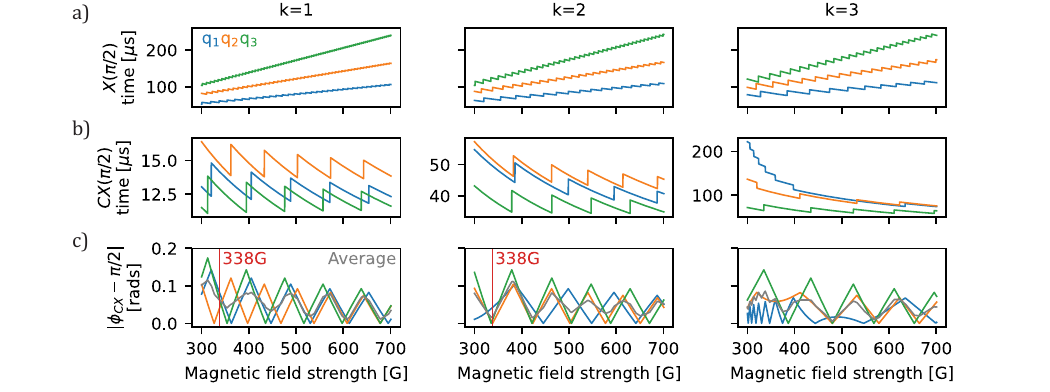}
\caption{
\textbf{Magnetic field effect on nuclear gate times}
\textbf{(a)} Total gate time for $X_{\pi/2}$ rotation of each nuclear qubit at first three resonance orders. 
Although the first and second order gates have nearly identical times, the second order gates use fewer pulse repeats, which means few operations on the electron, so these were used.
\textbf{(b)} Conditional version of (a).
Second order was again used, but this was to save time compared to third order.
\textbf{(c)} Absolute conditional angular error from $\pi/2$ to select specific field strength. 
}
\label{fig:B_opt_time}
\end{figure}

The particular magnetic field strength of 338G was chosen to minimize the nuclear gate times and errors. 
Based on the bandwidth of the microwave electronics, and the fact that both electron spin transitions are needed to align the field, the magnetic field strength was limited from above at approximately 700G. 
A sufficiently large field is also convenient for estimating DD resonance times made under a large field approximation \cite{taminiau2012detection,takou2023precise}. 

The magnetic field strength's effect on each nuclear qubit's total gate time was first considered (Fig. \ref{fig:B_opt_time}a,b). 
While the conditional gate times decrease slightly at larger fields, the unconditional gates increase linearly with field strength. 
This is due to the relative strengths of each qubit's Larmor frequency set by the field and their fixed hyperfine couplings. 
At larger fields (higher Larmor frequencies) the free evolution axes for each electron spin state $\hat{p}_j^{(\ell)}$ (Sec. \ref{sec:Ham}) become more aligned, thus requiring more DD unit pulse repeats to achieve a particular rotation angle.
Therefore, from an efficiency perspective, a smaller field should be used to decrease nuclear gate time.

Next, a particular field strength can be chosen to minimize the angular errors from the discrete nature of DD angular control. 
Since unconditional $X$ gates typically needed more repetitions than conditional gates, their angular resolution was sufficiently small which already led to small gate errors. 
On the other hand, at lower field strengths, the conditional gates do not need many repetitions, which could lead to considerable angular error in achieving a desired rotation angle.
Since all two-qubit conditional gates used $\pi/2$ rotations, the absolute angular error from this value was considered at the first three resonance orders (Fig. \ref{fig:B_opt_time}c). 
Each nuclear qubit's conditional gate time increased considerably between orders, which led us to not use any third order gates in this work. 
Based on the average error for the three strongest nuclear qubits, two field strengths were found that minimized both the first and second order conditional gate angular errors: 338G and 436G, of which the smaller was chosen to minimize unconditional gate time. 

%=====================================================
\section{Individual nuclear qubit control circuits}
%=====================================================

%=====================================================
\subsection{Nuclear qubit initialization and tomography}
%=====================================================

\begin{figure}[h!]
\centering
\includegraphics{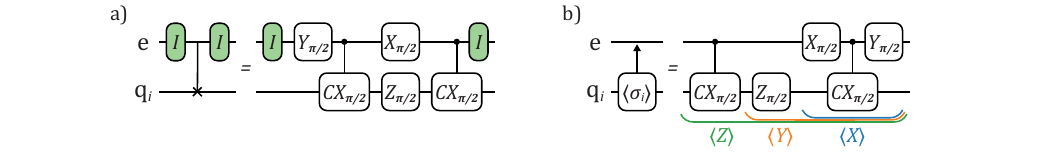}
\caption{
\textbf{Nuclear initialization and tomography circuits}
\textbf{(a)} Swap based initialization circuit of individual nuclear qubits. 
\textbf{(b)} State tomography circuit of individual nuclear qubits. 
Brackets below each subset of gates are color-coded to which nuclear expectation value they swap onto the electron for measurement.
}
\label{fig:nuc_init_and_tomo}
\end{figure}

To initialize nuclear qubits into partially polarized states, polarization from the electron was transferred to each nuclear qubit following the circuit \cite{taminiau2014universal} shown in Fig. \ref{fig:nuc_init_and_tomo}a. 
In \cite{taminiau2014universal}, initialization was verified via changes in the electron's $T_2^*$, but this method is not feasible here due to the shorter $T_2^*\approx 2\mu$s of this sample.
Instead, each nuclear qubit's initialization was verified with state tomography experiments (Fig. \ref{fig:nuc_init_and_tomo}b) as detailed in Sec. \ref{sec:exp gate calib}. 
However, this approach does not yield a direct measurement of the nuclear qubit initialization fidelity because there are also errors in the state tomography process that can not be separated. 
It is also useful to note that the $2\mu$s green laser pulse used to reinitialize the electron has very little effect on nuclear qubit polarization and coherence times once they've been initialized. 
The supplement of \cite{taminiau2014universal} contains detailed results of nuclear qubit lifetimes under green illumination and such experiments were not repeated here. 

Multiple nuclear qubits were initialized using the same sub-circuit sequentially with the electron reinitialized between each iteration.
This allowed for selective initialization of the quantum register. 
Alternative initialization schemes, such as PulsePol \cite{schwartz2018robust}, could have been used at room temperature, but simulation showed comparable fidelities and durations. 
Furthermore, such protocols require many more electron reinitialization steps than the sequential swap approach.
Without real-time electron charge state checks \cite{hopper2020real}, this introduces additional experimental noise and is already one of the leading causes of discrepancy between simulation and experimental contrast present in this work.

To readout complete information from each individual nuclear qubit, state tomography can be performed (Fig. \ref{fig:nuc_init_and_tomo}b). 
This process maps nuclear qubit expectation values onto the electronic state, which is then measured.
While individual nuclear qubit tomography is sufficiently high fidelity, multi-nuclear state tomography used to measure inter-nuclear correlations is much worse due to crosstalk between nuclear gates. 
Simulations showed a low ($\approx60\%$) average readout fidelity in the two nuclear qubit case, with the three nuclear qubit case behaving even worse. 

%=====================================================
\subsection{Experimental nuclear gate calibration}\label{sec:exp gate calib}
%=====================================================

\begin{figure}[th!]
\centering
\includegraphics{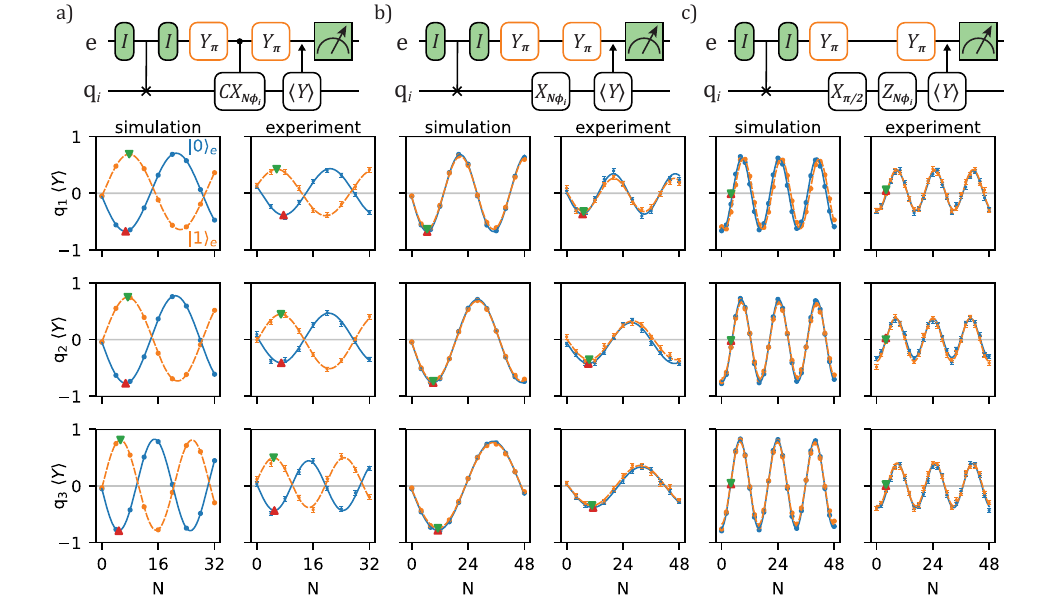}
\caption{
\textbf{Experimental nuclear gate calibration} \textbf{(a,b,c)} $CX$,$X$,$Z$ nuclear gate calibration by sweeping each unit pulse repeat (nuclear rotation angle), respectively.
Both data (points with error bars) and simulation (points) were fit (solid and dashed curves) to determine which number of repeats corresponds to a $\pi/2$ rotation (results in table \ref{tab:exp_NG}).
Both electron states ($\ket{0}_e$ blue and $\ket{1}_e$ orange) were used prior to sweeping each nuclear gate to highlight conditionality as in (a), or unconditionality in (b) and (c).
Red (green) triangles denote $N(\pi/2)$ for the electron state $\ket{0}_e$ ($\ket{1}_e$).
}
\label{fig:nuc_gate_calib}
\end{figure}

Using nuclear qubit state tomography, a variety of simple circuits (Fig. \ref{fig:nuc_gate_calib}) were developed to verify nuclear initialization and control of each nuclear qubit.
In each circuit, the number of unit pulse repeats for a particular gate was varied and partial state tomography along the $Y$ axis was performed. 
$Y$ axis tomography was chosen because most gates rotate about the $X$ axis, so an orthogonal direction will detect coherent rotations of nuclear qubits, and $Y$ tomography is shorter than $Z$ tomography. 
Additionally, the state of the electron was varied to observe the (un)conditionality of each gate. 
To highlight this effect, the electron was either kept in the initialized $\ket{0}$ state, or flipped to $\ket{1}$ prior to gate calibration and back again before tomography began. 
This highlights that all gates calibrated are indeed (un)conditional as expected. 
All experiments were run using simulated gate parameters from table \ref{tab:NG calib} and all of the fitting results for the repetitions needed for a $\pi/2$ (table \ref{tab:exp_NG}) nearly match the simulation results, with the exception of the $CX$ gate of nuclear qubit $q_1$. 
Even so, in that case the average between the results for the two electron states are in agreement for which $N$ creates a $\pi/2$ rotation.

\begin{table}[h!]
\centering
\begin{tabular}{c|c|c|c|c|c|c|c}
Nuclear qubit &  Electron State & Sim $CX$ $N_{\pi/2}$ & Exp $CX$ $N_{\pi/2}$ & Sim $X$ $N_{\pi/2}$ & Exp $X$ $N_{\pi/2}$ & Sim $Z$ $N_{\pi/2}$ & Exp $Z$ $N_{\pi/2}$ \\ 
\hline
\multirow{2}{*}{$q_1$} & 0 & 6.7 & 7.6 $\pm$ 0.1 & 6.6 & 6.8 $\pm$ 0.1 & 4.0 & 4.0 $\pm$ 0.0 \\
& 1 & 7.6 & 5.6 $\pm$ 0.1 & 6.5 & 7.2 $\pm$ 0.1 & 4.0 & 4.1 $\pm$ 0.0 \\
\hline
\multirow{2}{*}{$q_2$} & 0 & 6.8 & 7.0 $\pm$ 0.1 & 9.2 & 9.2 $\pm$ 0.2 & 4.0 & 4.0 $\pm$ 0.0 \\
& 1 & 7.0 & 6.9 $\pm$ 0.1 & 9.2 & 9.6 $\pm$ 0.2 & 4.0 & 4.0 $\pm$ 0.0 \\
\hline
\multirow{2}{*}{$q_3$} & 0 & 4.8 & 5.0 $\pm$ 0.1 & 11.1 & 11.2 $\pm$ 0.1 & 4.0 & 4.0 $\pm$ 0.0 \\
& 1 & 5.3 & 4.8 $\pm$ 0.1 & 11.1 & 10.8 $\pm$ 0.2 & 4.0 & 4.0 $\pm$ 0.0 \\

\end{tabular}
\caption{
\textbf{Experimental nuclear gate calibration fit results}
All values shown have been rounded to one decimal place and when rounded to the nearest integer match the calculated values in simulation table \ref{tab:NG calib}.
}
\label{tab:exp_NG}
\end{table}

%=====================================================
\section{Multiple Quantum Coherences}\label{sec:mqc_sup}
%=====================================================

%=====================================================
\subsection{Idealized theory}\label{sec:ideal_MQC}
%=====================================================

Here we derive the multiple quantum coherences (MQC) signal measured for the DD gate implementation of the method.
The main idealization made in the following analysis is the absence of nuclear gate crosstalk when considering more than one nuclear qubit.
This idealization means that when targeting the rotation of one of the nuclear qubits, others are transformed with the identity operator. 
As discussed extensively in the main text, this is not the case for DD, but in the single nuclear qubit case this analysis will be exact up to other errors. 
Even in the multinuclear case, the final result of this analysis holds given that we can correct for the crosstalk by optimizing GHZ state fidelity (more details in Sec. \ref{sec:mqc_opt}). 
In light of this, we consider only a single idealized form of the gate that creates multipartite entanglement, temporarily neglecting the complicated geometry of the parallel entanglement gates.
This simplification, and its implication will later be addressed in this section.

Consider the electron and $L$ initialized nuclear qubits in the state $\ket{\psi_0} = \ket{0}^{\otimes L+1}$.
To prepare for entanglement, rotate the electron with $Y_{\pi/2}$ and each nuclear qubit with $X_{\pi/2}$. This yields:
\begin{equation}
\begin{split}
    \ket{\psi_1} &= \ket{X} \otimes \ket{-Y}^{\otimes L} \\
    &= \frac{1}{\sqrt{2}} \big( \ket{0} \otimes \ket{-Y}^{\otimes L} + \ket{1} \otimes \ket{-Y}^{\otimes L} \big).
\end{split}
\end{equation}
Now apply an $L+1$-qubit entangling gate of the form:
\begin{equation}
\label{eq:ent_gate}
    C_eX_{\pi/2}^{\otimes L} = \ket{0}\bra{0} \otimes X_{\pi/2}^{\otimes L} + \ket{1}\bra{1} \otimes X_{-\pi/2}^{\otimes L}.
\end{equation}
Sequences of two-qubit gates takes this form in the absence of crosstalk and the parallel gate is locally equivalent to this. 
Transforming $\ket{\psi_1}$ gives the entangled state of:
\begin{equation}
\label{eq:GHZ_state}
\begin{split}
    C_eX_{\pi/2}^{\otimes L} \ket{\psi_1} &= \frac{1}{\sqrt{2}} \big( \ket{0} \otimes \big( X_{\pi/2}\ket{-Y} \big)^{\otimes L} + \ket{1} \otimes \big( X_{-\pi/2}\ket{-Y} \big)^{\otimes L} \big) \\
    &= \frac{1}{\sqrt{2}} \big( \ket{0} \otimes (-i)^L \ket{1} ^{\otimes L} + \ket{1} \otimes \ket{0}^{\otimes L} \big) \\
    &= \frac{1}{\sqrt{2}} \big( \ket{0} \otimes \ket{1} ^{\otimes L} + \frac{1}{(-i)^L} \ket{1} \otimes \ket{0}^{\otimes L} \big) = \ket{\text{GHZ}},
\end{split}
\end{equation}
where in the last line the global phase was adjusted. 
Let $1/(-i)^L = e^{i\alpha_L}$ be the relative phase of the two states in the entangled state due to the local operations. 
Now apply an equal phase gate $Z_{\phi}$ to each nuclear qubit:
\begin{equation}\label{eq:GHZ_phi}
\begin{split}
    \ket{\text{GHZ}_{\phi}} &= \frac{1}{\sqrt{2}}(e^{iL\phi/2} \ket{0}\otimes \ket{1}^{\otimes L} + e^{i\alpha_L} e^{-iL\phi/2} \ket{1}\otimes \ket{0}^{\otimes L}) \\
    &= \frac{1}{\sqrt{2}}(\ket{0}\otimes \ket{1}^{\otimes L} + e^{i(\alpha_L - L\phi)} \ket{1}\otimes \ket{0}^{\otimes L}). \\
\end{split}
\end{equation}
Now disentangle the system using the same entangling gate as before:
\begin{equation}
\begin{split}
    \ket{\text{Dis.}} &= \frac{1}{\sqrt{2}}(\ket{0}\otimes (X_{\pi/2} \ket{1})^{\otimes L} + e^{i(\alpha_L - L\phi)} \ket{1}\otimes (X_{-\pi/2}\ket{0})^{\otimes L}) \\
    &= \frac{1}{\sqrt{2}}(\ket{0}\otimes (-i \ket{Y})^{\otimes L} + e^{i(\alpha_L - L\phi)} \ket{1}\otimes  \ket{Y}^{\otimes L}) \\
    &= \frac{1}{\sqrt{2}}(\ket{0}\otimes \ket{Y}^{\otimes L} + e^{i(2\alpha_L - L\phi)} \ket{1}\otimes  \ket{Y}^{\otimes L}) \\ 
    &= \frac{1}{\sqrt{2}}(\ket{0} + e^{i(2\alpha_L - L\phi)} \ket{1}) \otimes  \ket{Y}^{\otimes L},
\end{split}
\end{equation}
where, as desired, the relative phase of the electron now carries information about the size of the entangled state. 
One could derive the expression for the electron probability to be in $\ket{0}$ by rotating it back with $Y^{-1}_{\pi/2}$ (which is done experimentally), or assuming a measurement in the $\ket{X}$ basis. 
Performing the latter here for simplicity, tracing out the nuclear qubits we have:
\begin{equation}
    \ket{\psi_e} = \frac{1}{\sqrt{2}}(\ket{0} + e^{i(2\alpha_L - L\phi)} \ket{1}), 
\end{equation}
and therefore:
\begin{equation}\label{eq:MQC_sig}
\begin{split}
    P_e(\ket{X}) &= |\innerproduct{X}{\psi_e}|^2 \\
    &= \frac{1}{2}(1+\cos(L\phi - \delta_L)),
\end{split}
\end{equation}
where $\delta_L = 2\alpha_L$ is the overall phase shift defined in the main text. 
Now let us verify a phase shift of $\pi$ in the $L=1$ case. 
Using $1/(-i)^L = e^{i \alpha_L}$, we see that for $L=1$ the LHS is $i$, which has a phase $\alpha_1 = \pi/2$. 
Thus $\delta_1 = 2\alpha_1 = \pi$ as desired.
For larger $L$, this phase shift could be calculated as well, but because the optimization was able to adjust local operations before entanglement, the exact phase shift may change, and appears to in many of the optimized simulations and experiments.
This is a harmless effect that maintains the frequency amplification of each nuclear state's phase onto the electron in the end. 

\begin{figure}[th!]
\centering
\includegraphics{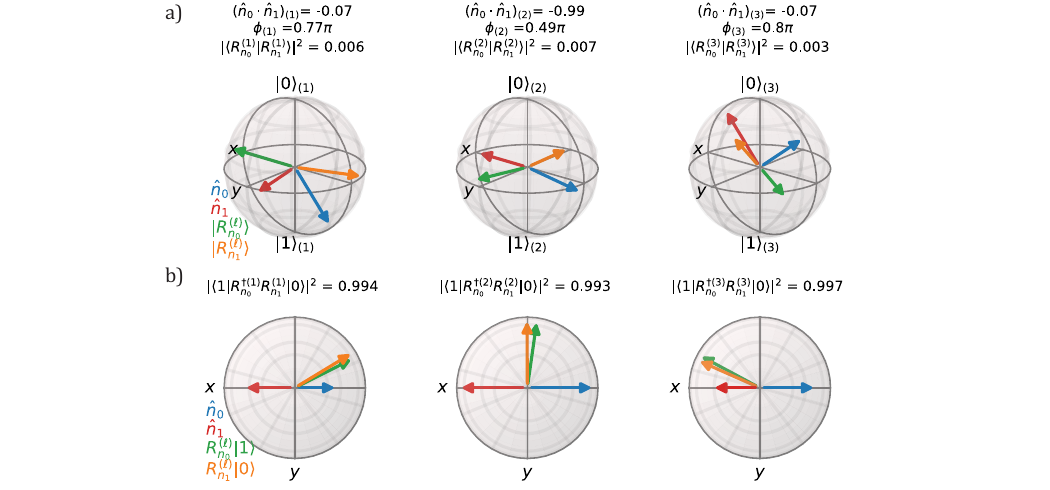}
\caption{
\textbf{Parallel entangling gate geometry} 
\textbf{(a)} Entangling transformation from an ideal nuclear qubit $\ket{0}$ initial state. 
Since the $L=3$ parallel gate unit pulse time was chosen approximately on resonance for nuclear qubit $q_2$, we expect $(\hat{n}_0 \cdot \hat{n})_{(2)}\approx-1$ and an angle of $\phi_{(2)}\approx\pi/2$. 
For the other off resonance nuclear qubits, since $\hat{n}_0 \cdot \hat{n}_1\neq-1$, a large rotation angle is expected to create anti-parallel states after rotation.
To quantify this behavior, one can consider the overlap $|\langle R_{n_0}^{(\ell)} | R_{n_1}^{(\ell)} \rangle |^2$ of the two quantum states, which should be near zero to generate a MME state from a geometric perspective.
\textbf{(b)} Disentangling transformations of the same gate, but applied to a GHZ state of equation (\ref{eq:GHZ_state}).
Here the transformed state overlap should be near one to properly disentangle the MME state. 
}
\label{fig:ent_axes}
\end{figure}

Now let us address the particular form of the entangling gate used in equation (\ref{eq:ent_gate}). 
For multi-qubit parallel entangling gates, each nuclear qubit rotation operator has a unique geometry, which can be generally written as:
\begin{equation}
    U_{\text{para}} = \ket{0}\bra{0}\bigotimes_{\ell}^L R_{\hat{n}_0^{(\ell)}}(\phi^{(\ell)}) + \ket{1}\bra{1}\bigotimes_{\ell}^L R_{\hat{n}_1^{(\ell)}}(\phi^{(\ell)}).
\end{equation}
In order to form a multipartite maximally entangled (MME) state, this operator must act on the proper initial separable state. 
Although finding such states could be analyzed analytically here, this is handled by the optimization of Sec. \ref{sec:mqc_opt} due to nuclear gate crosstalk making local rotations complicated.
For now, consider the action of this gate on an initial state $\ket{X} \otimes \ket{0}^{\otimes L}$, which gives an entangled state of the form:
\begin{equation}\label{eq:MME_general}
    \ket{\text{MME}} = \frac{1}{\sqrt{2}} \bigg( \ket{0} \bigotimes_{\ell}^L \ket{R_{n_0}^{(\ell)}} + \ket{1} \bigotimes_{\ell}^L \ket{R_{n_1}^{(\ell)}} \bigg),
\end{equation}
where $\ket{R_{n_j}^{(\ell)}}=R_{\hat{n}_j^{(\ell)}}(\phi^{(\ell)})\ket{0}$.
In order for this to be a MME state, $\ket{R_{n_0}^{(\ell)}}$ and $\ket{R_{n_1}^{(\ell)}}$ need to be anti-parallel, or $\ip{R_{n_1}^{(\ell)}}{R_{n_0}^{(\ell)}}=0$, for each nuclear qubit $q_{\ell}$.
This creates a state that can be transformed into a GHZ state with local rotations (\textit{i.e.}, locally equivalent). 
Equivalently, the optimization process sets these local rotations before entanglement so that a GHZ state is formed from the application of the parallel gate. 
As an example, Fig \ref{fig:ent_axes} considers the $L=3$ parallel entangling gate used in this work.
The top panel, Fig. \ref{fig:ent_axes}a, shows each of the states described in equation \ref{eq:MME_general} for each nuclear qubit, highlighting their orthogonality and thus the MME nature of the state created by the parallel gate.

Furthermore, the disentangling ability of the parallel gate needs to be carefully considered.
The relevant transformations to consider now are $R_{\hat{n}_0^{(\ell)}}(\phi^{(\ell)})\ket{1}$ and $R_{\hat{n}_1^{(\ell)}}(\phi^{(\ell)})\ket{0}$ for each nuclear qubit, based on equation (\ref{eq:GHZ_phi}).
As desired, each of these transformations leads to approximately identical states on each nuclear qubit's Bloch sphere, thus disentangling all of the nuclear qubits from the electron (Fig. \ref{fig:ent_axes}b).
Therefore, we can expect the conclusion (equation (\ref{eq:MQC_sig})) of the MQC circuit analysis to hold given that the optimization handles the creation of a GHZ state, which then allows the parallelized entangling gate to properly disentangle the system.

%=====================================================
\subsection{Circuit optimization}\label{sec:mqc_opt}
%=====================================================

As discussed in the main text and the previous section, in multinuclear implementations of the MQC method, the local operations before entanglement were optimized to maximize GHZ state fidelity using simulations. 
This optimization was performed using SciPy's dual annealing algorithm with equation (\ref{eq:GHZ_state}) defining the ideal target state. 
The exact state fidelity was computed using the QuTiP software package according to:
\begin{equation}
    F_s = \bigg(\trace\sqrt{\sqrt{\rho_1}\rho_2\sqrt{\rho_1}} \bigg)^2
\end{equation}
for two density matrices $\rho_{1,2}$ being compared.
During optimization, each nuclear qubit's rotation axis was set to $X$ or $Z$ by fixing the unit pulse times, and the total rotational angles were treated as free parameters set by the number of unit pulse repeats, as depicted in Fig. 3 of the main text. 
Using a targeted $X$ and $Z$ gate for each nuclear qubit allowed each qubit to be rotated anywhere in its Bloch sphere, with the annealing bounds set to rotations of $2\pi$ for each gate.
Dual annealing was chosen due to the highly non-convex loss landscape with many local optima, in addition to the fact that nuclear rotation angles can only change in discrete steps.
The optimization results that were used in each simulation and experiment are listed in Tables~\ref{tab:L3_opt_results}, \ref{tab:L2_opt_results}, and~\ref{tab:L2_res_opt_results}.
The unit pulse times used to set the rotation axis of each gate are the same as in Table \ref{tab:NG calib}.

\begin{figure}[h!]
\centering
\includegraphics{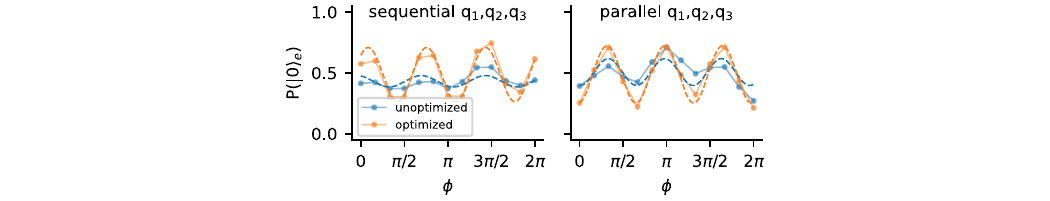}
\caption{
\textbf{Four-qubit entanglement optimization MQC comparison} 
Dashed lines are fits to simulations for visualization purposes.
A resolution of $\pi/6$ was used for the parallel phase gate here, as in experiments.
}
\label{fig:MQC_L3_opt}
\end{figure}

\begin{figure}[h!]\label{tab:L2_res_opt_results}
\centering
\includegraphics{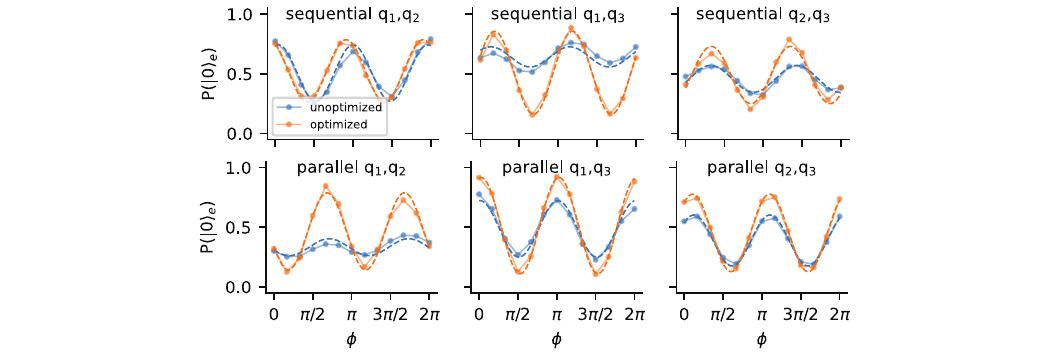}
\caption{
\textbf{Three-qubit entanglement optimization MQC comparison} 
}
\label{fig:MQC_L2_opt}
\end{figure}

\begin{figure}[h!]
\centering
\includegraphics{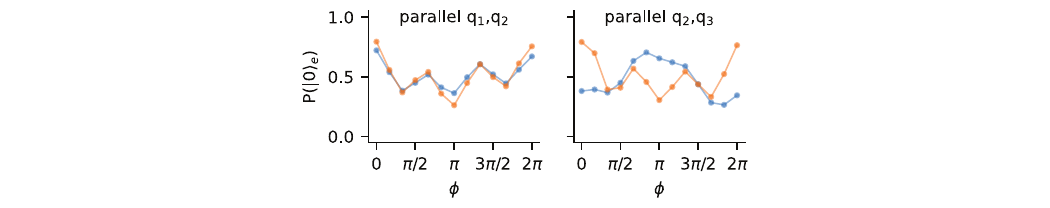}
\caption{
\textbf{Three-qubit entanglement, with residual entanglement, optimization MQC comparison} 
See Sec. \ref{sec:residual} for an explanation of the oscillations.
}
\label{fig:MQC_L2_residual_opt}
\end{figure}

Even in cases where a subset of nuclear qubits were not targeted for entanglement, their effects were still included in the simulated dynamics (including the fourth strongest nuclear qubit observed from Sec. \ref{sec:DD_spec}).
This allows any residual entanglement that was generated to manifest itself in the MQC signals.
Overall, the only deviation from the expected MQC behavior appeared when using parallel entangling gates that intentionally generated residual entanglement with nuclear qubits in mixed states (Fig. \ref{fig:MQC_L2_residual_opt}). 
These gates produced a beating pattern that is explained in the following section Sec. \ref{sec:residual}.
% This inclusion proved particularly helpful when interpreting some of the $L=2$ simulation and experimental results, while the overall dimensionality of the density matrices remained computationally tractable.
% See Sec. \ref{sec:residual} for an interpretation of the beating oscillations of the parallel $L=2$ simulation and experimental results.
When using the optimal parallel entangling gates that minimize residual entanglement (Fig. \ref{fig:MQC_L2_opt} and \ref{fig:MQC_L3_opt}), comparisons against the unoptimized rotations show drastic improvements in the MQC frequency mapped onto the electron as well as improvements in contrast, ultimately making this method experimentally viable.

%===================================================
% L=3 MQC opt
%===================================================

\begin{table}[h!]\label{tab:L3_opt_results}
\centering
\begin{tabular}{c|c|c|c|c}
\multicolumn{3}{c|}{Parallel} & \multicolumn{2}{c}{Sequential} \\
\hline
$q_{\ell}$ & $N_x$ & $N_z$ & $N_x$ & $N_z$ \\
\hline
$q_1$ & 10 & 6 & 8  & 6  \\
\hline
$q_2$ & 6  & 8 & 6  & 10 \\
\hline
$q_3$ & 14 & 1 & 32 & 4  \\
\end{tabular}
\caption{
\textbf{Four-qubit entanglement optimization results}
}
\end{table}

%===================================================
% L=2 MQC opt
%===================================================

\begin{table}[h!]\label{tab:L2_opt_results}
\centering
\begin{tabular}{c|c|c|c|c|c|c|c|c|c|c|c|c|c|c|c|c|c}
\multicolumn{9}{c|}{Parallel} & \multicolumn{9}{c}{Sequential} \\
\hline
$q_{\ell}$ & $N_x$ & $N_z$
& $q_{\ell}$ & $N_x$ & $N_z$
& $q_{\ell}$ & $N_x$ & $N_z$
& $q_{\ell}$ & $N_x$ & $N_z$
& $q_{\ell}$ & $N_x$ & $N_z$
& $q_{\ell}$ & $N_x$ & $N_z$ \\
\hline
$q_1$
& 22 & 4
& $q_1$ & 6 & 0
& $q_2$ & 9 & 4
& $q_1$ & 8 & 7
& $q_1$ & 8 & 11
& $q_2$ & 9 & 7 \\
\hline
$q_2$
& 8 & 3
& $q_3$ & 12 & 2
& $q_3$ & 9 & 3
& $q_2$ & 9 & 3
& $q_3$ & 14 & 3
& $q_3$ & 12 & 6 \\
\end{tabular}
\caption{
\textbf{Three-qubit entanglement optimization results}
}
\end{table}

%===================================================
% L=2 MQC opt with residual entanglement
%===================================================

\begin{table}[h!]
\centering
\begin{tabular}{c|c|c|c|c|c}
\multicolumn{6}{c}{Parallel} \\
\hline
$q_{\ell}$ & $N_x$ & $N_z$ & $q_{\ell}$ & $N_x$ & $N_z$ \\
\hline
$q_1$ & 8 & 0 & $q_2$ & 9 & 6 \\
\hline
$q_2$ & 9 & 5 & $q_3$ & 12 & 1 \\
\end{tabular}
\caption{
\textbf{Three-qubit entanglement, with residual entanglement, optimization results}
}
\end{table}

%=====================================================
\subsection{Residual entanglement effects on MQC signal}\label{sec:residual}
%=====================================================

In this section we prove that entanglement generated with mixed states effects the expected frequency measured in the MQC signal, as displayed in the $L=2$ simulations (Fig. \ref{fig:MQC_L2_residual_opt}). 
We begin this analysis similar to the idealized MQC case, but now include an additional mixed state nuclear qubit in the system:
\begin{equation}
    \rho_0 = \ket{0}\bra{0}^{\otimes L+1} \otimes \frac{1}{2}\mathbbm{1}.
\end{equation}
Assume no gate crosstalk between the initialized spins for simplicity, but track the crosstalk effects on the mixed state. 
The net preparation gate $U_p$ on the register is therefore given by:
\begin{equation}
    U_p = Y_{\pi/2} \otimes X_{\pi/2}^{\otimes L} \otimes U_m,
\end{equation}
where $U_m$ is unconditional crosstalk on the mixed state. Therefore:
\begin{equation}
\begin{split}
    \rho_1 &= U_p \rho_0 U_p^{\dagger}\\
    &= \ket{X}\bra{X} \otimes \ket{-Y}\bra{-Y}^{\otimes L} \otimes \frac{1}{2} U_m U_m^{\dagger} \\
    &= \ket{X}\bra{X} \otimes \ket{-Y}\bra{-Y}^{\otimes L} \otimes \frac{1}{2} \mathbbm{1}.
\end{split}
\end{equation}
However, the crosstalk during the parallel entangling operation is considerably (although not maximally) conditional on the electron state. 
Let the conditional rotation operators for the crosstalk on the mixed state be $R_0(\theta)$ and $R_1(\theta)$ giving a full operator of:
\begin{equation}
    U_E = \ket{0}\bra{0} \otimes X_{\pi/2}^{\otimes L} \otimes R_0(\theta) + \ket{1}\bra{1} \otimes X_{-\pi/2}^{\otimes L} \otimes R_1(\theta),
\end{equation}
which is the same as equation (\ref{eq:ent_gate}), but with conditional crosstalk considered. 
Applying this, we have:
\begin{equation}
\begin{split}
    \rho_E &= U_E \rho_1 U_E^{\dagger} \\ 
    &= \frac{1}{4}\bigg(
    \ket{0}\bra{0} \otimes \ket{1}\bra{1}^{\otimes L} \otimes \mathbbm{1} +
    \ket{0}\bra{1} \otimes e^{-i\alpha_L} \ket{1}\bra{0}^{\otimes L} \otimes R_0 R_1^{\dagger} \\
    &+
    \ket{1}\bra{0} \otimes e^{i\alpha_L} \ket{0}\bra{1}^{\otimes L} \otimes R_1 R_0^{\dagger} +
    \ket{1}\bra{1} \otimes \ket{0}\bra{0}^{\otimes L} \otimes \mathbbm{1} 
    \bigg).
\end{split}
\end{equation}
As discussed in Sec. \ref{sec:parallel_phase}, the MQC phase gate applied to the entangled state is naturally parallelized, which includes effects on the mixed state. 
Now that the system is entangled with the mixed state, the unconditional crosstalk effects of the phase gate no longer vanish like in $\rho_1$.
Therefore, with the form of the phase gate given by:
\begin{equation}
    Z_L(\phi) = \mathbbm{1}_e \otimes Z_{\phi}^{\otimes L+1},
\end{equation}
we have:
\begin{equation}
\begin{split}
    \rho_{E,\phi} &= Z_L(\phi) \rho_E Z_L^{\dagger}(\phi) \\ 
    &= \frac{1}{4}\bigg(
    \ket{0}\bra{0} \otimes \ket{1}\bra{1}^{\otimes L} \otimes \mathbbm{1} +
    \ket{0}\bra{1} \otimes e^{i(L\phi - \alpha_L)} \ket{1}\bra{0}^{\otimes L} \otimes Z_{\phi} R_0 R_1^{\dagger} Z_{\phi}^{\dagger} \\
    &+
    \ket{1}\bra{0} \otimes e^{-i(L\phi - \alpha_L)} \ket{0}\bra{1}^{\otimes L} \otimes Z_{\phi} R_1 R_0^{\dagger} Z_{\phi}^{\dagger} +
    \ket{1}\bra{1} \otimes \ket{0}\bra{0}^{\otimes L} \otimes \mathbbm{1} 
    \bigg).
\end{split}
\end{equation}
Now disentangle the system with the same entangling operator as before like usual:
\begin{equation}
\begin{split}
    \rho_{\text{dis}}  
    &= \frac{1}{4}\bigg(
    \ket{0}\bra{0} \otimes \ket{Y}\bra{Y}^{\otimes L} \otimes \mathbbm{1} +
    \ket{0}\bra{1} \otimes e^{i(L\phi - 2\alpha_L)} \ket{Y}\bra{Y}^{\otimes L} \otimes R_0 Z_{\phi} R_0 R_1^{\dagger} Z_{\phi}^{\dagger} R_1^{\dagger} \\
    &+
    \ket{1}\bra{0} \otimes e^{-i(L\phi - 2\alpha_L)} \ket{Y}\bra{Y}^{\otimes L} \otimes R_1 Z_{\phi} R_1 R_0^{\dagger} Z_{\phi}^{\dagger} R_0^{\dagger} +
    \ket{1}\bra{1} \otimes \ket{Y}\bra{Y}^{\otimes L} \otimes \mathbbm{1} 
    \bigg).
\end{split}
\end{equation}
Let the $\phi$ dependent operator appearing in the mixed state component be $A(\phi,\theta)=R_0 Z_{\phi} R_0 R_1^{\dagger} Z_{\phi}^{\dagger} R_1^{\dagger}$. 
Tracing out all nuclear qubits for just the electron state then gives:
\begin{equation}
\begin{split}
    \rho_{e} &= \trace_{\ell} \rho_{\text{dis}} \\ 
    &= \frac{1}{2}\bigg( \ket{0}\bra{0} 
    + \ket{1}\bra{1}
    + \ket{0}\bra{1} e^{i(L\phi - 2\alpha_L)} \trace[A(\phi,\theta) ]
    + \text{h.c.} \bigg).
\end{split}
\end{equation}
Thus neglecting the last electronic rotation and assuming a measurement in the $\ket{X}$ basis, we have:
\begin{equation}\label{eq:MQC_mixed_signal}
    P_e(\ket{X}) = \frac{1}{2}\bigg(1 + \frac{1}{2}\text{Re} \big( e^{i(L\phi - 2\alpha_L)} \trace[A(\phi,\theta)] \big) \bigg),
\end{equation}
which in the limit of no entangling crosstalk, $A=\mathbbm{1}$, returns equation (\ref{eq:MQC_sig}) as expected. 

\begin{figure}[h!]
\centering
\includegraphics{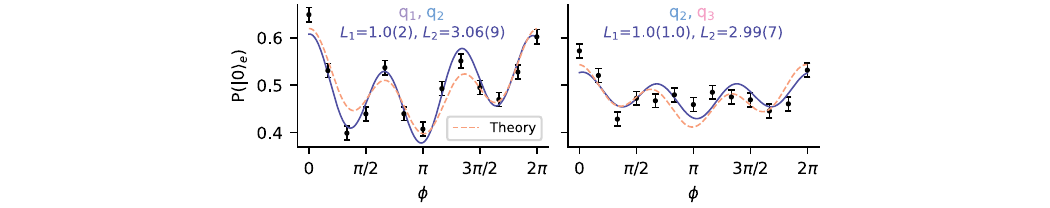}
\caption{
\textbf{Residual entanglement generated with $L=2$ parallel gates}
Theory curve corresponds to equation (\ref{eq:MQC_mixed_signal}) based on the parallel entangling gate parameters used from table \ref{tab:PE_calib}.
Only the contrast and vertical shift of the theory were adjusted to fit the data sets.
In the case of targeting entanglement with nuclear qubits $q_2$ and $q_3$, the lower contrast and only one full oscillation resulted in a large uncertainty on $L_1$.
More data could be taken to improve this in the future, but overall the fit (solid blue curve) matches the data well enough.
}
\label{fig:res_ent}
\end{figure}

Now let us assume the a generic form of the conditional crosstalk rotation operators $R_j(\theta)$.
Let the rotation axes be $\hat{n}_0 = (x_0, 0, z_0)$ and $\hat{n}_1 = (x_1, 0, z_1)$ and calculate $\trace[A(\phi,\theta)]$:
\begin{equation}\label{eq:tr_complicated_transformed}
\begin{split}
\trace[A(\phi,\theta)] = 
2 \cos^4 (\theta/2) 
+ 2 \big( (x_0 x_1 + z_0 z_1)^2 + (z_0 x_1 - x_0 z_1)^2 \cos(\phi) \big) 
  \sin^4(\theta/2)  \\
- \frac{1}{2} \big( z_0^2 - 2 x_0 x_1 + 4 z_0 z_1 + z_1^2 + (x_0 - x_1)^2 \cos(\phi) \big) 
  \sin^2(\theta) \\
+ 2 (x_0 x_1 + z_0 z_1)(x_0 z_1 - z_0 x_1) 
  \sin^2(\theta/2)  \sin(\theta) \sin(\phi).
\end{split}
\end{equation}
Combining this expression with $e^{iL\phi}$ in equation (\ref{eq:MQC_mixed_signal}) will result in sinusoidal terms with frequencies of the sum and difference between $L\phi$ and $\phi$; $(L\pm1)\phi$. 
Thus, in the case of $L=2$, we can expected two frequency components of $L_1=1$ and $L_2=3$. 
Together, this analysis provides a theoretical expression to compare simulation and data against, as well as motivation for the use of a two tone fit with frequencies $L_1$ and $L_2$ as the expected outcome (Fig. \ref{fig:res_ent}). 

As mentioned in Sec. \ref{sec:Parallel entanglement}, when targeting entanglement with two nuclear qubits, the next strongest is expected to have the largest residual entanglement generated. 
In the case of targeting $q_1$ and $q_2$, nuclear qubit $q_3$ acts as the mixed state and when targeting, $q_2$ and $q_3$, $q_1$ acts as the mixed state.
Given this and precise knowledge of the hyperfine parameters of each residual nuclear qubit, the crosstalk rotation operators $R_j(\theta)$ can be fixed for each parallel gate unit pulse time and repeats used.
This fully specifies equation (\ref{eq:MQC_sig}) (Theory curve of Fig. \ref{fig:res_ent}) as a function of only $\phi$ and thus can serve as a prediction for the beating pattern observed in simulation and data.
To better visualize this prediction, the contrast and vertical shift of equation (\ref{eq:MQC_sig}) were adjusted to match the data in each case.
The loss of contrast between the two experiments, and larger uncertainties in the fitting results can be accounted for by the fact that the unit pulse time used for targeting nuclear qubits $q_2$ and $q_3$ was farther into the spin bath region.
Therefore, additional small amounts of residual entanglement were likely generated with other weaker nuclear qubits that have yet to be measured using DD spectroscopy, or are full unresolvable, and were not accounted for in this analysis. 

%=====================================================
\subsection{Additional data}
%=====================================================

\begin{figure}[h!]
\centering
\includegraphics{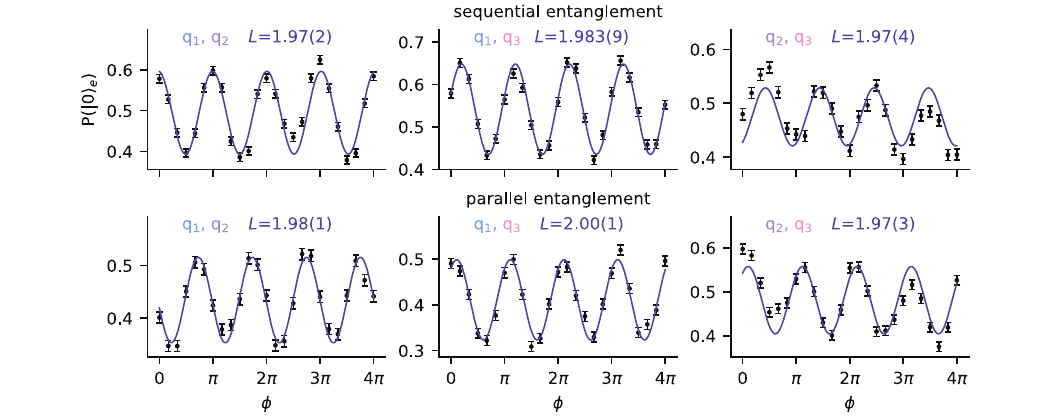}
\caption{
\textbf{Complete $L=2$ sequential and parallel entanglement data}
Entanglement with nuclear qubits $q_1$ and $q_2$ was shown in the main text from $\phi=$ 0 to $2\pi$.
}
\label{fig:L2_full_data}
\end{figure}

In the case of generating entanglement with two of the three possible nuclear qubits, there are three choices of which subset of qubits to target. 
The main text displayed sequential and parallel MQC results for entanglement with nuclear qubits $q_1$ and $q_2$ as an example, leaving pairs $q_1$, $q_3$ and $q_2$, $q_3$ left to consider (Fig. \ref{fig:L2_full_data}). 
% , while Sec. \ref{sec:residual} displayed results and theory for both the other choices of two nuclear qubits that could be maximally entangled in parallel. 
% Therefore, the only new pairs left to consider are sequential entanglement with $q_1$, $q_3$ and $q_2$, $q_3$ (Fig. \ref{fig:L2_full_data}).
Since additional data was taken out to $4\pi$ to improve frequency fitting results, data for nuclear qubits $q_1$, $q_2$ is again shown here in entirety.

%=====================================================
\section{Entangling gate fidelities}\label{sec:ent_gate_fid_sup} 
%=====================================================

%=====================================================
\subsection{State fidelity expression}\label{sec:fid_exp}
%=====================================================

The goal of this section is to derive an expression for the quantum state fidelity against $\ket{0}^{\otimes M}$ for an $M$ qubit system in terms of individual qubit measurements. 
Recall that $M=L+1$ where $L$ is the number of nuclear qubits in the system.
Ultimately, this approach will be an approximation given that correlator measurements need to be made to fully specify any entanglement present between qubits.
This point will be discussed at more length once an expression has been derived.

We begin by considering the trace inner product between the two density matrices, which serves as a measure of quantum state fidelity given by:
\begin{equation}
    F=\trace{(\rho_1 \rho_2)}.
\end{equation}
Since both $\rho_1$ and $\rho_2$ are Hermitian, this expression corresponds to their Frobenius inner product. 
Let $\rho_1=\rho$ be any density matrix and $\rho_2=(\ket{0}\bra{0})^{\otimes M}$. 
Since $\rho_2$ is a projector, we only need to consider a single matrix element of $\rho$:
\begin{equation}
\begin{split}
    F &= \trace{(\rho(\ket{0}\bra{0})^{\otimes M})} \\
      &= \trace(\bra{0}^{\otimes M} \rho \ket{0}^{\otimes M}) \\
      &= \bra{0}^{\otimes M} \rho \ket{0}^{\otimes M}.
\end{split}
\end{equation}
To proceed, we expand $\rho$ in terms of the $M$-qubit Pauli basis:
\begin{equation}
    \rho= \frac{1}{2^{M}} \sum_{i_1,...,i_M=0}^3 c_{i_1,...,i_M} \sigma_{i_{1}} \otimes \cdots \otimes \sigma_{i_{M}},
\end{equation}
where each coefficient is given by
 $c_{i_1,...,i_M} = \trace (\sigma_{i_{1}} \otimes \cdots \otimes \sigma_{i_{M}} \rho)$.
The matrix element of interest can be written as:
\begin{equation}
\begin{split}
    \bra{0}^{\otimes M} \rho \ket{0}^{\otimes M} &= \frac{1}{2^{M}} \sum_{i_1,...,i_M=0}^3 c_{i_1,...,i_M} \bra{0}\sigma_{i_{1}}\ket{0} \cdots \bra{0}\sigma_{i_{M}}\ket{0} \\ 
    &= \frac{1}{2^{M}} \sum_{i_1,...,i_M=0}^3 c_{i_1,...,i_M} \prod_{k=1}^M \bra{0}\sigma_{i_{k}}\ket{0}.
\end{split}
\end{equation}
Each $\bra{0}\sigma_{i_{k}}\ket{0}$ is only nonzero for the identity $\mathbbm{1}$ ($i=0$) and $Z$ ($i=3$), where each quantity is 1, this gives:
\begin{equation}
\label{eq:F_ex}
\begin{split}
    \bra{0}^{\otimes M} \rho \ket{0}^{\otimes M} &= \frac{1}{2^{M}} \sum_{i_1,...,i_M \in \{ 0,3\} } c_{i_1,...,i_M} \\
    &= \frac{1}{2^{M}} \sum_{i_1,...,i_M\in \{ 0,3\} } \trace (\sigma_{i_{1}} \otimes \cdots \otimes \sigma_{i_{M}} \rho).
\end{split}
\end{equation}
Hence, to measure this fidelity exactly, correlators involving different combinations of identity and $Z_m$ operators must be measured.
However, since electron-nuclear correlations cannot be measured without single-shot read out, correlator terms such as $\langle Z_e\otimes Z_1...Z_L\rangle$ have to be avoided to accommodate room temperature conditions. 
Similarly, swapping inter-nuclear correlations onto the electron is experimentally time consuming and would not provide useful information given the low readout fidelity found in simulation.
Given these constraints, we instead derive an approximation for the exact trace fidelity. 
To this end, we approximate that the state $\rho$ is separable: $\rho=\rho_e\otimes\rho_1\otimes...\otimes\rho_M$. 
The Pauli decomposition then takes a simpler form now with independent coefficients:
\begin{equation}
    \rho = \frac{1}{2^M}\bigotimes_{m=1}^M\sum_{i=0}^3\langle \sigma_{i_m} \rangle \sigma_{i_m}.
\end{equation}
Considering only the $\bra{0}^{\otimes M} \rho \ket{0}^{\otimes M}$ component for the fidelity expression, we have:
\begin{equation}
\label{eq:F_approx}
\begin{split}
    F &= \bra{0}^{\otimes M} \rho \ket{0}^{\otimes M} \\
    &= \frac{1}{2^M}\prod_{m=1}^M\sum_{i=0}^3\langle \sigma_{i_m} \rangle \langle0|\sigma_{i_m}|0\rangle \\
    &= \frac{1}{2^M}\prod_{m=1}^M (1+\langle{Z_m}\rangle),
\end{split}
\end{equation}
which means each qubit's $Z$ projection needs to be independently measured. 

The difference between the exact state fidelity and the approximate form is determined by correlations within the system. 
To motivate this, consider the bipartite case where $M=2$.
Let $F_{ex}$ be the exact expression of equation (\ref{eq:F_ex}) and $F_{a}$ be the approximate expression of equation (\ref{eq:F_approx}).
For this small system size, these are given by:
\begin{equation}
    F_{ex} = \frac{1}{2^2}(1 + \langle Z_e \rangle + \langle Z_1 \rangle + \langle Z_e \otimes Z_1 \rangle),
\end{equation}
and
\begin{equation}
    F_{a} = \frac{1}{2^2}(1 + \langle Z_e \rangle + \langle Z_1 \rangle + \langle Z_e \rangle \langle Z_1 \rangle).
\end{equation}
Thus, the difference is given by:
\begin{equation}\label{eq:L1_F_diff}
    F_{ex} - F_a = \langle Z_e \otimes Z_1 \rangle - \langle Z_e \rangle \langle Z_1 \rangle
\end{equation}
Therefore, $F_{a} \approx F_{ex}$ if the product of the independent measurements is approximately equal to the correlator measurement: $\langle Z_e \rangle \langle Z_1 \rangle \approx \langle Z_e \otimes Z_1 \rangle$.
This is true as long as the covariance between the individual observables is near zero.
The covariance is defined as:
\begin{equation}
\begin{split}
    \text{Cov}(Z_e \otimes \mathbbm{1},\mathbbm{1} \otimes Z_1) &= \text{E}[Z_e \otimes Z_1] - \text{E}[Z_e \otimes \mathbbm{1}]\text{E}[\mathbbm{1} \otimes Z_1] \\
    &= \langle Z_e \otimes Z_1 \rangle - \langle Z_e \rangle \langle Z_1 \rangle,
\end{split}
\end{equation}
hence if we assume a covariance near zero, then $\langle Z_e \rangle \langle Z_1 \rangle \approx \langle Z_e \otimes Z_1 \rangle$ as desired.
This is exactly equal when there are no correlations, classical or quantum, between the qubits, as expected from the derivation of equation (\ref{eq:F_approx}) using a separable state.
Since our goal is to prepare the state $\ket{00}$ up to the small errors present in the entangling gate, the covariance between observables remains small and the approximation is well justified.

In the case of multipartite entanglement, similar arguments can be made but become more complicated. 
In order to relate individual measurements to measurements of correlations, one can consider $(L+1)$th order cumulant expansions of observables.
To do so, one needs to consider each possible partition of the qubits in the system and analyze the correlations that exist between them.
For example, with $L=2$, the difference in state fidelity is given by
\begin{equation}\label{eq:L2_F_diff}
\begin{split}
    F_{ex} - F_a =& \langle Z_e \otimes Z_1 \rangle + \langle Z_e \otimes Z_2 \rangle + \langle Z_1 \otimes Z_2 \rangle + \langle Z_e \otimes Z_1 \otimes Z_2 \rangle \\
    &- \langle Z_e \rangle \langle Z_1 \rangle - \langle Z_e \rangle \langle Z_2 \rangle - \langle Z_1 \rangle \langle Z_2 \rangle - \langle Z_e \rangle \langle Z_1 \rangle \langle Z_2 \rangle \\
    =& \text{Cov}(Z_e,Z_1) + \text{Cov}(Z_e,Z_2) + \text{Cov}(Z_1,Z_2) + \langle Z_e \otimes Z_1 \otimes Z_2 \rangle - \langle Z_e \rangle \langle Z_1 \rangle \langle Z_2 \rangle.
\end{split}
\end{equation}
To relate $\langle Z_e \otimes Z_1 \otimes Z_2 \rangle $ and $\langle Z_e \rangle \langle Z_1 \rangle \langle Z_2 \rangle$, one can consider the third order cumulant given by:
\begin{equation}
    \text{Cum}(Z_e,Z_1,Z_2) = \langle Z_e \otimes Z_1 \otimes Z_2 \rangle - \left( \langle Z_e \rangle \langle Z_1 \otimes Z_2 \rangle + \langle Z_1 \rangle \langle Z_e \otimes Z_2 \rangle + \langle Z_2 \rangle \langle Z_e \otimes Z_1 \rangle \right) + 2 \langle Z_e \rangle \langle Z_1 \rangle \langle Z_2 \rangle,
\end{equation}
where rearranging gives:
\begin{equation}
\label{eq:L2_cumulant}
\begin{split}
    \langle Z_e \otimes Z_1 \otimes Z_2 \rangle - \langle Z_e \rangle \langle Z_1 \rangle \langle Z_2 \rangle  
    &= \text{Cum}(Z_e,Z_1,Z_2) + \big[ \langle Z_e \rangle \langle Z_1 \otimes Z_2 \rangle + \langle Z_1 \rangle \langle Z_e \otimes Z_2 \rangle + \langle Z_2 \rangle \langle Z_e \otimes Z_1 \rangle \big] \\
    &- 3\langle Z_e \rangle \langle Z_1 \rangle \langle Z_2 \rangle.
\end{split}
\end{equation}
Now recast each pairwise correlator $\langle Z_j \otimes Z_k \rangle$ in terms of covariance:
\begin{equation}
    \langle Z_j \otimes Z_k \rangle = \text{Cov}(Z_j,Z_k) + \langle Z_j \rangle \langle Z_k \rangle.
\end{equation}
Therefore:
\begin{equation}
    \langle Z_i \rangle \langle Z_j \otimes Z_k \rangle = \langle Z_i \rangle\text{Cov}(Z_j,Z_k) + \langle Z_i \rangle\langle Z_j \rangle \langle Z_k \rangle.
\end{equation}
Substituting this into equation (\ref{eq:L2_cumulant}) gives:
\begin{equation}
\begin{split}
    \langle Z_e \otimes Z_1 \otimes Z_2 \rangle - \langle Z_e \rangle \langle Z_1 \rangle \langle Z_2 \rangle &= \text{Cum}(Z_e,Z_1,Z_2)  + \big[ \langle Z_e \rangle \text{Cov}(Z_1,Z_2) + \langle Z_1 \rangle \text{Cov}(Z_e,Z_2) + \langle Z_2 \rangle \text{Cov}(Z_e,Z_1) \\
    &+ 3\langle Z_e \rangle \langle Z_1 \rangle \langle Z_2 \rangle\big] - 3\langle Z_e \rangle \langle Z_1 \rangle \langle Z_2 \rangle \\
    &= \text{Cum}(Z_e,Z_1,Z_2)  + \langle Z_e \rangle \text{Cov}(Z_1,Z_2) + \langle Z_1 \rangle \text{Cov}(Z_e,Z_2) + \langle Z_2 \rangle \text{Cov}(Z_e,Z_1).
\end{split}
\end{equation}
Hence, if each covariance and the three way cumulant are approximately zero, we have our desired outcome:
\begin{equation}
    \langle Z_e \otimes Z_1 \otimes Z_2 \rangle \approx \langle Z_e \rangle \langle Z_1 \rangle \langle Z_2 \rangle.
\end{equation}
Again, since the target state contains no correlations, if this is prepared up to small errors present in the system, then each covariance and the three way cumulant are approximately zero and the fidelity approximation holds well.

Proceeding to larger $L$ system sizes, the argument extends similarly, but requires that each $(L-1)$th order cumulant remain near zero. 
Ultimately, as long as we can prepare states close to $\ket{0}^{\otimes M}$ without inducing strong correlations in the system, we can expect the individual measurement state fidelity (equation (\ref{eq:F_approx})) to serve as an approximation for the exact state fidelity (equation (\ref{eq:F_ex})). 

%=====================================================
\subsection{Entangling gate repeat sampling}
%=====================================================

Now that we have an expression for the state fidelity in terms of independent qubit measurements that can readily be made experimentally, we need to determine which number of repeats of the entangling gates $N_E$ are expected to return the system to $\ket{0}^{\otimes M}$, up to the error associated with the gate.
This will reveal which number of repeats we should measure at and then use equation (\ref{eq:F_approx}) on the outcomes.
From these state fidelities, the decay of state fidelity will be fit to an noisy quantum channel model to extract the error per gate, or equivalently the entangling gate fidelity (more details in Sec. \ref{sec:Kraus})

%=====================================================
\subsubsection{Bipartite and sequential gates}
%=====================================================

\begin{figure}[h!]
\centering
\includegraphics{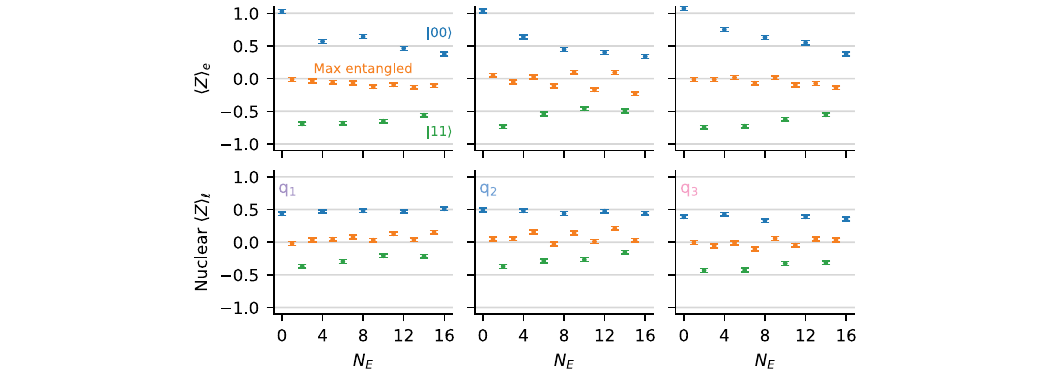}
\caption{
\textbf{Full bipartite entangling gate data}
Each plot is a single, distinct data set with the individual points color coded to highlight their meaning and role in the analysis.
Each column is a distinct bipartite entangling gate repeated $N_E$ times and each row corresponds to whether the electron or targeted nuclear qubit was measured.
Only the blue "$\ket{00}$" points where $N_E\equiv0(\text{mod4})$ are used in calculating each bipartite entangling gate fidelity in the main text.
}
\label{fig:bipartite_N_E_Zs}
\end{figure}

In the bipartite entanglement case, determining the number of entangling gate repetitions, $N_E$, required to return the initialized system to $\ket{00}$ is straightforward.
Since the gate geometry is $CX_{\pi/2}$, then multiples of four repeats create $CX_{2\pi k} = \mathbbm{1} \otimes \mathbbm{1}$ for $k\in\mathbbm{Z}$. 
More generally, odd values of $N_E$ produce maximally entangled states, while even values results in separable states. 
Although $N_E$ values of resolution one are not needed to calculate the entangling gate's fidelity, full data was taken to highlight the gate's geometry described above and is shown in Fig. \ref{fig:bipartite_N_E_Zs}. 
From these data sets, only $N_E$ values divisible by four were used to calculate the $\ket{00}$ state fidelity according to equation (\ref{eq:F_approx}) and are shown in the main text. 
When the pair of qubits is maximally entangled, each qubit's $Z$ projection is expected to be zero since each qubit is in a maximally mixed state when traced out (orange "Max entangled" points of Fig. \ref{fig:bipartite_N_E_Zs}).

Similarly, when using sequences of two-qubit $CX_{\pi/2}$ gates to generate multipartite entanglement, it is expected that $N_E$ values divisible by four should return the system to the state $\ket{0}^{\otimes M}$.
Sampling at these values compares the $M$-qubit sequential entangling gate against the ideal version of $CX_{\pi/2}^{\otimes L}$.
As discussed, crosstalk is a significant factor in preventing the sequential approach from behaving like this ideal gate with high fidelity. 
Thus using $N_E$ values divisible by four directly probe the crosstalk errors associated with it that do not affect the bipartite entangling cases.
% Given the sequential entangling gate's large duration of $\approx$150$\mu$s, at most $N_E=12$ repeats could be performed before the electron reached its $T_2$ decoherence limit.  

\begin{figure}[h!]
\centering
\includegraphics{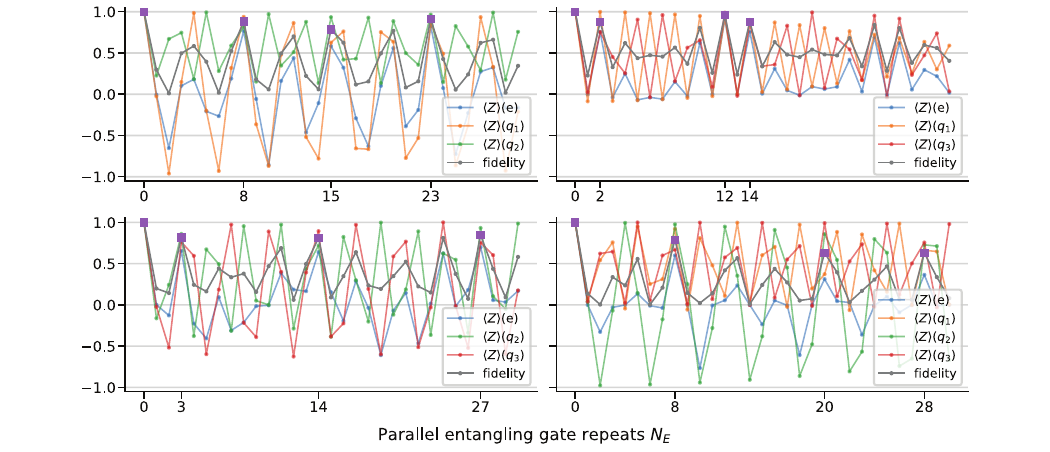}
\caption{
\textbf{Idealized parallel entangling gates repeats} 
$Z$ projections of each qubit, and exact state fidelity, from an ideal simulation of the $M=3$ and $M=4$ parallel entangling gates repeated $N_E$ times. 
The four values of $N_E$ that came closest to returning the state of the system to $\ket{0}^{\otimes M}$ (purple square markers) were chosen as points to experimentally sample at.
}
\label{fig:multipartite_N_E_Zs}
\end{figure}

%=====================================================
\subsubsection{Parallel gates}\label{sec:ideal_parallel_repeats}
%=====================================================

In the parallel entanglement case, determining the number of entangling gate repeats required to return the initialized system to $\ket{0}^{\otimes M}$ is more challenging given each gate's complicated geometry.
In order to determine these values, an ideal simulation of each parallel gate was used to examine the role of its geometry is isolation.
This simulation idealized register initialization, electron gate errors, measurement of correlators and only included the nuclear qubits there were targeted for entanglement (Fig. \ref{fig:multipartite_N_E_Zs}). 
From this, the four values of $N_E$ that came closet to return the system to $\ket{0}^{\otimes M}$ were chosen as sample points.
To account for the small difference in the exact states produced, the ideally produced density matrices at each $N_E$ were used an initial states in the fitting procedure described in the following section. 

\subsection{Kraus operator error model}\label{sec:Kraus}

In order to extract the entangling gate fidelity from the decay of the state fidelity measurements, a bit-flip quantum channel was developed and fit to the data.
In this quantum channel, the probability of an individual bit-flip caused by the entangling gate is given by $\varepsilon\in[0,1]$ and thus the probability of maintaining the bit without error is given by $1-\varepsilon$.
For $M$ qubits, there are $2^M$ total possible bit-flip combinations, each represented with their own Kraus operator $K_i$.
Thus, for $k$ bit-flips where $0 \leq k \leq M$, the probability amplitude of the corresponding Kraus operator is given by:
\begin{equation}\label{eq:Kraus_coeffs}
    c_{M,k}(\varepsilon) = \sqrt{\varepsilon^k (1-\varepsilon)^{M-k}}.
\end{equation}
This is the coefficient of the Kraus operator with $k$ Pauli $X$ operators, representing the bit-flips, tensored with $M-k$ identity operators. 
For $k$ bit-flips, there are $n_{M,k}=\binom{M}{k}$ many Kraus operators with $k$ Pauli $X$ operators present among the $M$ components of the Hilbert space. 
This fact is useful in proving the trace preservation of this channel $\sum_i K_i^{\dagger}K_i=I$, which relies on the sum of the squares of the probability amplitudes to be 1: 
\begin{equation}
\begin{split}
    \sum_{\text{Kraus ops}}^{2^M} c_{M,k}^2(\varepsilon) &= \sum_{k=0}^M n_{M,k}c_{M,k}^2(\varepsilon) \\
    &= \sum_{k=0}^M\binom{M}{k}\varepsilon^k (1-\varepsilon)^{M-k} \\
    & = (1-\varepsilon + \varepsilon)^M \\
    & = 1
\end{split}
\end{equation}
For example, in the $M=2$ bipartite case, the four Kraus operators are given by:
\begin{equation}
\begin{split}
    & K_{1} = \sqrt{(1-\varepsilon)^2}I\otimes I \\
    & K_{2} = \sqrt{\varepsilon(1-\varepsilon)}I\otimes X \\
    & K_{3} = \sqrt{\varepsilon(1-\varepsilon)}X\otimes I \\
    & K_{4} = \sqrt{\varepsilon^2} X\otimes X
\end{split}
\end{equation}
In all $M$ qubit cases, the squared coefficient of the Kraus operator containing only identities is what provides the gate fidelity measurement. 
Based on equation (\ref{eq:Kraus_coeffs}), the gate fidelity $G_M$ is therefore given by:
\begin{equation}\label{eq:gate_fid_error_rate}
\begin{split}
    G_M &= c^2_{M,0}(\varepsilon)\\
    &= (1-\varepsilon)^M.
\end{split}
\end{equation}
Thus, based on the least squares fitting of the gate's error rate $\varepsilon_{\text{gate}}$ of this quantum channel to the state fidelities measured, the entangling gate fidelities were calculated with equation (\ref{eq:gate_fid_error_rate}). 

To account for constant SPAM errors, one application of this quantum channel was applied to an initial of state with a free parameter error rate $\varepsilon_{\text{SPAM}}$.
This initial state was $\ket{0}^{\otimes M}$ for the bipartite and sequential entangling gates, and the ideally simulated states nearest $\ket{0}^{\otimes M}$ from Fig. \ref{fig:multipartite_N_E_Zs} for the parallel entangling gates.
The use of the ideally simulation states for the parallel gate allowed the channel to not penalize the gate for producing states that it was unable to due to its geometry. 
Then, using these imperfect initial density matrices, the quantum channel with a free parameter $\varepsilon_{\text{gate}}$ was applied $N_E$ times recursively for each entangling gate repeat sampled experimentally. 
Based on the model's final density matrix, the state fidelity against $\ket{0}^{\otimes M}$ was then calculated for each $N_E$ and fit to the state fidelities measured using least squares.

\begin{figure}[h!]
\centering
\includegraphics{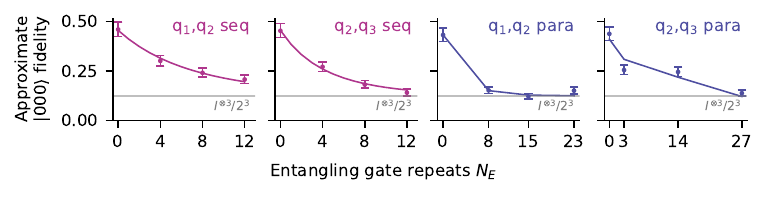}
\caption{
\textbf{Additional $L=2$ entangling gate fidelity data} 
Solid lines denote fits to noisy quantum channel described in Sec. \ref{sec:Kraus}.
}
\label{fig:L2_ent_gate_fid_data}
\end{figure}

The only data sets not shown in the main text were the other $L=2$ nuclear qubit combinations of $q_1,q_2$ and $q_2,q_3$ (Fig. \ref{fig:L2_ent_gate_fid_data})
Note, the sampling of the fits is larger (resolution of 1 repeat) for the sequential entangling gates since a constant initial density matrix of $\ket{000}$ was used. 
In the parallel case, since a unique ideal density matrix nearest $\ket{000}$ was taken from simulation for each $N_E$, a finer resolution could not be meaningfully used in displaying the fitting results. 
The entirety of the fitting results from the data sets in the main text and Fig. \ref{fig:L2_ent_gate_fid_data}, including the SPAM fidelity, is shown in Table \ref{tab:gate_fids}.

\begin{table}[h]
\label{tab:gate_fids} 
\centering
\begin{tabular}{c|c|c}
\textbf{Entangling gate} & \textbf{SPAM fidelity} & \textbf{Gate fidelity} \\
\hline
$q_0$ & $0.70 \pm 0.04$ & $0.972 \pm 0.008$ \\
$q_1$ & $0.73 \pm 0.03$ & $0.956 \pm 0.008$ \\
$q_2$ & $0.71 \pm 0.02$ & $0.963 \pm 0.004$ \\
\hline
$q_1,q_2$ sequential & $0.46 \pm 0.01$ & $0.86 \pm 0.01$ \\
$q_1,q_2$ parallel   & $0.43 \pm 0.02$ & $0.74 \pm 0.08$ \\
\hline
$q_1,q_3$ sequential & $0.48 \pm 0.03$ & $0.69 \pm 0.06$ \\
$q_1,q_3$ parallel   & $0.48 \pm 0.02$ & $0.80 \pm 0.04$ \\
\hline
$q_2,q_3$ sequential & $0.46 \pm 0.01$ & $0.79 \pm 0.02$ \\
$q_2,q_3$ parallel   & $0.41 \pm 0.04$ & $0.94 \pm 0.04$ \\
\hline
$q_1,q_2,q_3$ sequential & $0.31 \pm 0.01$ & $0.69 \pm 0.03$ \\
$q_1,q_2,q_3$ parallel   & $0.32 \pm 0.01$ & $0.92 \pm 0.04$ \\
\end{tabular}
\caption{
\textbf{Gate and SPAM fidelity fit results.}
The SPAM fidelity decreases for larger registers due to the combined effects of reduced nuclear initialization and measurement fidelity.
}
\end{table}

\begin{figure}[h!]
\centering
\includegraphics{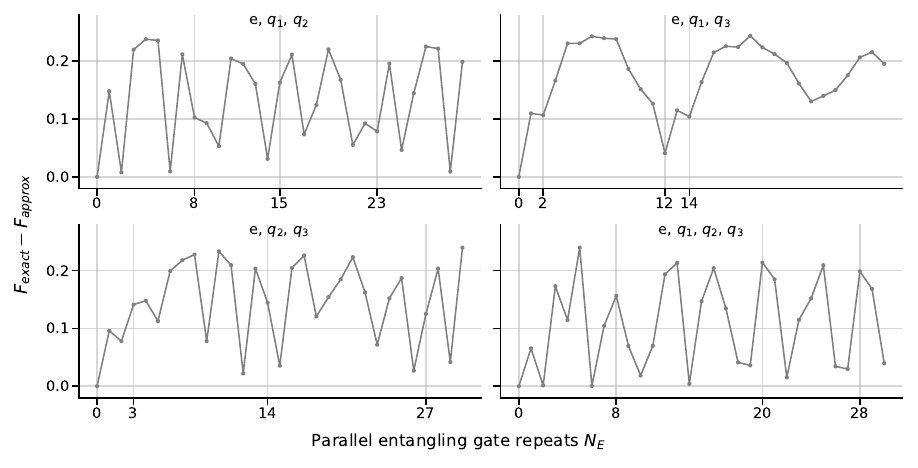}
\caption{
\textbf{Difference between the exact and approximate state fidelities.} 
$F_{exact}$ is calculated based on equation equation (\ref{eq:F_ex}) and $F_{approx}$ is calculated based on equation (\ref{eq:F_approx}). 
}
\label{fig:F_diff}
\end{figure}

To give some insight into the nature of using approximate state fidelities to calculate the gate fidelities, the state fidelities created by the parallel entangling gates were calculated using both the exact trace fidelity as in equation (\ref{eq:F_ex}) and the approximate fidelity in equation (\ref{eq:F_approx}).
The difference in fidelities depends on the extent to which correlations exist within the system and subsystems (see equations (\ref{eq:L1_F_diff}) and (\ref{eq:L2_F_diff}) for exact expressions).
Fig. \ref{fig:F_diff} reveals that the approximate state fidelity is always less than the exact, with the difference approximately growing larger as a function of $N_E$.
A similar behavior holds for the same behavior holding for the sequential entangling gates. 
Thus, the state fidelities calculated experimentally according to the approximate fidelity decrease faster as a function of $N_E$ than the exact values and therefore the quantum channel fit to the data produces a lower entangling gate fidelity. 
This observation suggests that the experimental entangling gate fidelities we obtain actually underestimate the exact values.

% To give some insight into the nature of using approximate state fidelities to calculate the gate fidelities, the parallel entangling gate fitting process was considering two ways. 
% One in which the state fidelities were calculated according to the exact trace fidelity as in equation (\ref{eq:F_ex}), and one in which they were calculating according to the approximate fidelity in equation (\ref{eq:F_approx}).
% This case in particular was examined further because the initial density matrices that the SPAM error channel was applied to contained small amounts of correlation between qubits that persisted through the fitting process.
% Since the approximate state fidelity neglects these, comparing the fitting outcomes is useful in determining how large of a difference the approximation makes. 
% For the $L=3$ parallel gate, the gate fidelity measured with the exact trace fidelity calculation is 0.92(4), while the approximate fidelity calculation is 0.93(4).
% Given the closeness of the two values, especially with respect to each uncertainty, we conclude that the approximate state fidelities provide an accurate determination of the gate fidelity. 
% The value from the exact calculation was quoted in the main text and table \ref{tab:gate_fids} to reflect the small role that correlations can play in using these methods.

\section{Applications to quantum error correction}

\begin{figure}[th!]
\centering
\includegraphics{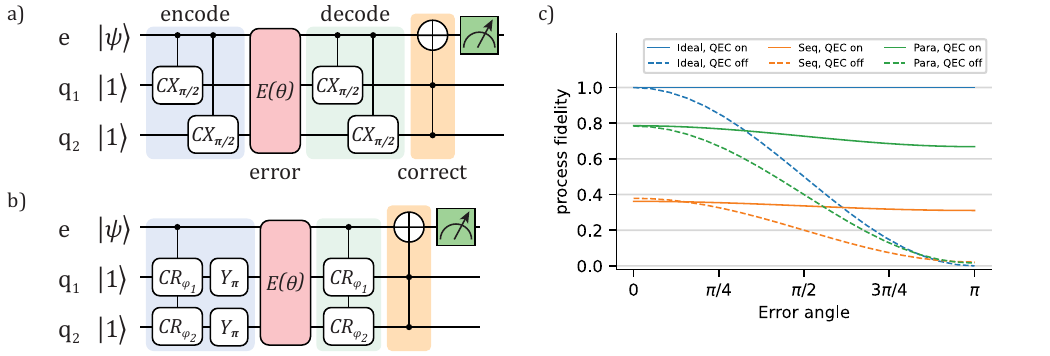}
\caption{
\textbf{Three-qubit bit-flip code} 
\textbf{(a)} bit-flip code using sequences of two-qubit gates.
\textbf{(b)} bit-flip code using parallel three-qubit gate. 
Local nuclear $Y_{\pi}$ gates are needed to compensate for entangling gate geometry.
\textbf{(c)} Process fidelity of correcting bit-flip error on the electron qubit. 
"QEC on" refers to the initial nuclear state $\ket{11}$, while "QEC off" refers to the initial nuclear state $\ket{01}$ (equivalently, $\ket{10}$). 
Since the code cannot correct two bit-flips, the flipped nuclear qubit deactivates the code.
}
\label{fig:QEC}
\end{figure} 

Following the quantum circuits of Takou \textit{et al.} \cite{takou2023precise} and the experimental techniques of Taminiau \textit{et al.} \cite{taminiau2014universal}, we simulate how the parallel entangling gates used in this work could improve quantum error correction protocols. 
In this example, a three-qubit bit-flip code is simulated that can detect and correct any single qubit bit-flip errors by majority voting.
The protocol is simulated two ways --- one using a sequence of two-qubit gates in the encoding and decoding steps (Fig. \ref{fig:QEC}a), and the other using a single parallel entangling gate (Fig. \ref{fig:QEC}b).
The correction is implemented using a doubly controlled not (Toffoli) gate, which is idealized in the following simulations to accentuate only the differences in the entangling gates used during encoding and decoding. 
The Toffoli gate can be implemented with a sequence of single and two-qubit DD gates, but is therefore subject to their crosstalk. 
Similarly, register initialization is idealized in all cases.
Single qubit errors are induced on the electron as a variable $X_{\theta}$ rotation.
The correction is turned off by flipping one of the nuclear qubits at the start of the circuit (keeping it in $\ket{0}$ at the start).
In the parallel case, additional local $Y_{\pi}$ gates are needed on the nuclear qubits to compensate for the complicated geometry of the parallel gate \cite{takou2023precise}, also idealized in the following simulation.
The electron is prepared in 6 initial states $\ket{\psi}=\ket{\pm X},\ket{\pm Y},\ket{\pm Z}$, from which the process fidelity as a function of error angle is calculated according to \cite{taminiau2014universal} (Fig. \ref{fig:QEC}c).
Notably, the parallel gate maintains approximately double the process fidelity than the sequential entangling gates over all error angles --- which is directly related to its higher fidelity.
With no bit-flip error on the electron, the parallel gate still does not reach unit process fidelity with idealized local nuclear gates due to residual entanglement generated by the next strongest nuclear qubit. 
Parallel gates with lower amounts of residual entanglement will behave even closer to the idealized blue curve of Fig. \ref{fig:QEC}c.
Improvements to such QEC circuits with the parallel gate can immediately benefit recent studies of hybrid entanglement between the electron, multiple $^{13}$C nuclear qubits, and photons at low temperatures for error corrected quantum networking applications \cite{chang2025hybrid}. 

While three-qubit bit (and phase) flip codes are useful for single qubit errors, they fail to correct multiple errors and are overall not fault-tolerant. 
With additional nuclear qubits, minimal fault-tolerant protocols can be achieved with parallel entangling gates using a similar approach as in Fig. \ref{fig:QEC}b.
The first fault-tolerant operation of a NV-nuclear qubit register was demonstrated using $L=5$ weakly coupled $^{13}$C nuclear qubits, as well as the $^{14}$N qubit as a flag qubit \cite{abobeih2022fault}.
The encoding and decoding steps to form the logical qubit required a sequence of two-qubit gates, dominating the circuit depth. 
Based on the simulations of Sec. \ref{sec:generality}, $L=5$ parallel entangling gates with properties similar to the ones demonstrated in this work are possible in a small, but reasonable, fraction of NV centers. 
Searching for such an NV center could drastically reduce circuit durations and errors, potentially enabling fault tolerance at room temperature for the first time. 

\bibliography{SM.bib}